\documentclass[aps,prd,superscriptaddress,nofootinbib,amsfonts,amssymb,amsmath,notitlepage,twocolumn,10pt,floatfix]{revtex4-1}

\usepackage[dvipdfmx]{graphicx}
\usepackage{adjustbox}
\usepackage{amsmath}
\usepackage{amssymb}
\usepackage{wrapfig}
\usepackage{bm}
\usepackage{color}
\usepackage{xcolor}
\usepackage{xfrac}
\usepackage{hyperref}
\usepackage[nameinlink]{cleveref}
\hypersetup{
	colorlinks,
	linkcolor={blue!75!black},
	citecolor={blue!75!black},
	urlcolor={blue!75!black}
}

\crefname{section}{Sec.}{Secs.}
\crefname{subsection}{Sec.}{Secs.}
\crefname{equation}{Eq.}{Eqs.}
\crefname{figure}{Fig.}{Figs.}
\crefname{table}{Tab.}{Tabs.}

\Crefname{section}{Sec.}{Secs.}
\Crefname{subsection}{Sec.}{Secs.}
\Crefname{equation}{Eq.}{Eqs.}
\Crefname{figure}{Fig.}{Figs.}
\Crefname{table}{Tab.}{Tabs.}

\allowdisplaybreaks[4]

\newcommand{\gsim}{\gtrsim}
\newcommand{\Slash}[1]{{\ooalign{\hfil/\hfil\crcr$#1$}}} 

\newcommand{\p}{\partial}
\newcommand{\df}{\text{d}}
\newcommand{\Tr}{{\rm Tr}\,}

\newcommand{\pmat}[1]{\begin{pmatrix}#1\end{pmatrix}}

\graphicspath{{./figs/}}
\newbox{\ORCIDicon}
\sbox{\ORCIDicon}{\large
                  \includegraphics[width=0.8em]{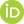}}

\begin{document}

\title{
\texorpdfstring{$CP$}{} phase structure of QCD from functional renormalization group
}

\author{Yuepeng \surname{Guan}\,\href{https://orcid.org/0009-0007-8571-0931}{\usebox{\ORCIDicon}}}
\affiliation{Center for Theoretical Physics and College of Physics, Jilin University, Changchun 130012, China}
\author{Shinya \surname{Matsuzaki}\,\href{https://orcid.org/0000-0003-4531-0363}{\usebox{\ORCIDicon}}}
\affiliation{Center for Theoretical Physics and College of Physics, Jilin University, Changchun 130012, China}
\author{Masatoshi \surname{Yamada}\,\href{https://orcid.org/0000-0002-1013-8631}{\usebox{\ORCIDicon}}\,}
\affiliation{Department of Physics and Astronomy, Kwansei Gakuin University, Sanda, Hyogo, 669-1330, Japan}

\begin{abstract}
We study the low-energy properties of QCD-like theories in the presence of a $P$-odd and $U(1)$ axial breaking four-fermion operator $\left( \bar{\psi} \psi \right) \left( \bar{\psi} i \gamma_5 \psi \right)$. 
We apply the functional renormalization group for a low-energy effective theory involving the $CP$-violating operator. 
We find that allowing for the running gauge coupling, the $CP$-violating four-fermion interaction 
becomes relevant in the chirally broken phase.
In the presence of a finite quark mass, 
the RG running of the $\theta$-parameter is shown to be strongly suppressed toward the infrared. 
The present work clarifies how strong-$CP$ effects generated at UV can non-trivially be transferred to the infrared physics in QCD-like theories.
\end{abstract}
\maketitle

\section{Introduction}
\label{sec:introduction}

In quantum chromodynamics (QCD), the size of the charge conjugation ($C$) and parity ($P$) violation induced from the $\bar \theta$-parameter is permitted, in principle, to be of the order of one.
However, experimental bounds on the neutron electric dipole moment \cite{Abel:2020pzs, Liang:2023jfj} suggests a tiny value, $\bar \theta < 10^{-11}$. 
This mystery motivates us not only to call for beyond the Standard Model (SM), but also to explore yet-uncovered dynamical and theoretical aspects of QCD with $CP$ violation. 

On theoretical grounds, the physical $\bar \theta$-parameter is constructed from the original $\theta$-parameter associated with the QCD topological vacuum and 
the phase of quark masses: $\bar \theta = \theta + {\rm arg}\det (M)$ where $M$ denotes the quark mass matrix. 
The former is believed to be renormalization group (RG) invariant, while the latter generically is affected by 
the RG running effects. Hence, the total $\bar \theta$-parameter may receive RG corrections. 

In fact, the RG running can induce a shift in $\bar{\theta}$.
Within the SM effective field theory framework, the perturbative RG evolution of loop-induced $CP$- and $P$-violating operators has been studied \cite{Jenkins:2017dyc,Hisano:2012cc,deVries:2021sxz,Banno:2023yrd}. 
It has been clarified that the RG-induced shift in $\bar{\theta}$ remains small down to the intrinsic QCD scale $\sim 1$ GeV, which is comparable to the neutron mass scale. 
This can be understood as a consequence of the irrelevance of higher-dimensional operators in the sense of perturbative renormalizability.

However, this expectation may not hold at scales below $1$ GeV. Certain induced higher-dimensional operators, including four-fermion interactions, can play a crucial role in dynamical chiral symmetry breaking. Their effective couplings tend to grow and eventually diverge at a critical scale, signaling the onset of chiral symmetry breaking, often characterized as a second-order phase transition; see, e.g., Refs.~\cite{Aoki:1996fh,Aoki:2000dh,Aoki:2012mj}. In such a strongly interacting regime, perturbative approaches are no longer justified, and nonperturbative methods are required to reliably capture the RG evolution of $CP$-violating phases.

As a low-energy effective description of QCD, Nambu--Jona-Lasinio (NJL) type models provide a useful framework for studying the chiral dynamics in the infrared (IR) regime. 
Within such approaches, four-fermion operators play a central role as effective interactions describing the low-energy dynamics.
In particular, the flow of $C_4$, which is a coupling of the $P$-odd four-fermion operator $(\bar\psi\psi)(\bar\psi i\gamma_5\psi)$, is of special interest in the presence of $CP$ violation. 
In Ref.~\cite{Huang:2024ypj}, a low-energy effective theory involving such four-fermion interactions was studied, and its phase structure was analyzed by means of the functional renormalization group (fRG). 
It was shown that the $\bar \theta$-parameter and the coupling of the $P$-odd four-fermion operator become relevant in the chirally broken phase, as long as $U(1)_A$ is anomalous. 


In this paper, we extend the analysis of Ref.~\cite{Huang:2024ypj} by incorporating $SU(3)$ gauge interactions into the low-energy fermionic RG framework. 
We show that, once QCD gluon fluctuations are included, the $P$-odd four-fermion coupling $C_4$ acquires an additional gauge-induced contribution to its beta function and can become dynamically relevant along the RG flow. 
Although the running of the $\theta$ parameter itself is found to be numerically weak in the present truncation, it still leaves a nontrivial influence on changing the dominance of the four-fermion interactions.
The present work, therefore, provides a further step toward clarifying how strong-$CP$ effects generated at ultraviolet (UV) are transferred to the IR properties of QCD-like theories.

The paper is organized as follows.
In \Cref{sec:SpontaneousCPviolationWithThetaParameter}, we briefly introduce the low-energy description in the presence of the $CP$-breaking source within the SM.
In \Cref{sec:RenormalizationGroupStudy}, we summarize the formulations of the fRG as well as the approximation scheme, which is the main method applied in this paper.
Besides, we present the flow equations for the couplings in the system.
In \Cref{sec:NumericalAnalysis}, we perform numerical analysis to explore the phase structure and its RG flows.
\Cref{sec:Conclusions} is devoted to summarizing and discussing the results obtained in this work.
The detailed derivations of the flow equations are exhibited in \Cref{app:DerivationOfTheBetaFunctions}.

\section{General construction of CP-violation with \texorpdfstring{$\theta$}{}-parameter}
\label{sec:SpontaneousCPviolationWithThetaParameter}

In this section, we study the general construction of the $CP$ violation in QCD with the presence of the $\theta$-term, for which the classical Lagrangian at UV reads
\begin{align}
    \mathcal L =& -\frac{1}{4} F^a_{\mu\nu} F^{a\mu\nu} + \theta^g \frac{g_s^2}{64\pi^2} \epsilon^{\mu\nu\rho\sigma} F^a_{\mu\nu} F^a_{\rho\sigma}  \nonumber\\
    &+ \bar \psi \Big( i \gamma^\mu \partial_\mu + g_s\gamma^\mu A_\mu^a T^a - m e^{i\gamma_5 \theta^q/2} \Big) \psi \, ,
    \label{eq:UVActionOfQCD}
\end{align}
where $F^a_{\mu\nu} = \partial_\mu A_\nu^a - \partial_\nu A_\mu^a + g_s f^{abc} A_\mu^b A_\nu^c$ is the field strength of the gluon field $A_\mu^a$ ($a=1,\cdots, N_c^2-1$); $T^a$ are the generators of $SU(N_c)$ normalized as ${\rm tr}[T^a T^b]=\delta^{ab}/2$ and $f^{abc}$ is the structure constant with 
the normalization $f^{123}=1$; $\psi$ and $\bar \psi$ are the quark fields; 
$\gamma^\mu$ 
stands for the Dirac's gamma matrices in 4-dimensional Minkowski space;  
$m$ denotes the quark mass matrix; 
$\theta^g$ denotes the coefficient of the topological $\theta$-term in the gluon sector, while $\theta^q$ arises from the complex Yukawa phases in the electroweak sector.

In the generating functional, the fermionic fields can be redefined such that the topological $\theta$-term is canceled.
To see this point, consider the axial rotation of the quark fields
\begin{align}
    &\psi \rightarrow \psi^\prime = e^{i \alpha \gamma_5} \, \psi \,,&
    &\bar \psi \rightarrow \bar \psi^\prime = \bar \psi \, e^{i \alpha \gamma_5} \, .
    \label{eq:UoneAxialRotation}
\end{align}
%
%
%
This axial rotation is anomalous, which generates the shift 
for the Lagrangian as
\begin{align}
   & 
   - \bar \psi m e^{i\gamma_5 \theta^q/2} \psi + \frac{\theta^g}{64\pi^2} \epsilon^{\mu\nu\rho\sigma} F^a_{\mu\nu} F^a_{\rho\sigma}
    \nonumber\\[1ex]
    \rightarrow & 
 - \bar \psi' m e^{i\gamma_5 (\theta^q/2 + 2 \alpha)} \psi' 
    \nonumber\\
    &\qquad\qquad + \big(\theta^g - 2 N_f \alpha\big) \frac{g_s^2}{64\pi^2} \epsilon^{\mu\nu\rho\sigma} F^a_{\mu\nu} F^a_{\rho\sigma}.
\end{align}
%
Since this rotational degree of freedom $\alpha$ is redundant, the angle parameters ($\theta^g$ and $\theta^q$) are linearly dependent on each other, leaving only one physical parameter, denoted as $\theta$. 
One can always fix $\alpha$ so that the topological $\theta$-term of QCD is not presented in the Lagrangian, namely $\alpha=\theta^g/(2N_f)$. 
The resulting Lagrangian in the $(\psi^\prime,\bar \psi^\prime)$-basis is thus transformed from \cref{eq:UVActionOfQCD} into 
\begin{align}
    \mathcal L^\prime =& -\frac{1}{4} F^a_{\mu\nu} F^{a,\mu\nu}  \nonumber\\
    &+ \bar \psi^\prime \Big( i \gamma^\mu \partial_\mu + g_s\gamma^\mu A_\mu^a T^a - m e^{i\gamma_5 \theta/2} \Big) \psi^\prime \, ,
    \label{eq:UVActionOfQCDInPrimeBase}
\end{align}
where $\theta = \theta^q + 2 \theta^g/N_f$. Hereafter, we will drop the prime symbol from $\psi$s. 
%
Regarding \Cref{eq:UVActionOfQCDInPrimeBase} as the classical UV theory, the $U(1)_A$ symmetry, the $P$ and thus $CP$ symmetry are explicitly broken by a nonzero $\theta$.  
Consequently, $P$- and $U(1)_A$-violating operators may be generated during the RG evolution down to the energy scale around the QCD intrinsic scale $\Lambda_{\rm QCD}$.  

In this work, we focus on a one-flavor ($N_f=1$) four-fermion model coupled to a nonabelian gauge field. Extension to multi-flavor cases will be briefly addressed later. 
Up to the four-fermion interactions, the corresponding low-energy description then reads~\cite{Huang:2024ypj}\footnote{
The coupling notations follow the literature~\cite{Hisano:2012cc,Huang:2024ypj}.
}
\begin{align}
    \mathcal L_{\rm eff}&(k \gsim \Lambda_{\rm QCD}) \nonumber\\
    =& -\frac{1}{4} F^a_{\mu\nu} F^{a,\mu\nu} + \bar \psi \Big( i \gamma^\mu \partial_\mu - g_s\gamma^\mu A_\mu^a T^a - m e^{i\gamma_5 \theta/2} \Big) \psi \nonumber\\
    &+ \frac{G_S}{2} (\bar \psi \psi)^2 + \frac{G_P}{2} (\bar \psi i\gamma_5 \psi)^2 + \cdots \nonumber\\
    &+ C_4 (\bar \psi \psi) (\bar \psi i \gamma_5 \psi) + C_5 (\bar \psi \sigma^{\mu\nu} \psi) (\bar \psi i \sigma^{\mu\nu} \gamma_5 \psi) \nonumber\\
    &+ C_2^{\prime} \operatorname{tr} \big[ \bar \psi \psi \, \bar \psi i \gamma_5 \psi \big] + C_4^{\prime} \operatorname{tr} \big[ \bar \psi \sigma^{\mu\nu} \psi \, \bar \psi i\sigma^{\mu\nu} \gamma_5 \psi \big],
    \label{eq:LEFTforQCDfromNB}
\end{align}
where the trace in the last line acts in the color space.
In the current work, we study the interplay between the $CP$ phase, the gauge sector, and the $P$-violating four-fermion interactions in a minimal parameter subspace, namely $(\theta, \, g_s, \, C_4)$.
We will later give a brief discussion on the contributions from other $P$-odd operators carrying tensorial structures different from $C_4$, such as 
$C_5, C_2'$, and $C_4'$. 
%
%
%
%
Thus, in this parameter subspace, we may have the action at $k\gtrsim \Lambda_{\rm QCD}$:  
\begin{align}
    &S[\Phi]
    = \int_x \biggl[ -\frac{1}{4} F^a_{\mu\nu} F^{a,\mu\nu} - \frac{1}{2 \xi} (\partial^\mu A_\mu^a)^2 \nonumber\\
    &  + \bar{\psi} \biggl( i \gamma^\mu \partial_\mu - g_s\gamma^\mu A_\mu^a T^a - m e^{i\gamma_5 \theta/2} \biggl) \psi \nonumber\\
    & + \frac{G_S}{2} \left( \bar{\psi} \psi \right)^2 + \frac{G_P}{2} \left( \bar{\psi} i \gamma_5 \psi \right)^2 + C_4 \left( \bar{\psi} \psi \right) \left( \bar{\psi} i \gamma_5 \psi \right) \biggl]\,,
    \label{eq:UVQCDaction}
\end{align}
with $\int_x = \int \df^4 x$.
Here, $\Phi$ represents the superfield composed of
\begin{align}
    \Phi(p) = \pmat{
         \big[ A_\mu^a(p) \big]^{\rm T}, &
         \big[ \psi(p) \big]^{\rm T}, &
         \bar{\psi}(-p)
         }^{\rm T}.
\end{align}
The quark fields $\psi$ and $\bar\psi$ transform in the fundamental representation of the color group $SU(N_c)$.
Among the four-fermion interactions in \cref{eq:UVQCDaction}, the $G_S$ and $G_P$ coupling terms are $P$-even, whereas the mixed operator proportional to $C_4$ violates the $P$ and $CP$. 
Accordingly, the model is $CP$-symmetric when both $C_4$ and $\theta$ vanish.
Furthermore, if one imposes $G_S = G_P$ together with $C_4 = 0$, the theory exhibits the $U(1)_A$ symmetry.
In such a case, the phase $\theta$ becomes unphysical because it can be rotated away through a $U(1)_A$ transformation of the quark fields. 
Note that the set of ($G_S,\,G_P,\,C_4$) takes a closed form under the $U(1)_A$ transformation.

For a multi-flavor system, the four-fermion sector in \cref{eq:UVQCDaction} may be extended to a global $SU(N_f)_A$ invariant system.
For instance, the four-fermion interactions in the two-flavor system associated with $G_S$ and $G_P$ are extended to
\begin{align}
\label{eq:fourfermi2flavor}
&G_S \left[ (\bar{\psi}\psi)^2 + (\bar{\psi} i \gamma_5 \tau^a \psi)^2\right] \nonumber\\
&\quad + G_P \left[ (\bar{\psi} \tau^a \psi)^2 + (\bar{\psi} i \gamma_5 \psi)^2\right],
\end{align}  
where $\psi$ denotes an $SU(2)$-flavor doublet and $\tau^a$ ($a=1,2,3$) are the Pauli matrices.
We see here that the four-fermion operators \labelcref{eq:fourfermi2flavor} is invariant under the $SU(2)_A$ transformation ($\psi\to e^{i\tau^a\alpha^a \gamma_5}\psi$), while the $U(1)_A$ symmetry remains explicitly broken when $G_S-G_P\neq 0$. 
From this point of view, the difference between $G_S$ and $G_P$ can be interpreted as a phenomenological parametrization of $U(1)_A$ anomaly effects in effective QCD models. By contrast, we note that spontaneous breaking of the $SU(2)_A$ symmetry is triggered once the four-fermion coupling(s) become sufficiently strong, in close analogy with the NJL model, and does not rely on explicit $U(1)_A$ breaking (i.e., it does not require $G_S \neq G_P$). 

\section{Renormalization group study}
\label{sec:RenormalizationGroupStudy}

\subsection{Functional renormalization group and truncation scheme}

To study the model \labelcref{eq:UVQCDaction} in the presence of the $CP$-breaking sources, we employ the fRG method.
In this section, we briefly summarize the necessary ingredients in the formulation.
The detailed derivation of the $\beta$-functions is found in \Cref{app:DerivationOfTheBetaFunctions}.

A central object is the one-particle irreducible (1PI) effective average action
\begin{align}
    \Gamma_k[\Phi] = \sup_J \left( J \cdot \Phi - W_k[J] \right) - \Delta S_k[\Phi]\,,
    \label{eq:defineGammak}
\end{align}
where $W_k[J]$ denotes the Schwinger's functional defined as
\begin{align}
    e^{W_k[J]} = \int \mathcal{D}\Phi \, e^{-S_{\rm bare} - \Delta S_k + J \cdot \Phi}\,,
    \label{eq:IRSchwingerFunctional}
\end{align}
and $\Delta S_k[\Phi]$ is the regulator term quadratic in fields, $\Delta S_k = \int_x \Phi^T \cdot \mathcal{R}_k(\partial) \cdot \Phi$, which suppresses the propagation of the field with a lower momentum mode $|p|<k$, realizing the Wilsonian coarse-grainning process.
The regulator matrix $\mathcal{R}_k$ takes the following structure in the momentum space within the current superfield content
\begin{align}
    \mathcal{R}_k(p) = \pmat{
    R_k^A(p) & 0 & 0 \\
    0 & 0 & (R_k^{\psi}(p))^\text{T} \\
    0 & R_k^{\psi}(p) & 0
    }\,.
    \label{eq:regulatorMatrix}
\end{align}
The effective average action $\Gamma_k$ is thus defined at the renormalization scale $k\in [0,\Lambda]$ with an UV boundary scale $\Lambda$, and, by construnction, satisfies
\begin{align}
    \Gamma_k \rightarrow 
    \begin{cases}
        S & \quad k \rightarrow \Lambda= \infty\,, \\[1ex]
        \Gamma & \quad k \rightarrow 0\,,
    \end{cases}
\end{align}
where $\Gamma$ is the full 1PI effective action which contains all fluctuations in the duration $|p|\in [0,\Lambda]$.

The evolution of $\Gamma_k$ is goverend by the Wetterich equation~\cite{Wetterich:1992yh} (see also \cite{Ellwanger:1993mw, Morris:1993qb}),  
\begin{align}
    \partial_t \Gamma_k[\Phi] &= \frac{1}{2} \operatorname{STr} \left[ \left( \Gamma_k^{(2)} + \mathcal{R}_k \right)^{-1} \partial_t \mathcal{R}_k \right],
    \label{eq:WetterichEq}
\end{align}
where $\partial_t = k \partial_k$ is the derivative with respect to the dimensionless scale $t = \log(k/\Lambda)$.
In \Cref{eq:WetterichEq}, ``$\operatorname{STr}$'' denotes the supertrace acting on all field spaces and the momentum space, and $\Gamma_k^{(n)}$ stands for the $n$-point function obtained by the $n$-th functional derivative with respect to external fields.
In particular, $\Gamma_k^{(2)}$ is the full-inverse propagator (two-point function)
\begin{align}
    \Gamma_{k,I,J}^{(2)}(q,p) = \frac{\overset{\rightarrow}{\delta}}{\delta \Phi^T_I(-p)} \Gamma_k  \frac{\overset{\leftarrow}{\delta}}{\delta \Phi_J(q)},
    \label{eq:defineHessian}
\end{align}
where the indices $I$ and $J$ stand for labels of the superfield content.

Now, we assume the 1PI effective average action in Euclidean spacetime\footnote{
We employ the notation in Euclidean spacetime used in Ref.~\cite{Huang:2024ypj}.
} as
\begin{align}
    \Gamma_k&[\Phi] 
    \simeq \int_x \biggl[ \frac{Z_A}{4} F^a_{\mu\nu} F_{\mu\nu}^{a} + \frac{1}{2 \xi} (\partial_\mu A_\mu^a)^2 \nonumber\\
    & \quad + \bar{\psi} \biggl( Z_\psi\gamma_\mu \partial_\mu + i g_s\gamma_\mu A_\mu^a T^a + m e^{i\gamma_5 \theta/2} \biggl) \psi \nonumber\\
    & \quad - \frac{G_S}{2} \left( \bar{\psi} \psi \right)^2 - \frac{G_P}{2} \left( \bar{\psi} i \gamma_5 \psi \right)^2 - C_4 \left( \bar{\psi} \psi \right) \left( \bar{\psi} i \gamma_5 \psi \right) \biggl]\nonumber\\
    &\qquad
    + \Gamma_\text{gh},
    \label{eq:IRQCDaction}
\end{align}
where $\Gamma_\text{gh}$ is the ghost action and $\xi$ is the gauge-fixing parameter.
In the present work, we employ the local potential approximation (LPA), i.e., $ Z_A = Z_\psi =1$ for any scale, which corresponds to the leading order of the derivative expansion of the effective action and retains only the external-momentum-independent part. We truncate the effective action to include operators up to four-fermion couplings at the classical level.
The couplings ($m$, $\theta$, $G_S$, $G_P$, $C_4$, and $g_s$) are understood as scale-dependent quantities that run with the RG scale $k$.
We then investigate the flow properties of the effective action \labelcref{eq:IRQCDaction} by following the RG evolution of the set of couplings $\{\lambda_i\} = \{ m, \theta, G_S, G_P, C_4, g_s \}$.

For the effective average action \labelcref{eq:IRQCDaction}, we choose the regulator matrix \labelcref{eq:regulatorMatrix} as
\begin{align}
    \mathcal{R}_k(p) =  \pmat{  p^2r^A_k(p) \,\Pi^{1/\xi}_{\mu\nu}(p) & 0 & 0 \\ 0 & 0 &  i\Slash{p}^\text{T}r^\psi_k(p) \\ 0 &  i\Slash{p}r^\psi_k(p) & 0 },
    \label{eq:generalRegulator}
\end{align}
where we have defined the tensor structure
\begin{align}
    &\Pi^{1/\xi}_{\mu\nu}(p) = \Pi^{\perp}_{\mu\nu}(p) + \frac{1}{\xi}\Pi^{\parallel}_{\mu\nu}(p),
\end{align}
with the transverse and longitudinal projectors
\begin{align}
    &\Pi^{\perp}_{\mu\nu}(p) = \biggl( \delta_{\mu\nu} - \frac{p_\mu p_\nu}{p^2} \biggl),\\
    &\Pi^{\parallel}_{\mu\nu}(p) = \frac{p_\mu p_\nu}{p^2}.
\end{align}
For the regulators, we employ the Litim-type regulator functions~\cite{Litim:2001up}
\begin{align}
    & r^A_k(p) = \left( \frac{k^2}{p^2} - 1 \right) \Theta\left( k^2 - p^2 \right), \\
    & r^\psi_k(p) = \left( \sqrt{\frac{k^2}{p^2}} - 1 \right) \Theta\left( k^2 - p^2 \right).
    \label{eq:LitimRegulatorFunctions}
\end{align}
Here $\Theta(x)$ is the step function defined as 
\begin{align}
    \Theta(x) = \begin{cases}
        1, &  x \geq 0\,, \\[1ex]
        0, &  x < 0\,.
    \end{cases}
    \label{eq:Step}
\end{align}

Here we comment on the validity of the truncated effective average action \labelcref{eq:IRQCDaction}.
Other possible four-fermion interactions such as vector type $(\bar\psi \gamma_\mu\psi)^2$, $(\bar\psi \gamma_\mu \gamma^5\psi)^2$ and tensor type $(\bar\psi \sigma_{\mu\nu}\psi)^2$ also contribute to the fermionic dynamics at the same order in canonical dimension as the scalar and psudescalar four-fermion terms retained in the effective action \eqref{eq:IRQCDaction}.
From this perspective, an analysis based on a Fierz-complete operator basis is more comprehensive; see Refs.~\cite{Aoki:2009zza,Braun:2017srn,Braun:2018bik,Braun:2019aow}.
Nevertheless, these additional channels tend to still be irrelevant, even in the vicinity of a nontrivial fixed point~\cite{Aoki:1999dv,Braun:2017srn}, so that we omit them in the present work.
Likewise, higher-dimensional operators such as $(\bar\psi\psi)^3$ are expected to remain irrelevant in the IR, even after incorporating nonperturbative effects, owing to their strongly negative canonical dimensions.
Their strongly negative canonical mass dimensions make it implausible that anomalous dimensions could become large enough to change their relevance.
Consequently, such operators are expected not to play a significant role in the onset of dynamical chiral symmetry breaking.

\subsection{Flow equations}

Using the Wetterich equation \labelcref{eq:WetterichEq} for the truncated effective average action \labelcref{eq:IRQCDaction}, we obtain the flow equations for the couplings $\{\partial_t \lambda_i\}$ as a coupled system of ordinary differential equations.
The detailed derivations of the $\beta$-functions is presented in \Cref{app:DerivationOfTheBetaFunctions}.
We note here that we adopt the large-$N$ leading approximation for the four-fermion sector induced self-energy terms [see \Cref{eq:largeNApprox}], while we take into account the so-called ``nonladder'' contributions from gauge-field exchanges.
To study the fixed point structure and to explore the phase diagram in the system, we define the dimensionless couplings for the dimensionful couplings ($m$, $G_S$, $G_P$ and $C_4$) as
\begin{align}
    & \tilde m = k^{-1} m ,
    & \tilde G_S = k^2 G_S, \nonumber\\[2ex]
    & \tilde G_P = k^2 G_P,
    & \tilde C_4 = k^2 C_4.
\end{align}

The resulting flow equations are: 
\begin{widetext}
\begin{subequations}
    \label{eq:flowequationsinmaintext}
\begin{align}
    \partial_t \tilde{m} &= - \tilde{m} - 4 N_c \biggl[ (\tilde{G}_S + \tilde{G}_P) + 2 \tilde{C}_4 \sin{\theta} + (\tilde{G}_S - \tilde{G}_P)\cos{\theta} \biggl]\tilde{m} \tilde{\mathcal{M}}_{(2)} \nonumber\\
    &\quad\quad - 2(3+\xi) C_2 g_s^2 \tilde{m} \left( \tilde{\mathcal{M}}_{(2)} + \tilde{\mathcal{M}}_{(1)} \right), 
    \\
    \partial_t \theta &= 8 N_c \biggl[ (\tilde{G}_S - \tilde{G}_P) \sin{\theta} - 2 \tilde{C}_4 \cos{\theta} \biggl] \tilde{\mathcal{M}}_{(2)}, 
    \\
    \partial_t \tilde G_S &= 2 \tilde G_S - 8 N_c  \Big( (\tilde G_S^2 + \tilde C_4^2) (1 - \tilde m^2) - 2 \tilde m^2 (\tilde G_S^2 - \tilde C_4^2) \cos{\theta} - 4 \tilde m^2 \tilde G_S \tilde C_4 \sin{\theta} \Big) \tilde{\mathcal{I}}_{(3)} \nonumber\\
    &\quad\quad + \mathcal A_S(\tilde m, \theta) \ g_s^2 \ \tilde G_S + \mathcal B_S(\tilde m, \theta) \ g_s^2 \ \tilde C_4 + \mathcal C_S(\tilde m, \theta) \ g_s^4,
    \\
    \partial_t \tilde G_P &= 2 \tilde G_P - 8 N_c \Big( (\tilde G_P^2 + \tilde C_4^2) (1 - \tilde m^2) + 2 \tilde m^2 (\tilde G_P^2 - \tilde C_4^2) \cos{\theta} - 4 \tilde m^2 \tilde G_P \tilde C_4 \sin{\theta} \Big) \tilde{\mathcal{I}}_{(3)} \nonumber\\
    &\quad\quad + \mathcal A_P(\tilde m, \theta) \ g_s^2 \ \tilde G_P + \mathcal B_P(\tilde m, \theta) \ g_s^2 \ \tilde C_4 + \mathcal C_P(\tilde m, \theta) \ g_s^4,
    \\
    \partial_t \tilde C_4 &= 2 \tilde C_4 - 8 N_c \Big( (\tilde G_S + \tilde G_P) \tilde C_4 (1 - \tilde m^2) - 2 m^2 (\tilde G_S - \tilde G_P) \tilde C_4 \cos{\theta} - 2 \tilde m^2 (\tilde G_S \tilde G_P + \tilde C_4^2) \sin{\theta} \Big) \tilde{\mathcal{I}}_{(3)} \nonumber\\
    &\quad\quad + \mathcal A_4(\tilde m, \theta) \ g_s^2 \ \tilde C_4 + \mathcal B_4(\tilde m, \theta) \ g_s^2 \ (\tilde G_S + \tilde G_P) + \mathcal C_4(\tilde m, \theta) \ g_s^4,
    \label{eq:betaFunctionOfC4}
\end{align}
\end{subequations}
\end{widetext}
where the explicit forms of $\mathcal A$s, $\mathcal B$s, and $\mathcal C$s, and the threshold functions ($\tilde{\mathcal M}$s and $\tilde{\mathcal I}$s) are shown in \cref{eq:coefficientfunctions,eq:thresholdfunction1,eq:thresholdfunction2}.
We use the perturbative result for the flow equation of the gauge coupling $\beta_{g_s}$. See \Cref{sec:NumericalAnalysis} for details.

We see from \cref{eq:betaFunctionOfC4} that the flow of the $P$-odd four-fermion coupling $C_4$ receives a contribution proportional to $g_s^4$ with $\mathcal C_4(\tilde m,\theta)\propto \sin\theta$.
Thus, QCD dynamics induces a nontrivial RG flow of $C_4$ within the SM in the presence of a physical $\theta$-parameter. 
This implies that $P$-odd fermionic operators naturally arise as effective interactions even without introducing physics beyond the SM.

It is worth noting that the flow equation of the $\theta$-parameter is not affected directly by the gauge interaction.
Meanwhile, the functions $\mathcal B_4$ and $\mathcal C_4$ are proportional to $\sin \theta$, and thus $\mathcal B_4(\tilde m, 0) = \mathcal C_4(\tilde m, 2n\pi) = 0$ with $n\in \mathbb{Z}$, causing the flow of the $P$-violating coupling $\partial_t \tilde C_4 \propto \tilde C_4$.
These features arise from the fact that the gauge coupling preserves $P$ as well as $C$ symmetry in the low-energy description.
However, the interplay between the parameter spaces of QCD and the $CP$-violating sectors needs further justification, as has been discussed in part in Ref.~\cite{Huang:2024ypj}.
In the following subsections, we shall discuss this point in different aspects.

\section{Numerical analysis}
\label{sec:NumericalAnalysis}

\subsection{Phase diagrams}

In this section, we study the phase structure of the coupled system \labelcref{eq:flowequationsinmaintext}.
For simplicity, we momentarily restrict ourselves to the chiral limit with $\tilde m = 0$.
This situation is protected since the flow equation of the mass parameter is proportional to itself, reflecting the concept of technical naturalness~\cite{tHooft:1979rat, Wells:2013tta}, until the critical scale where the chiral susceptibility becomes divergent.
We discuss two scenarios regarding the different running behaviors of the gauge coupling. One is the fixed gauge-coupling case. Another is the QCD case, namely, running effects of the gauge coupling are taken into account.
Hereafter, we choose the Landau gauge, i.e., $\xi=0$.

\subsubsection{Fixed gauge-coupling case}
\label{sec:FixedGaugeCouplingCase}

\begin{figure*}
    \centering
    \includegraphics[width=0.3\linewidth]{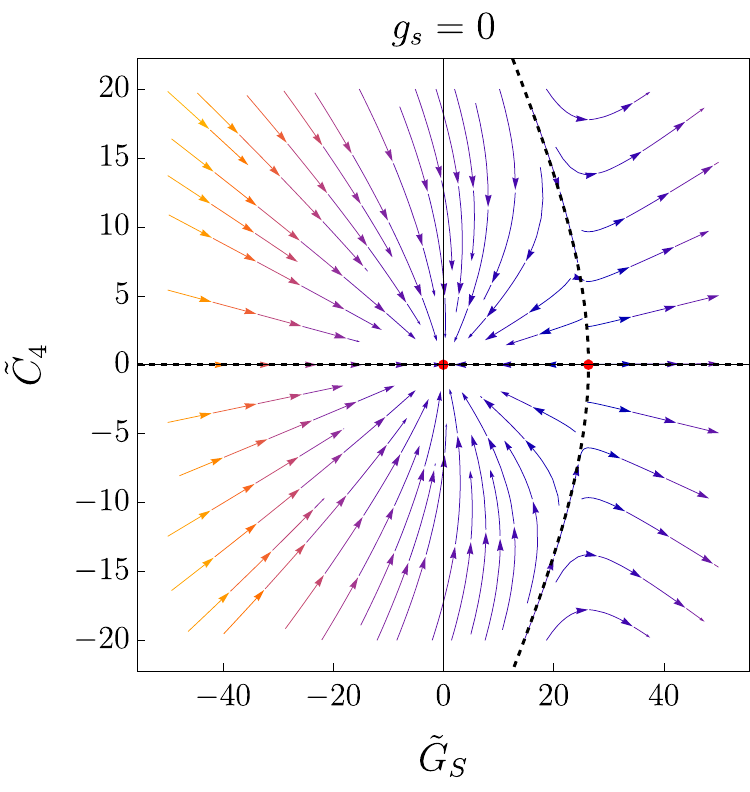}
    \hspace{5ex}
    \includegraphics[width=0.3\linewidth]{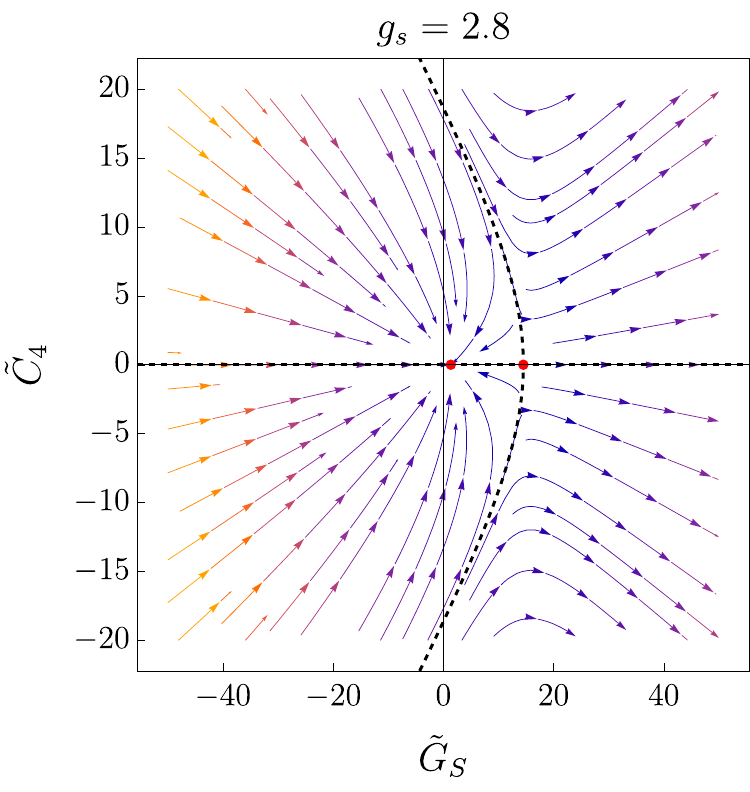}
    \hspace{5ex}
    \includegraphics[width=0.3\linewidth]{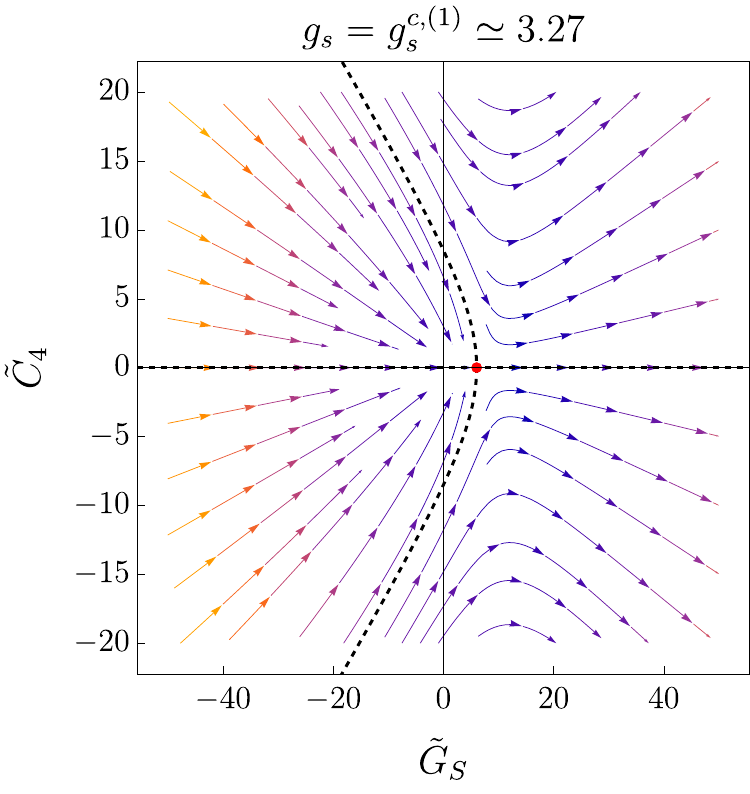}
    \\
    \includegraphics[width=0.3\linewidth]{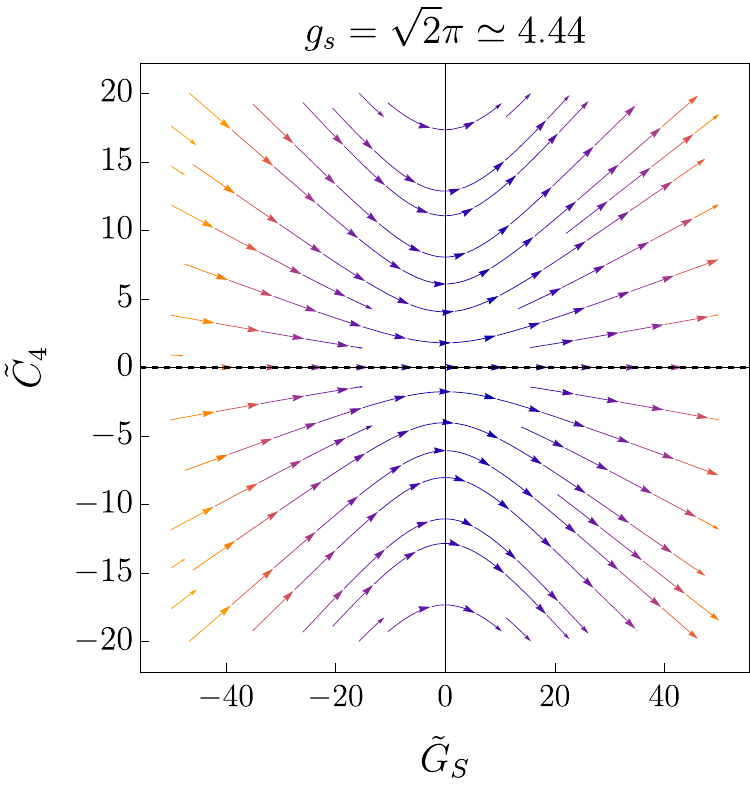}
    \hspace{5ex}
    \includegraphics[width=0.3\linewidth]{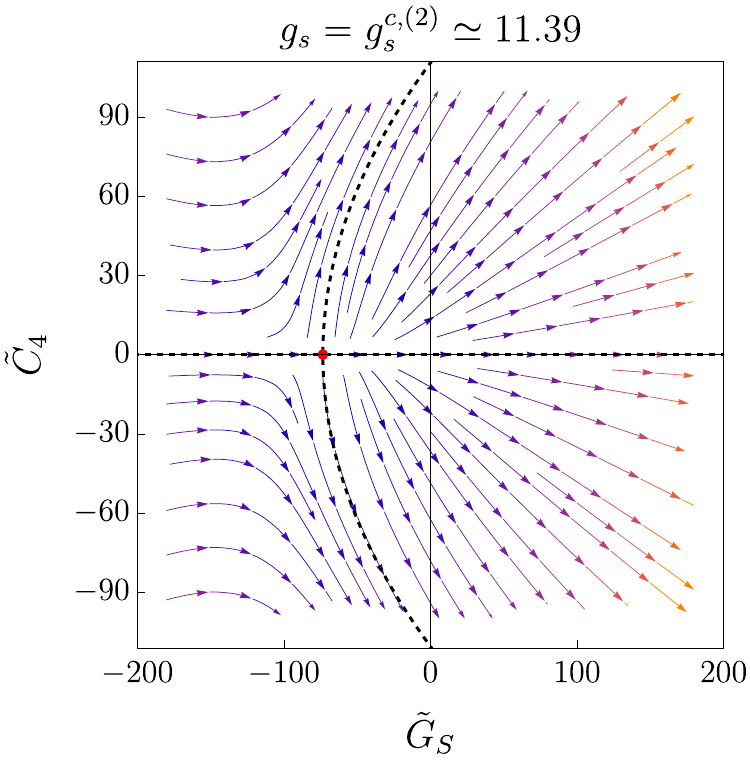}
    \hspace{5ex}
    \includegraphics[width=0.3\linewidth]{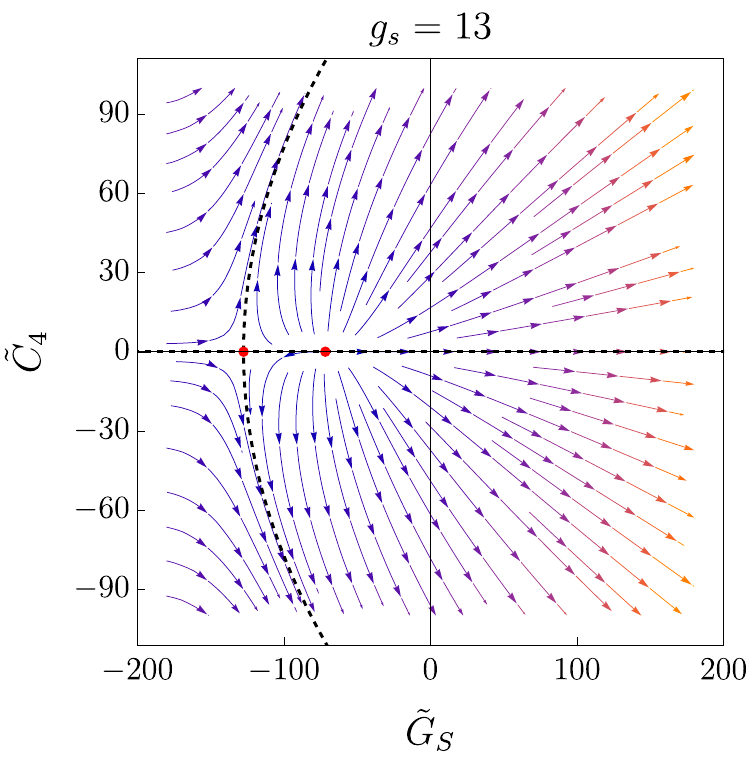}
    \caption{
    Phase diagrams on the $(\tilde G_S, \tilde C_4)$-plane in the chiral limit ($\tilde{m}=0$) with different values of the fixed gauge coupling $g_s$.
    We set $N_c = 3$, $N_f = 1$, and the flow equations are evaluated in the subspace where $\tilde m = \theta = \tilde G_P = 0$, i.e., \Cref{eq:reducedFlowsGaugedNJL1,eq:reducedFlowsGaugedNJL}.
    Red points exhibit the fixed points, and the dashed black line denotes the critical surface.
    }
    \label{fig:GaugedNJLPhaseDiagrams}
\end{figure*}

First, we consider the case where the running of the gauge coupling $\partial_t g_s$ is turned off: $\p_t g_s = 0$. 
In \Cref{fig:GaugedNJLPhaseDiagrams}, we show the phase diagrams on the $(\tilde G_S, \tilde C_4)$-plane with different fixed values of the gauge coupling $g_s$ and setting $\tilde m = \theta = \tilde G_P = 0$.
In this case, the flow equations of the couplings $\tilde G_S$ and $\tilde C_4$ are reduced to
\begin{align}
    \partial_t \tilde G_S &= 2 \tilde G_S - 8 N_c  \Big( \tilde G_S^2 + \tilde C_4^2\Big) \tilde{\mathcal{I}}_{(3)}(\tilde m = 0) \nonumber\\
    &\quad\quad + \mathcal A_S(0, 0) \ g_s^2 \ \tilde G_S + \mathcal C_S(0, 0) \ g_s^4,
    \label{eq:reducedFlowsGaugedNJL1}
    \\
    \partial_t \tilde C_4 &= 2 \tilde C_4 - 8 N_c \Big( \tilde G_S + \tilde G_P \Big) \tilde C_4 \tilde{\mathcal{I}}_{(3)}(\tilde m = 0) \nonumber\\
    &\quad\quad + \mathcal A_4(0, 0) \ g_s^2 \ \tilde C_4.
    \label{eq:reducedFlowsGaugedNJL}
\end{align}

The fixed point structure of the scalar 4-fermion coupling $G_S$ is equivalent to the one obtained in the conventional gauged-NJL setup in the positive-$\tilde G_S$ plane (see, e.g., Equation (11) in Ref.~\cite{Aoki:1999dv}).
With $N_c = 3$, the fixed points in $(\tilde G_S, \tilde C_4)$ plane are found to be
\begin{align}
    \tilde G_S^*(g_s) &= \frac{4 \pi^2}{3} - \frac{2}{3}g_s^2 \pm \frac{\sqrt{2}}{12}\sqrt{9 g_s^4 - 128 \pi^2 g_s^2+ 128 \pi^4} \, ,
        \label{eq:fixedPointsInGs}
\\
    \tilde C_4^* &= 0 \, .
    \label{eq:fixedPointsInGsC4Space}
\end{align}
From \Cref{eq:fixedPointsInGsC4Space}, we see that the fixed points have no real solution when
\begin{align}
    9 g_s^4 - 128 \pi^2 g_s^2+ 128 \pi^4 < 0 \,, 
\end{align}
which indicates the critical values of the gauge coupling at
\begin{align}
    g_s^{c,(1)} &= \frac{2\pi}{3} \sqrt{2 (8  - \sqrt{46})} \simeq 3.27,\\
    g_s^{c,(2)} &= \frac{2\pi}{3} \sqrt{2 (8 + \sqrt{46} )} \simeq 11.39.
\end{align}
The phase diagrams at the above critical values of the gauge coupling $g_s$ are shown in the top-right and the bottom-middle panels of \Cref{fig:GaugedNJLPhaseDiagrams}.

The $\tilde C_4$-flow on the $\tilde C_4$-axis, i.e., $\tilde G_S = 0$, reads
\begin{align}
    \partial_t \tilde C_4 \biggl|_{\tilde G_S = 0} =& 2 \tilde C_4 - \frac{16}{(4\pi)^2} \tilde C_4 g_s^2,
\end{align}
which implies the condition of the vanishing flow of $\tilde C_4$ coupling as
\begin{align}
    g_s^* = \sqrt{2} \pi.
    \label{eq:typicalValueOfRelevantC4}
\end{align}
Since $\mathcal A_S(0,0) = \mathcal A_4(0,0)$, the canonical scaling of the flow equations is canceled by the contribution from the triangle diagrams ($\propto g_s^2 \tilde G_S$ or $\propto g_s^2 \tilde C_4$).
Thus, the flow of the coupling $\tilde G_S$ ($\tilde C_4$) is symmetric (antisymmetric) under the transformation of $\tilde G_S \leftrightarrow -\tilde G_S$ and $\tilde C_4 \leftrightarrow -\tilde C_4$.
We show the corresponding flow structure for this case in the bottom-left panel of \Cref{fig:GaugedNJLPhaseDiagrams}.

From \Cref{fig:GaugedNJLPhaseDiagrams}, we see that the phase structure undergoes a sequence of qualitative changes as the gauge coupling $g_s$ increases.
At $g_s=0$, the phase diagram in the $(\tilde G_S,\tilde C_4)$ plane contains a nontrivial fixed point in addition to the Gaussian fixed point, and the two phases are separated by a distinct critical boundary.
For $0<g_s<g_s^{c,(1)}$, these two fixed points move closer to each other and eventually merge at $g_s=g_s^{c,(1)}$.
Once $g_s$ exceeds this critical value, the phase boundary disappears, leaving only the broken phase throughout the parameter space.
At $g_s=g_s^{c,(2)}$, a new fixed point and its associated critical surface emerge in the negative-$\tilde G_S$ region.
For $g_s>g_s^{c,(2)}$, the fixed point at $g_s=g_s^{c,(2)}$ is split into two branches. 
One of them remains attached to the critical surface, while the other moves away from there and becomes a UV fixed point. 
 Nevertheless, the phase diagram still contains no chirally symmetric phase, and the entire parameter space is occupied by the broken phase.

\subsubsection{QCD case}

\begin{figure*}[t]
    \centering
    \includegraphics[width=0.25\linewidth]{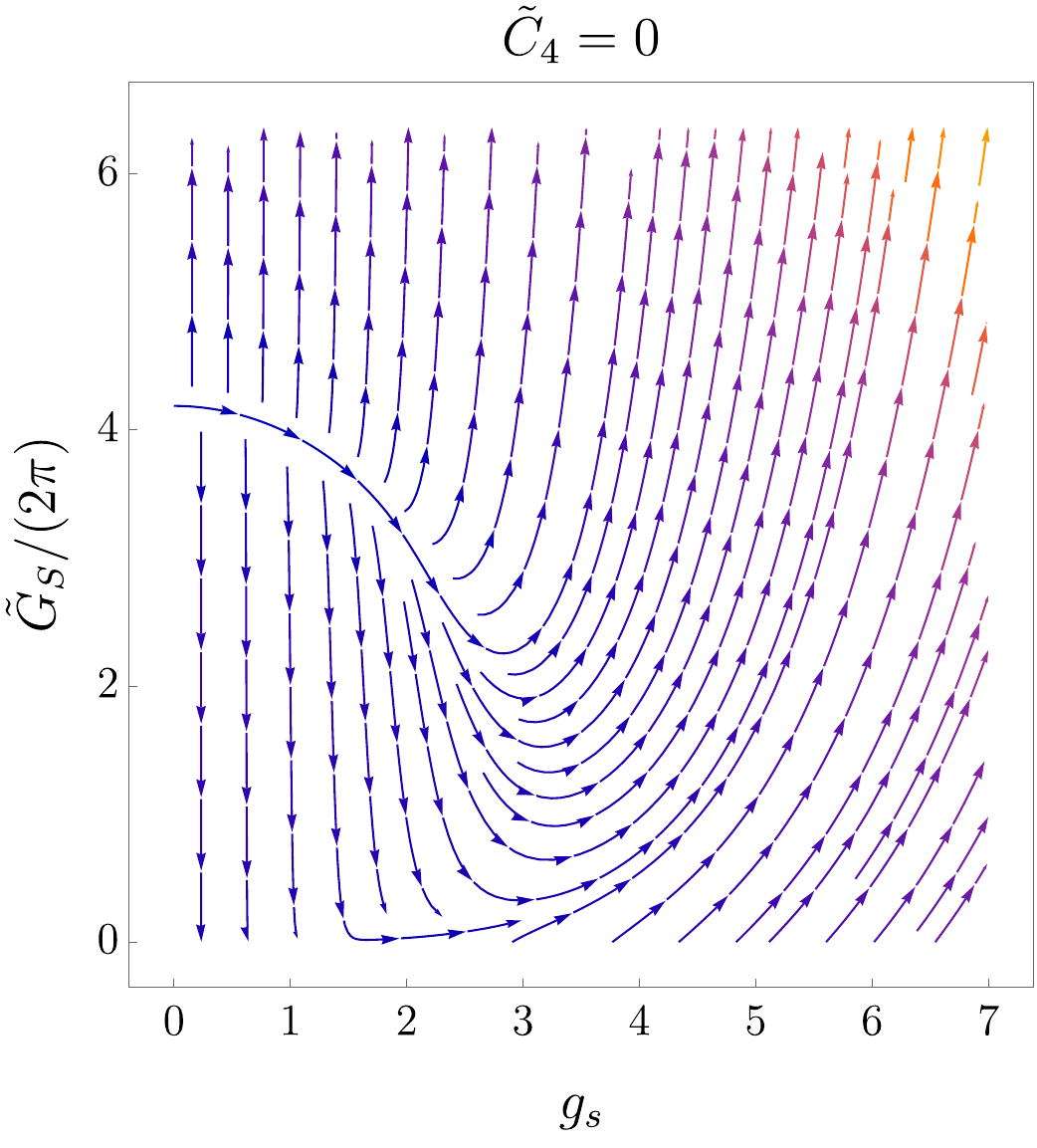}
    \hspace{5ex}
    \includegraphics[width=0.25\linewidth]{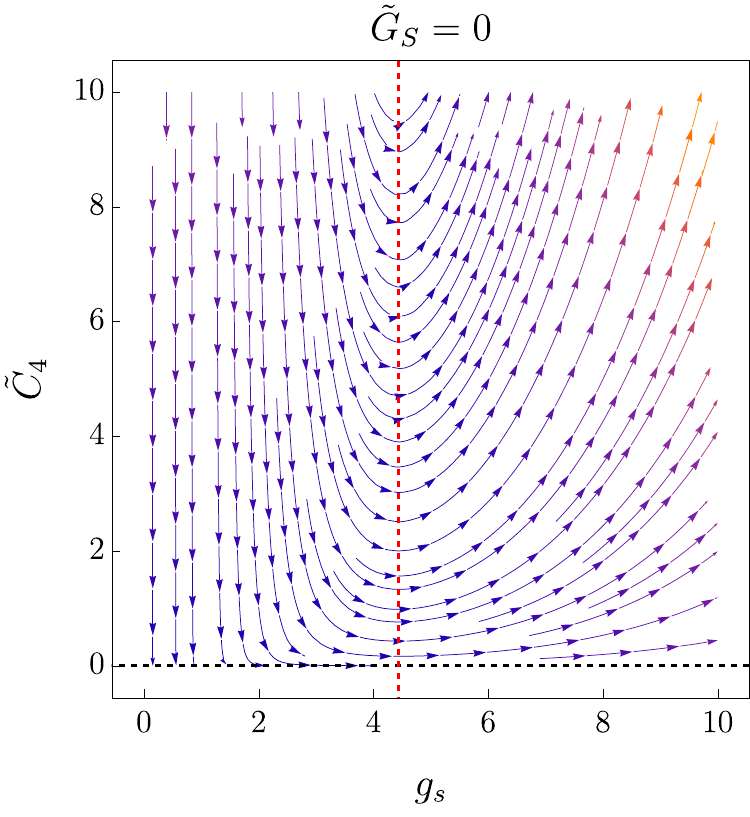}
    \hspace{5ex}
    \includegraphics[width=0.3\linewidth]{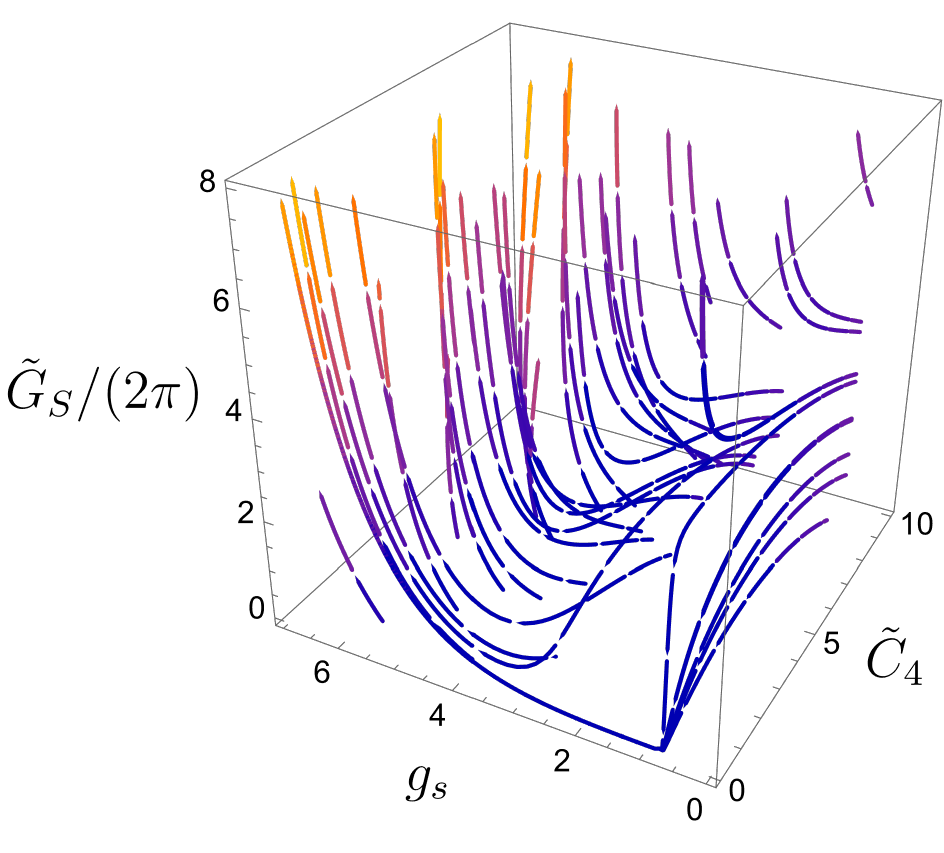}
    \caption{
    Phase diagrams in the $(g_s,\,\tilde G_S)$ plane (left panel) and $(g_s, \tilde C_4)$ plane (middle panel) in the chiral limit ($\tilde{m}=0$).
    The streamline diagrams of the flow equations in the $(g_s, \,\tilde C_4,\, \tilde G_S)$ space are shown in the right panel, where the couplings are flowing from the blue part to the orange part on the streamlines.
    In the left and the right panel, we rescaled the coupling $\tilde G_S \to \tilde G_S/(2 \pi)$ for the sake of readability.
    The red-dashed line in the bottom panel denotes the critical value of the gauge coupling $g_s = \sqrt{2}\pi$, such that the flow of the $\tilde C_4$ Coupling flips its sign; see the discussion around \Cref{eq:typicalValueOfRelevantC4}.
    }
    \label{fig:QCDPhaseDiagrams2D}
\end{figure*}

In QCD-like theories, the gauge coupling is no longer an external control parameter but an additional running coupling, the RG flow of which is governed by asymptotic freedom. 
Since our present discussion is focused on the vicinity of the Gaussian fixed point, it is sufficient to approximate the RG flow of the gauge coupling by the perturbative one-loop QCD beta function,
\begin{align}
    \beta_{g_s} \simeq \beta_\text{QCD}^\text{1-loop} = - \frac{g_s^3}{(4\pi)^2} \left( \frac{11}{3} N_c - \frac{2}{3} N_f \right).
\end{align}

As discussed in \Cref{sec:FixedGaugeCouplingCase} as well as in Ref.~\cite{Aoki:1999dv}, in the gauged-NJL model with a fixed gauge coupling, there exists a chiral symmetric phase 
for $g_s < g_s^{c,(1)}$, which is 
separated from the broken phase by a nontrivial critical boundary.
Once the running of $g_s$ is switched on, however, the symmetric phase disappears, which implies no chiral symmetric phase in the full QCD-type system. 
This can be seen explicitly in the left panel of \Cref{fig:QCDPhaseDiagrams2D}, where we show the flow in the $(g_s,\tilde G_S)$ plane at $\tilde C_4 = 0$, with rescaling $\tilde G_S \to \tilde G_S/(2\pi)$.

A similar phenomenon can be observed in the $\tilde C_4$ direction.
From the flow equation~\labelcref{eq:betaFunctionOfC4}, one finds that the $\tilde C_4$ interaction becomes relevant at IR once the gauge coupling exceeds the critical value, 
\begin{align}
    g_s^{*} = \sqrt{2}\,\pi .
\end{align}
This behavior is shown in the middle panel of \Cref{fig:QCDPhaseDiagrams2D}.
Hence, the $\tilde C_4 = 0$ hypersurface is no longer IR attractive beyond this threshold, so that even an arbitrarily small perturbation in the $\tilde C_4$ direction is amplified along the RG flow.

The full flow structure is summarized in the right panel of \Cref{fig:QCDPhaseDiagrams2D}, where we show the flow in the three-dimensional subspace $(g_s,\tilde C_4,\tilde G_S)$.
The critical surface appearing in the top-left panel of \Cref{fig:GaugedNJLPhaseDiagrams} no longer persists when the running effect of the gauge coupling is taken into account,  
which stays merely as a line-like structure and hence cannot divide the theory space into distinct chiral symmetric and broken phases.
Although in the fixed-$g_s$ plane we still see a remnant of critical behavior, 
a distinct phase boundary is not formed in the full coupling space due to the RG flow of the gauge coupling.  


Thus, the IR dynamics is not governed by the fixed-point structure in the symmetry realization regime, but rather by the instabilities caused by dynamical chiral symmetry breaking and the dynamics of bosonized fields. 
In particular, when $\tilde C_4 \neq 0$ at UV, the $P$-even subspace becomes unstable, and RG trajectories generically flow toward a $P$-violating strong-coupling regime. 

\subsection{Relevance of the \textit{P}-violating coupling with \texorpdfstring{$\theta$}{}}

\begin{figure*}[t]
    \centering
    \includegraphics[width=0.3\linewidth]{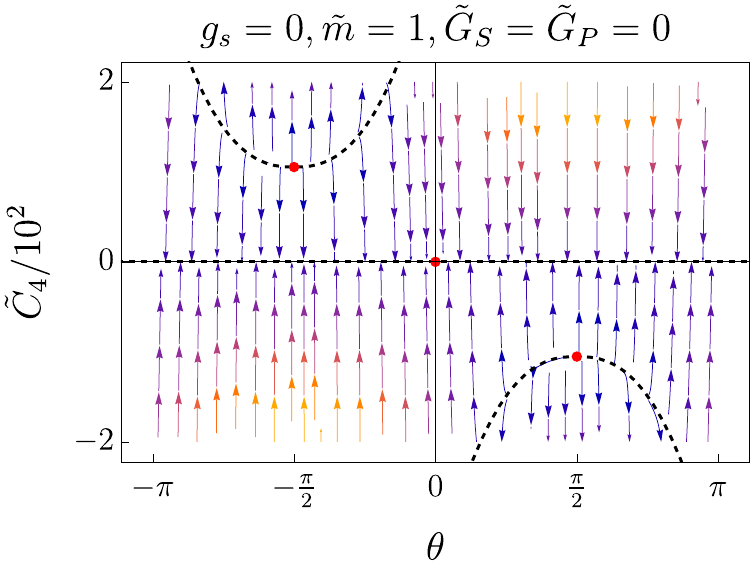}
    \hspace{5ex}
    \includegraphics[width=0.3\linewidth]{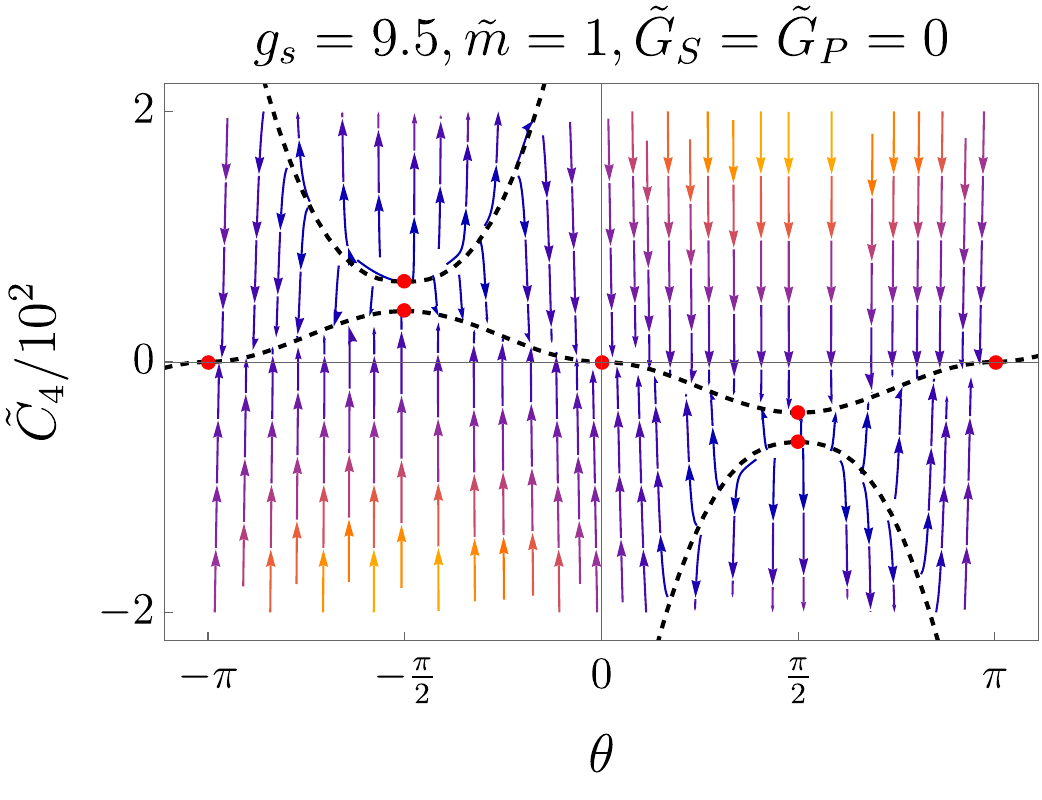}
    \\
    \includegraphics[width=0.3\linewidth]{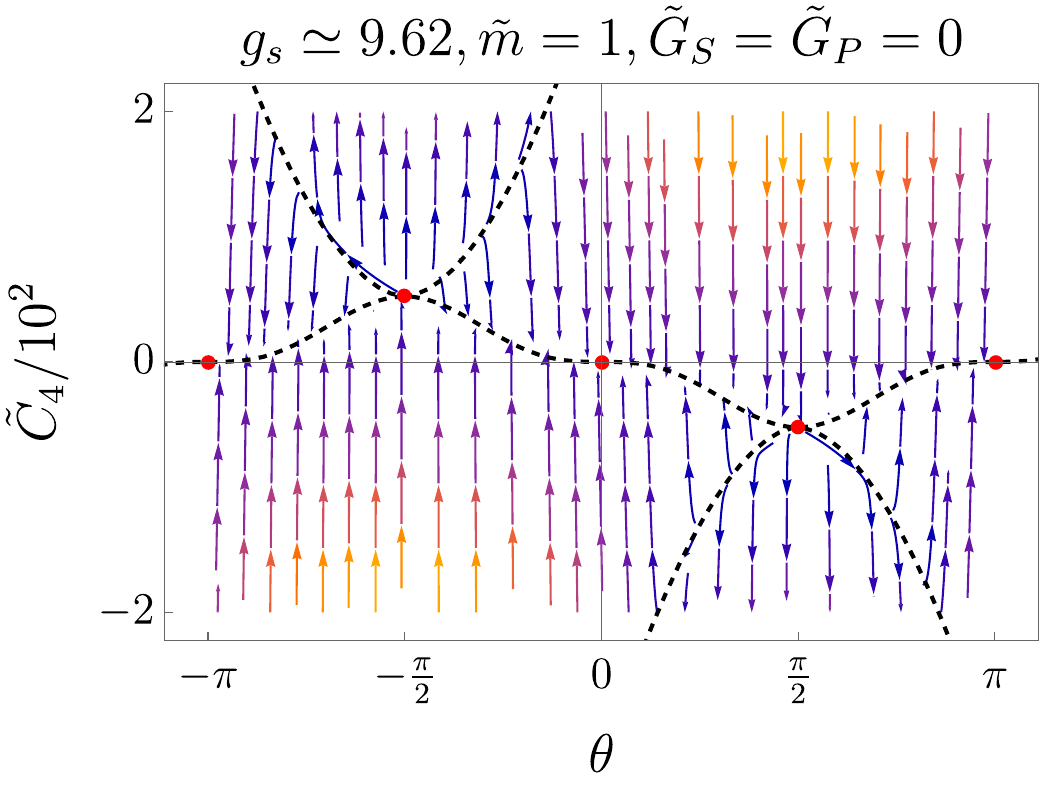}
    \hspace{5ex}
    \includegraphics[width=0.3\linewidth]{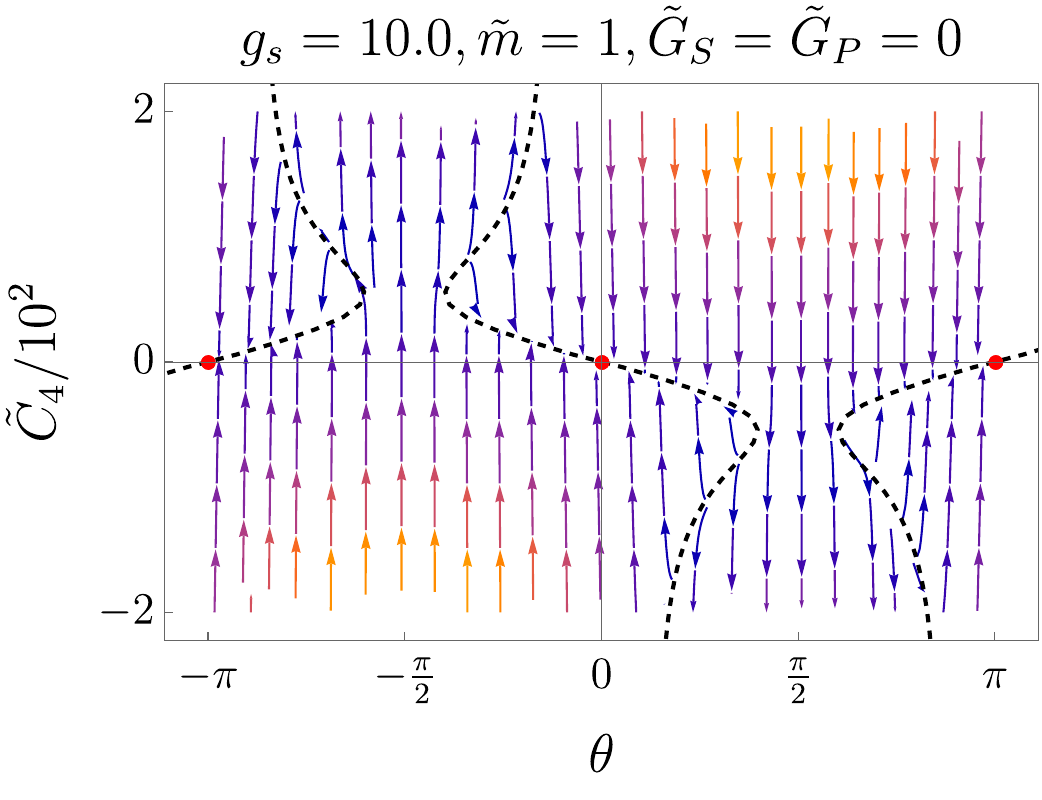}
    \caption{
    Phase structure of the flow equations~\labelcref{eq:flowequationsinmaintext} in $(\theta, \tilde C_4)$-plane, with different values of the gauge coupling $g_s$.
    The red dots denote the fixed points, and the black-dashed lines represent the phase boundaries.
    Streaming lines represent the flow equations projected onto the $(\theta, \tilde C_4)$-plane, flowing from UV to IR direction.
    }
    \label{fig:thetaC4PD}
\end{figure*}

As discussed in \Cref{sec:SpontaneousCPviolationWithThetaParameter}, the CP phase parameter $\theta$ is always coupled to the mass parameter in the form of $m e^{i \gamma_5 \theta}$; thus, it becomes relevant at IR when the (dimensionless) mass parameter $\tilde m$ is nontrivially relevant.
In this section, we study the interplay between the $P$-violating four-fermion interaction and the $\theta$ parameter in massive QCD. 

In \Cref{fig:thetaC4PD}, we show the phase structure of massive QCD theory in the presence of the $P$-violating four-fermion coupling $\tilde C_4$, in the slice of $\theta$ and $\tilde C_4$ coupling.
For demonstration purposes, we set $\tilde G_S = \tilde G_P = 0$, and $\tilde m = 1$.
From the top-left panel of \Cref{fig:thetaC4PD}, which corresponds to the NJL-like theory, we observe a fixed line at $\tilde C_4 = 0$ covering the Gaussian fixed point at $\theta=0$, together with two phase boundaries separating the symmetric and broken phases.
From the remaining panels in \Cref{fig:thetaC4PD}, we see that as $g_s$ increases, the phase boundaries are drastically deformed to get closer to each other, eventually merging into a point.
Thus, the phase structure of $\tilde C_4$ is highly affected by $\theta$ along with the RG running of $g_s$.

\subsection{RG evolution of the four-fermion couplings}

\begin{figure*}[t]
    \centering
    \includegraphics[width=0.3\linewidth]{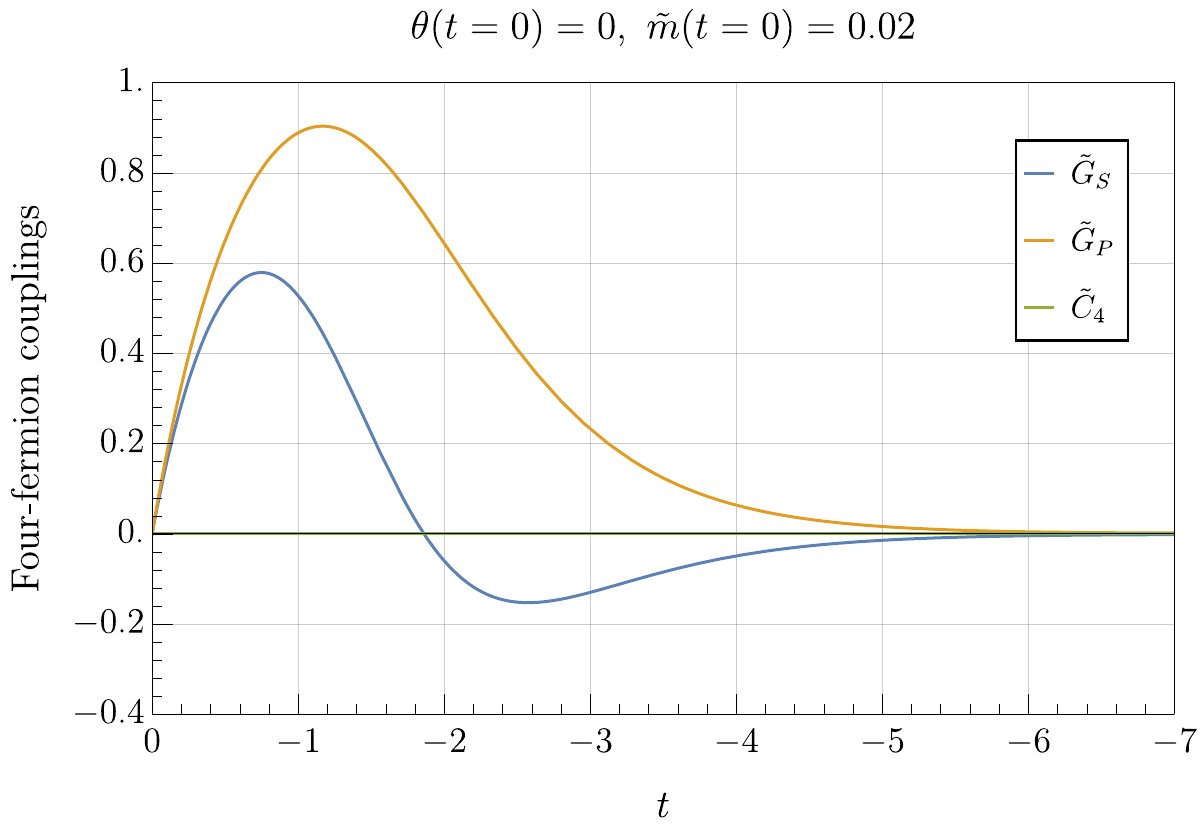}
    \hspace{5ex}
    \includegraphics[width=0.3\linewidth]{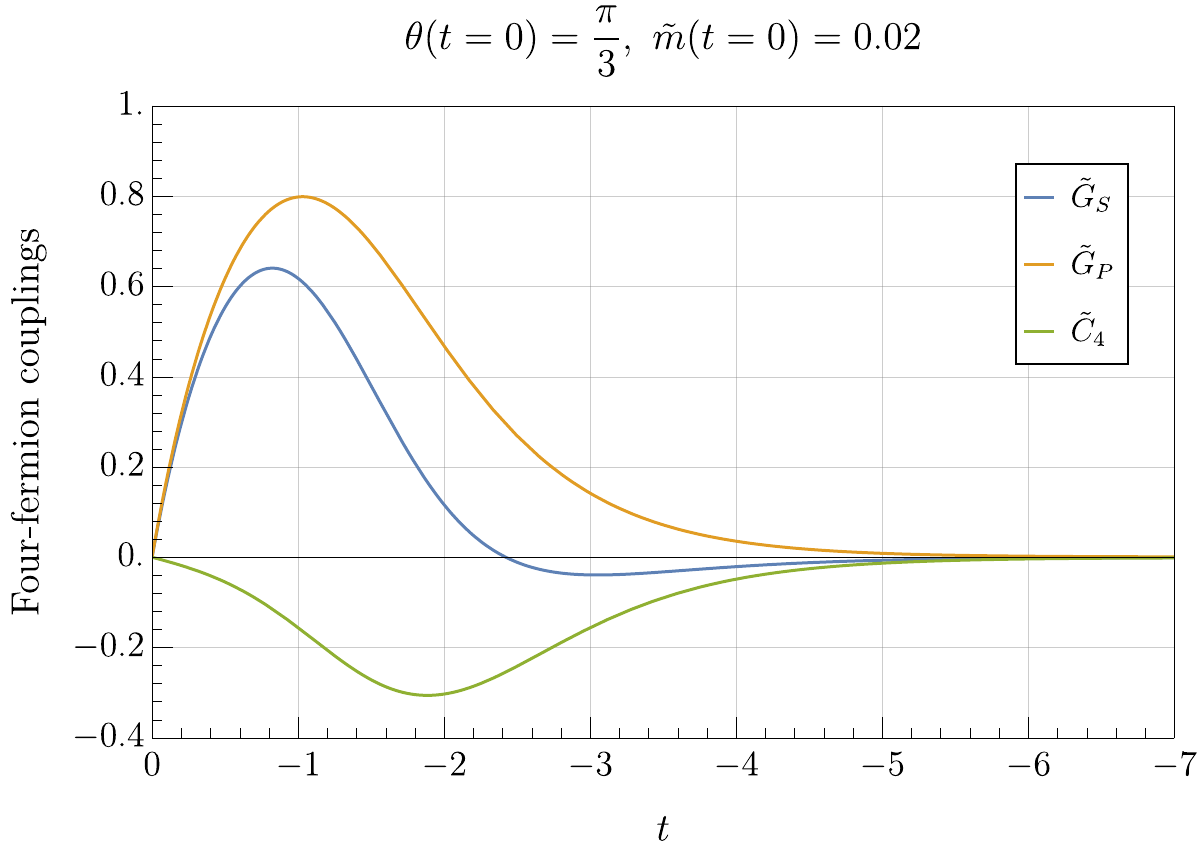}
    \hspace{5ex}
    \includegraphics[width=0.3\linewidth]{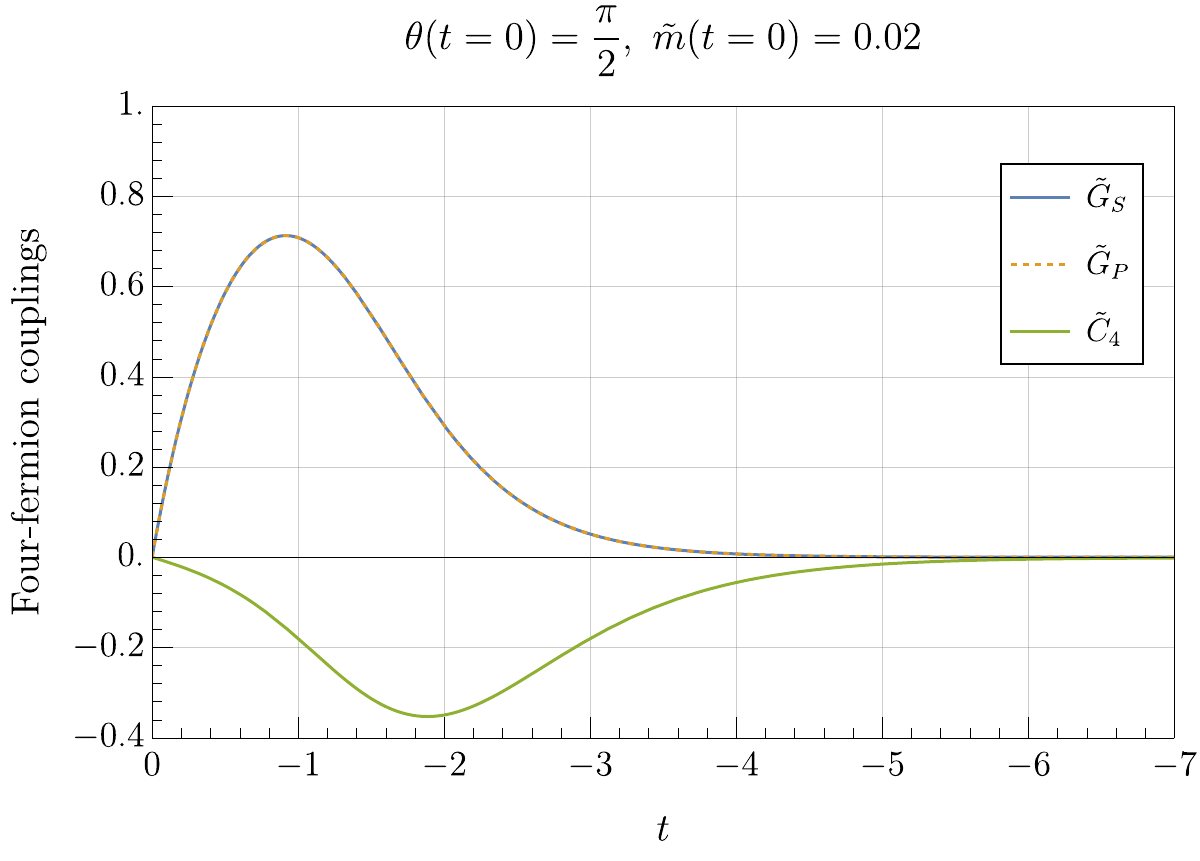}
    \\
    \includegraphics[width=0.3\linewidth]{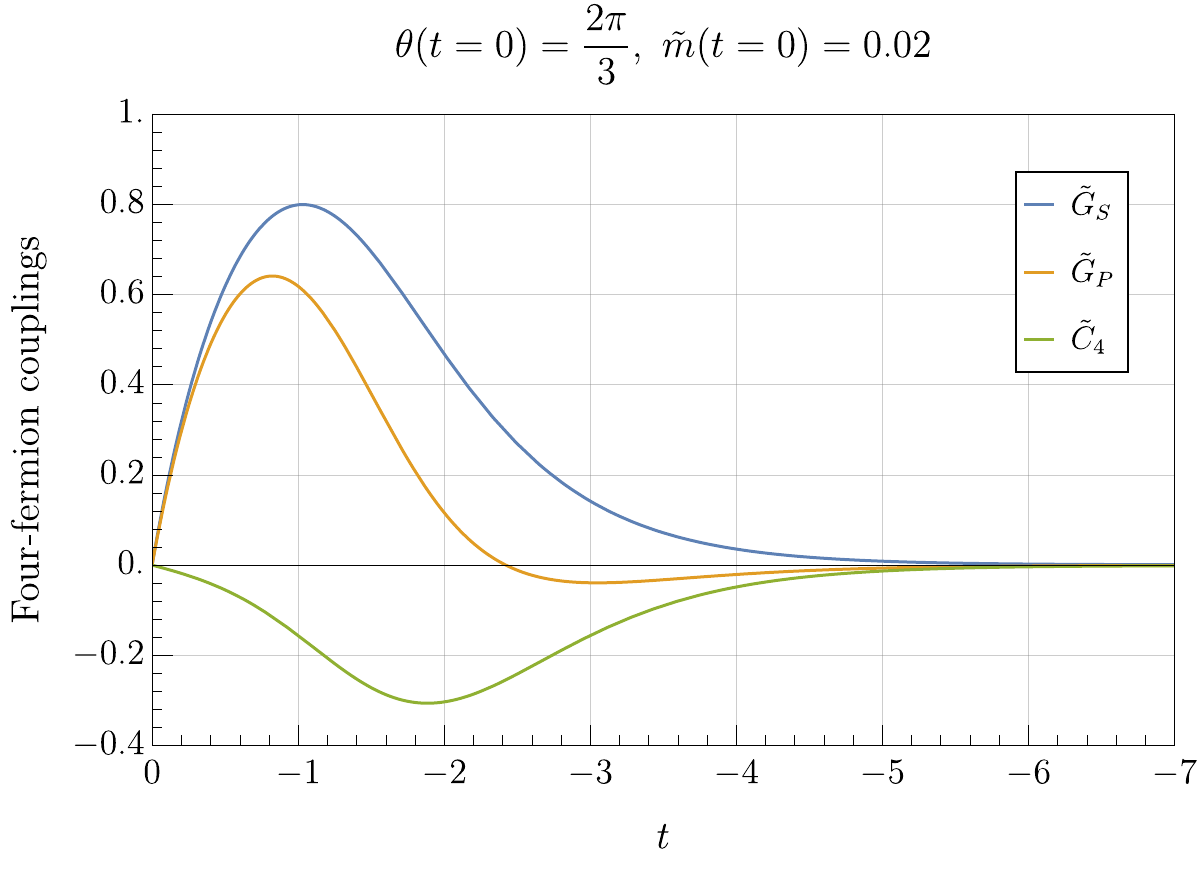}
    \hspace{5ex}
    \includegraphics[width=0.3\linewidth]{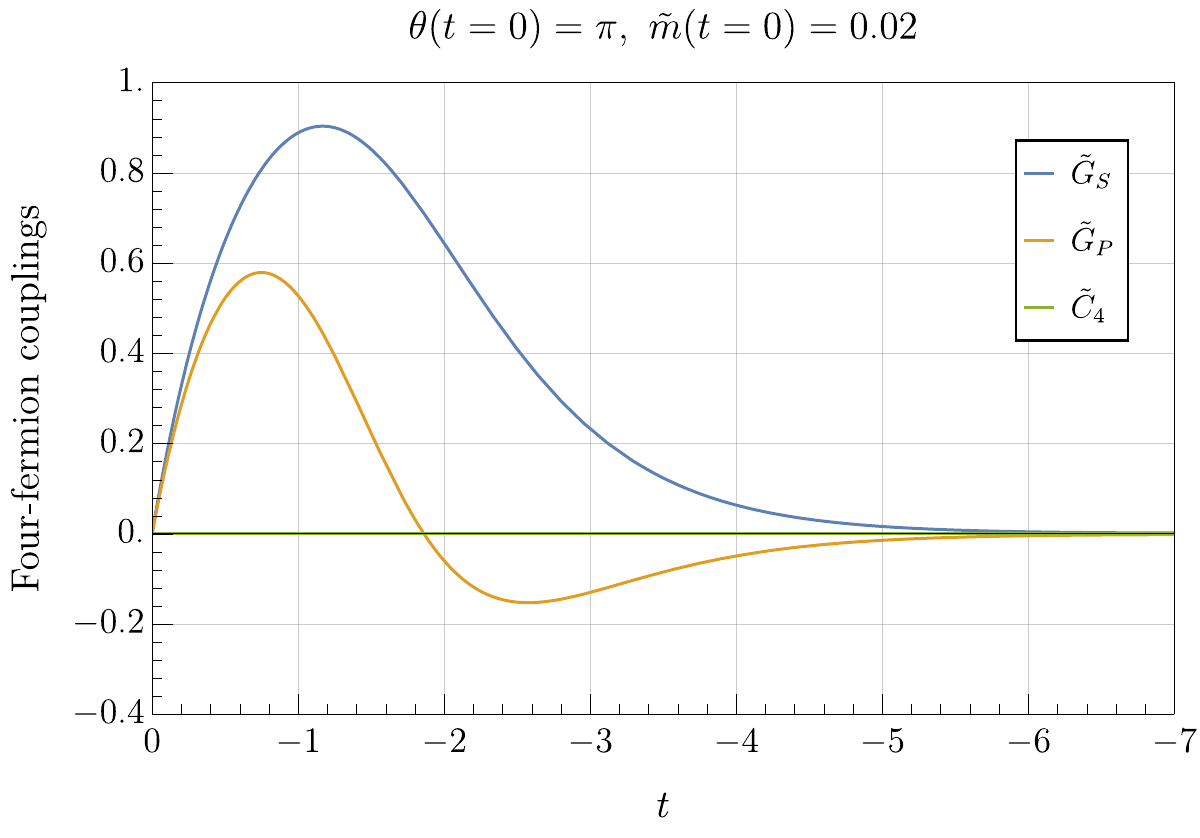}
    \caption{
    RG evolutions of the four-fermion couplings in the $P$-violating QCD, with trivial initial conditions $\tilde G_S = \tilde G_P = \tilde C_4 = 0$.
    In different panels, the initial condition of $\theta$ has been set to different values, as labeled on the top of each panel.
    }
    \label{fig:evoluting4Fermion}
\end{figure*}

As a further step, we study the RG evolution of the coupled system in ~\labelcref{eq:flowequationsinmaintext}. 
In \Cref{fig:evoluting4Fermion}, we show the solutions of the coupled $\beta$-functions~\labelcref{eq:flowequationsinmaintext} for the dimensionless four-fermion couplings.
To monitor a low-energy effective description of QCD, we choose the initial conditions
at UV: 
\begin{gather}
    \tilde m (t = 0) = 0.02, 
    \qquad
    g_s (t = 0) = 1, \nonumber\\
    \tilde G_S (t = 0) = \tilde G_P (t = 0) = \tilde C_4 (t = 0) = 0,
\end{gather}
and consider several initial values of $\theta$ in the range $0\le \theta \le \pi$. 
The RG evolution generates a nontrivial operator mixing between the scalar and pseudoscalar sectors involving $\tilde C_4$ and $\theta$. 
The gauge interaction radiatively induces effective fermion self-interactions and drives the system to go away from the Gaussian fixed point regime.

The flow exhibits a characteristic ``three-stage behavior".
First, the gauge interaction generates nonzero four-fermion couplings at intermediate scales.
Second, over a substantial interval of RG time, the dimensionless couplings become small again. 
This trend should be understood mainly as a canonical scaling effect: once threshold decoupling sets in, the corresponding dimensionful four-fermion couplings are approximately stabilized, while the dimensionless couplings continue to scale according to their canonical mass dimension.
Hence, the temporal suppression of $\tilde G_S$, $\tilde G_P$, and $\tilde C_4$ does not indicate that the interaction itself disappears but rather reflects the decoupling of the running scale from the physical IR couplings.
Finally, the perturbative RG running of the gauge coupling reaches an IR Landau pole at around $t\sim -7.64$.
Consequently, the four-fermion couplings are also driven to diverge. 
Since this region lies beyond the physical validity of the present framework, it is not displayed in \Cref{fig:evoluting4Fermion}.

The dependence on the initial value of $\theta$ 
controls which attractive interaction channels are dominant at IR. 
In particular, we see a symmetry associated with the following reflection: 
\begin{align}
    \tilde G_S \leftrightarrow \tilde G_P,
    \qquad
    \theta \leftrightarrow \pi-\theta .
\end{align}
Thus, the RG flows for $\theta$ and $\pi-\theta$ are related by exchanging the scalar and pseudoscalar interaction channels. 
For $\theta(0)=0$ and $\theta(0)=\pi$, the scalar-pseudoscalar mixing coupling $\tilde C_4$ remains zero throughout the RG running toward IR, while for generic values such as $\theta(0)=\pi/3$, $\pi/2$, and $2\pi/3$, a nonzero $\tilde C_4$ is generated dynamically. 
This would lead to an implication for the phase-diagram analysis: the $\tilde C_4$ direction is not only relevant in the linearized sense, but is radiatively populated and further enhanced by the growing gauge coupling toward IR.
Therefore, the $P$-even subspace is generically unstable in the QCD-type flow.

\begin{figure}
    \centering
    \includegraphics[width=0.8\linewidth]{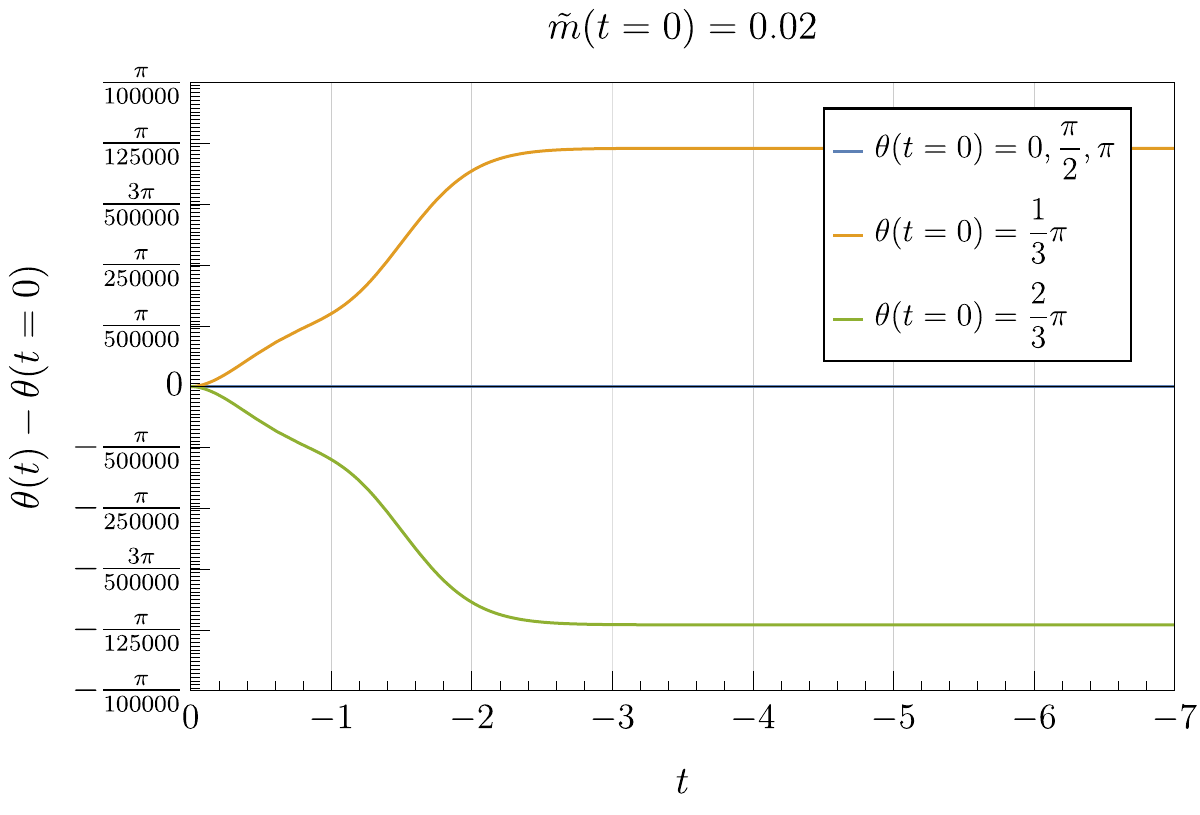}
    \caption{
    RG evolution of the $\theta$ parameter in \Cref{fig:evoluting4Fermion}, with different initial conditions.
    The values of the $\theta$-parameter have been subtracted from their initial conditions for the sake of clarity of comparisons.
    }
    \label{fig:ThetaFlows}
\end{figure}

$\theta$ itself hardly runs. 
As shown in \Cref{fig:ThetaFlows}, the change of $\theta$ along the flow is numerically small for all initial conditions considered here. 
Within the present truncation, $\theta$ therefore acts mainly as an external parameter that selects the orientation of whether the infrared instability is realized dominantly by the $P$-even or -odd subspace, while the dominant nonperturbative dynamics is anyhow governed by the four-fermion sector.

The case $\theta=\pi/2$ gets particularly peculiar: 
Introducing the scalar and pseudoscalar bilinears as
\begin{align}
    S = \bar\psi\psi,
    \qquad
    P = \bar\psi i\gamma_5\psi ,
\end{align}
and the rotated combinations as
\begin{align}
    S_\pm = \frac{1}{\sqrt{2}}(S \pm P) ,
\end{align}
one finds that the fermion mass term takes the following form:
\begin{align}
    m \bar\psi e^{i\gamma_5\pi/4}\psi = m\,S_+ .
\end{align}
Thus, at $\theta=\pi/2$, the quark mass couples only to $S_+$.
The four-fermion sector can be rewritten as
\begin{align}
    \mathcal L_{4f}
    &=
    -\frac14\left(G_S+G_P+2C_4\right)S_+^2
    -\frac14\left(G_S+G_P-2C_4\right)S_-^2
    \nonumber\\
    &\hspace{5ex} -\frac12\left(G_S-G_P\right)S_+S_- \,.
\end{align}
The numerical solution shows that for $\theta=\pi/2$, the two diagonal couplings coincide in the RG evolution:  
\begin{align}
    G_S = G_P ,
\end{align}
so that the mixed term $S_+S_-$ vanishes identically.
The interaction then becomes diagonal in the rotated basis,
\begin{align}
    \mathcal L_{4f}\big|_{\theta=\pi/2}
    =
    -\frac12(G_S+C_4)S_+^2
    -\frac12(G_S-C_4)S_-^2 \,, 
\end{align}
which is invariant under the interchange $S \leftrightarrow P$. 
This implies that $\theta=\pi/2$ corresponds to a self-dual (symmetric) point for the scalar-pseudoscalar sector.
The quark mass $m$ selects the $S_+$ channel, while the coupling $C_4$ controls the splitting between the $S_+$ and $S_-$ channels.
Therefore, the IR instability is more adequately described in the $(S_+,S_-)$ basis rather than in terms of the original $(S, P)$ basis. 


\subsection{Towards bosonization}

We remark on the framework of bosonization within the present model.
First, for the present model, we introduce bosonic auxiliary fields, $\sigma$ and $\eta$, into the path integral via the Gaussian integral
\begin{widetext}
\begin{align}
\label{eq:HStrans}
1 = \mathcal N \int \mathcal D \sigma \mathcal D \eta\, e^{ - \int_x \left[\frac{1}{2G_S}(\sigma + G_S\bar\psi\psi + c_S)^2+ \frac{1}{2G_P}(\eta + G_P\bar\psi i\gamma^5\psi+ c_P)^2 +\frac{C_4}{G_SG_P}(\sigma + G_S\bar\psi\psi +   c_S)(\eta + G_P\bar\psi i\gamma^5\psi +c_P)\right]}\,,
\end{align}
where $\mathcal N$ is a field-independent normalization factor.
The four-fermion interactions in the action \labelcref{eq:UVQCDaction} cancel and then we recast the four-fermion sector as a Higgs-Yukawa type theory
\begin{align}
\mathcal L_{4f} &= \frac{m\cos\theta}{G_S} \sigma  + \frac{m\sin\theta}{G_P} \eta
+ \frac{C_4}{G_SG_P} \sigma \eta
+ \frac{1}{2G_S}\sigma^2 + \frac{1}{2G_P}\eta^2 
+ \left( \sigma +  \frac{C_4}{G_P}\eta \right)\bar\psi \psi + \left( \eta + \frac{C_4}{G_S} \sigma \right) \bar\psi i\gamma^5 \psi\,,
\end{align}
\end{widetext}
where $c_S$ and $c_P$ are constants set to be
\begin{align}
c_S&= m\frac{G_S(G_P \cos\theta -C_4 \sin\theta)}{G_SG_P-C_4^2}\,,\\
c_P&= m\frac{G_P(G_S \sin\theta -C_4 \cos\theta)}{G_SG_P-C_4^2}\,.
\end{align}
Note that the equations of motion for $\sigma$ and $\eta$ read as
\begin{align}
&m\cos\theta + \frac{C_4}{G_P} \eta + \sigma + G_S\bar\psi\psi + C_4\bar\psi i\gamma^5 \psi=0\,,\\
&m\sin\theta + \frac{C_4}{G_S} \sigma + \eta + C_4 \bar\psi\psi +  G_P\bar\psi i\gamma^5 \psi=0\,.
\end{align}

Ignoring the gauge interaction in the Euclidean version of the starting action, \cref{eq:UVQCDaction}, and integrating out the fermion fields, one obtains the effective potential for $\sigma$ and $\eta$ within the mean-field approximation. Such a mean-field analysis is presented in \Cref{sec:meanfield}.

In the bosonized description, the effective action $\Gamma_k$ for the Higgs--Yukawa sector is given by
\begin{align}
&\Gamma_k^\mathrm{HY}
\simeq
\int_x
\bigg[
\bar{\psi}
\biggl(
Z_\psi \gamma_\mu \partial_\mu
+i g_s \gamma_\mu A_\mu^a T^a
+m e^{i\gamma_5\theta/2}
\biggr)
\psi
\nonumber\\
&
+y_\sigma \sigma \bar{\psi}\psi
+y_\eta \eta \bar{\psi} i\gamma^5 \psi
+h_\sigma \sigma \bar{\psi} i\gamma^5 \psi
+h_\eta \eta \bar{\psi}\psi
\nonumber\\
&
+\frac{Z_\sigma}{2}(\partial_\mu \sigma)^2
+\frac{Z_\eta}{2}(\partial_\mu \eta)^2
+\frac{Z_{\sigma\eta}}{2}
(\partial_\mu \sigma)(\partial_\mu \eta)
+V(\sigma,\eta)
\bigg] .
\label{eq:IRQCDaction}
\end{align}
Here, its initial (matching) considtion is given at $k=\Lambda$ as $Z_\psi =1$, $Z_\sigma=Z_\eta=Z_{\sigma\eta}=0$, $y_\sigma=y_\eta=1$, $h_\sigma=C_4/G_S$, $h_\eta=C_4/G_P$, and $V(\sigma,\eta)=\frac{m\cos\theta}{G_S} \sigma  + \frac{m\sin\theta}{G_P} \eta+ \frac{C_4}{G_SG_P} \sigma \eta+ \frac{1}{2G_S}\sigma^2 + \frac{1}{2G_P}\eta^2$. 
It is noteworthy that the Yukawa interactions can generate a mixed kinetic term between \(\sigma\) and \(\eta\). Therefore, one has to redefine the bosonic fields, \(\sigma'\) and \(\eta'\), such that the kinetic terms are brought into canonical diagonal form.

Within the framework of the fRG, however, the Yukawa interactions regenerate four-fermion interactions even if the four-fermion terms are bosonized in the bare UV action. This complicates the analysis of the IR dynamics. To address this issue, one may employ dynamical bosonization \cite{Gies:2001nw, Gies:2002hq, Pawlowski:2005xe, Floerchinger:2009uf, Braun:2014ata, Mitter:2014wpa, Cyrol:2017qkl, Cyrol:2017ewj, Alkofer:2018guy, Denz:2019ogb, Fu:2019hdw, Goertz:2024dnz}. This procedure is analogous to performing a Hubbard--Stratonovich transformation at each infinitesimal RG scale. A detailed analysis along these lines will be presented in future work.


\section{Conclusions}
\label{sec:Conclusions}

In this paper, we have investigated the phase structure of QCD-like theories in the presence of the $P$-odd operator $\left( \bar{\psi} \psi \right) \left( \bar{\psi} i \gamma_5 \psi \right)$, which is motivated by the topological $\theta$ term in QCD as a possible source of $CP$ violation.
From the broader perspective of the strong $CP$ problem, our aim has been to explore the low-energy properties of such $P$- and $CP$-violating theories when coupled to QCD, thereby extending the analysis of Ref.~\cite{Huang:2024ypj}.

To this end, we constructed a minimal low-energy truncation within the LPA framework and derived the corresponding RG flow equations for the running couplings.
By combining flow-diagram analyses with explicit solutions of the coupled $\beta$ functions, we obtained a global picture of the infrared behavior.
Compared with Ref.~\cite{Huang:2024ypj}, the phase structure in the $(\tilde G_S,\tilde C_4)$ plane is substantially modified once gauge interactions are included.
In particular, after allowing the gauge coupling to run, the $\tilde C_4$ direction becomes relevant in the chirally broken regime, in close analogy with the extension of the phase structure in gauged NJL models~\cite{Aoki:1999dv} without $CP$ violation.
The coupled RG trajectories further show that, for generic $\theta$, the $P$-violating coupling $\tilde C_4$ is not only relevant in the linearized sense, but is generated radiatively and enhanced toward the IR together with the other four-fermion couplings.

For finite $\tilde m$, we find that the running effect of the $\theta$ parameter itself is numerically small.
In contrast, its effect is efficiently transferred to the fermionic sector through induced four-fermion interactions:
it still leaves a nontrivial influence on changing the dominance of the $P$-even or -odd four-fermion couplings.
Finally, an appropriate IR formulation in terms of dynamical bosonization is necessary for a systematic treatment of the RG flows in the chirally broken phase.

This picture should be contrasted with approaches that emphasize the IR fate of the topological charge in pure Yang-Mills theory, where the effective $\theta$ parameter is argued to be attracted toward the Gaussian fixed point in the deep infrared~\cite{Nakamura:2021meh}. 
This is the consequence of the deep IR running of the topological charge. 
On the other hand, the running of the $\theta$ parameter in the present work is understood as the RG flow projection onto the phase of the fermionic mass.
A more complete treatment would be required to clarify how the topological and fermionic descriptions are matched in the deep IR.

In the context of constructing low-energy effective models for massive QCD with a $CP$ phase $\bar\theta$, our results indicate that the $P$-odd coupling $\tilde C_4$ is an unavoidable operator generated along the RG flow.
This suggests that $CP$-violating four-fermion interactions may play a nontrivial role in the IR dynamics of QCD-like theories and provides a concrete framework for systematically incorporating strong-$CP$ effects into low-energy effective descriptions.

\begin{acknowledgments} 
We thank Akio Tomiya for helpful discussions.
The work of S.\,M. was supported in part by the National Science Foundation of China (NSFC) under Grant Nos.\,11747308, 11975108, 12047569, and the Seeds Funding of Jilin University. 
The work of M.\,Y. was supported by the National Science Foundation of China (NSFC) under Grant No.\,12205116 and the Seeds Funding of Jilin University.
\end{acknowledgments}

\onecolumngrid
\appendix
\section{Derivation of the \texorpdfstring{$\beta$}{}-functions}
\label[appendix]{app:DerivationOfTheBetaFunctions}

The scale evolution of the 1PI effective average action $\Gamma_k$ is described by the functional partial differential equation, i.e., the Wetterich equation \labelcref{eq:WetterichEq}.
The properties and applications of the flow equation in many research fields are discussed in Refs.~\cite{Reuter:1993kw,Morris:1998da,Berges:2000ew,Aoki:2000wm,Bagnuls:2000ae,%
  Polonyi:2001se,Pawlowski:2005xe,Gies:2006wv,Delamotte:2007pf,Sonoda:2007av,Igarashi:2009tj,%
  Rosten:2010vm,Braun:2011pp,Dupuis:2020fhh}.
In this Appendix, we present the details for deriving the flow equations in the model \labelcref{eq:IRQCDaction}.

\subsection{Hessian}

For the 1PI effective average action \labelcref{eq:IRQCDaction}, we start by obtaining the Hessian, which is the second-order functional derivative for the effective action
\begin{align}
    \Gamma_{k,ab}^{(2)} = \frac{\overset{\rightarrow}{\delta}}{\delta \Phi^\text{T}_a(-q)} \Gamma_k  \frac{\overset{\leftarrow}{\delta}}{\delta \Phi_b(p)} = \pmat{
    H_{BB} & H_{BF} \\
    H_{FB} & H_{FF}
    },
\end{align}
where $B/F$ denotes the bosonic/fermionic component of the superfield.
For the gluonic sector, we have
\begin{align}
    \left( H_{BB} \right)_{\mu\nu}^{ab} = (2\pi)^4 \delta^{(4)}(p-q) \, p^2 \biggl( \delta_{\mu\nu} - ( 1 - \frac{1}{\xi} ) \frac{p_\mu p_\nu}{p^2} \biggl) \delta^{ab} = (2\pi)^4 \delta^{(4)}(p-q) \delta^{ab} \, D_{\mu\nu}^{-1}(p),
\end{align}
where we have expanded the flow around the vanishing background gauge field $A_\mu^a = 0$ for simplicity, and we have defined the projection operators
\begin{align}
    &\Pi^{1/\xi}_{\mu\nu}(p) = \Pi^{\perp}_{\mu\nu}(p) + \frac{1}{\xi}\Pi^{\parallel}_{\mu\nu}(p), \nonumber\\
    &\Pi^{\perp}_{\mu\nu}(p) = \biggl( \delta_{\mu\nu} - \frac{p_\mu p_\nu}{p^2} \biggl),\nonumber\\
    &\Pi^{\parallel}_{\mu\nu}(p) = \frac{p_\mu p_\nu}{p^2}.
\end{align}

\begin{figure}[t]
    \centering
    \includegraphics[width=0.7\linewidth]{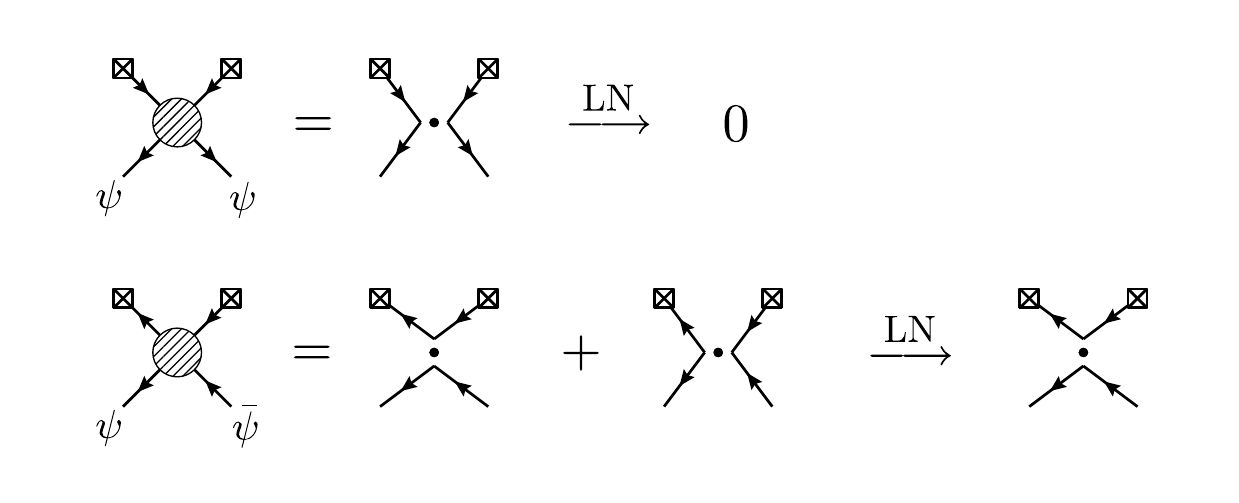}
    \caption{
    Sketch of the large-$N$ approximation of $\mathcal{V}_k^{4f}$.
    The top line corresponds to the $(1,1)$ conponent of $\mathcal{V}_k^{4f}$, and the bottom line corresponds to the $(1,2)$ conponent of $\mathcal{V}_k^{4f}$, respectively.
    The bare-arrowed line denotes the amputated external legs, and the crossed box denotes the fermionic mean-field background.
    }
    \label{fig:LargeNFourFermion}
\end{figure}

For the fermionic sector, we may split the Hessian into a field-independent and a field-dependent part as
\begin{align}
    H_{FF} = \mathcal{K}_k + \mathcal{V}_k^{4f},
\end{align}
where we have defined
\begin{align}
    \mathcal{K}_k = (2\pi)^4 \delta^{(4)}(p-q) \pmat{
    0 & i \Slash{p}^\text{T} - m e^{i \gamma_5 \theta / 2} \\
    i \Slash{p} + m e^{i \gamma_5 \theta / 2} & 0
    },
\end{align}
and
\begin{align}
    \mathcal{V}_k^{4f} = \int_x \, e^{-i(q-p)\cdot x} \pmat{
    V^{4f}_{11} & V^{4f}_{12} \\
    V^{4f}_{21} & V^{4f}_{22}
    }_x.
\end{align}
Here, the superscript ``$4f$'' denotes contributions from the four-fermion interaction, and the matrix elements read
\begin{align}
    V^{4f}_{11}(x) &= \biggl[ G_S \left( \bar{\psi}^\text{T} \bar{\psi} \right)_x + G_P i \gamma_5 \left( \bar{\psi}^\text{T} \bar{\psi} \right)_x i \gamma_5 + C_4 \left\{ \left( \bar{\psi}^\text{T} \bar{\psi} \right)_x, i\gamma_5 \right\} \biggl], \nonumber\\
    V^{4f}_{22}(x) &= \biggl[ G_S \left( \psi \psi^\text{T} \right)_x + G_P i \gamma_5 \left( \psi \psi^\text{T} \right)_x i \gamma_5 + C_4 \left\{ \left( \psi \psi^\text{T} \right)_x, i\gamma_5 \right\} \biggl], \nonumber\\
    V^{4f}_{12}(x) &= \biggl[ -G_S \left( \bar{\psi}^\text{T} \psi^\text{T} \right)_x - G_P i \gamma_5 \left( \bar{\psi}^\text{T} \psi^\text{T} \right)_x i \gamma_5 - C_4 \left\{ \left( \bar{\psi}^\text{T} \psi^\text{T} \right)_x, i\gamma_5 \right\} \biggl] \nonumber\\
    &\quad + \biggl[ G_S \textbf{1}_{\rm total} \left( \bar{\psi} \psi \right)_x + G_P i \gamma_5 \left( \bar{\psi} i \gamma_5 \psi \right)_x + C_4 \textbf{1}_{\rm total} \left( \bar{\psi} i \gamma_5 \psi \right)_x + C_4 i \gamma_5 \left( \bar{\psi} \psi \right)_x \biggl], \nonumber\\
    V^{4f}_{21}(x) &= \biggl[ -G_S \left( \psi \bar{\psi} \right)_x - G_P i \gamma_5 \left( \psi \bar{\psi} \right)_x i \gamma_5 - C_4 \left\{ \left( \psi \bar{\psi} \right)_x, i\gamma_5 \right\} \biggl] \nonumber\\
    &\quad - \biggl[ G_S \textbf{1}_{\rm total} \left( \bar{\psi} \psi \right)_x + G_P i \gamma_5 \left( \bar{\psi} i \gamma_5 \psi \right)_x + C_4 \textbf{1}_{\rm total} \left( \bar{\psi} i \gamma_5 \psi \right)_x + C_4 i \gamma_5 \left( \bar{\psi} \psi \right)_x \biggl],
\end{align}
where the singlet tensorial structure is defined as $\textbf{1}_{\rm total} = \textbf{1}_{\rm Dirac} \otimes \textbf{1}_{\rm flavor} \otimes \textbf{1}_{\rm color}$.
In the current work, we only adopt the large-$N$ leading terms in $\mathcal{V}_k^{4f}$ such that
\begin{align}
    \mathcal{V}_K^{4f,\rm LN} = \int_x \, e^{-i(q-p)\cdot x} \pmat{
    V^{4f,\rm LN}_{11} & V^{4f,\rm LN}_{12} \\[2ex]
    V^{4f,\rm LN}_{21} & V^{4f,\rm LN}_{22}
    }_x,
\end{align}
with
\begin{align}
    V^{4f,\rm LN}_{11}(x) &\approx 0, \nonumber\\
    V^{4f,\rm LN}_{22}(x) &\approx 0, \nonumber\\
    V^{4f,\rm LN}_{12}(x) &= + \biggl[ G_S \textbf{1}_{\rm total} \left( \bar{\psi} \psi \right)_x + G_P i \gamma_5 \left( \bar{\psi} i \gamma_5 \psi \right)_x + C_4 \textbf{1}_{\rm total} \left( \bar{\psi} i \gamma_5 \psi \right)_x + C_4 i \gamma_5 \left( \bar{\psi} \psi \right)_x \biggl], \nonumber\\
    V^{4f,\rm LN}_{21}(x) &= - \biggl[ G_S \textbf{1}_{\rm total} \left( \bar{\psi} \psi \right)_x + G_P i \gamma_5 \left( \bar{\psi} i \gamma_5 \psi \right)_x + C_4 \textbf{1}_{\rm total} \left( \bar{\psi} i \gamma_5 \psi \right)_x + C_4 i \gamma_5 \left( \bar{\psi} \psi \right)_x \biggl].
    \label{eq:largeNApprox}
\end{align}
The schematic sketch of the large-$N$ approximation is shown in \Cref{fig:LargeNFourFermion}.

The fermion-gluon mixing terms with the vanished gluon background field are given by
\begin{align}
    &\left( H_{BF} \right)_\mu^a = i g_s \int_x \, e^{-i(q-p)\cdot x} \pmat{ \bar{\psi}_x \gamma_\mu T^a & - \psi^\text{T}_x T^{a}{}^\text{T} \gamma_\mu^\text{T} } , \nonumber\\
    &\left( H_{FB} \right)_\nu^b = i g_s \int_x \, e^{-i(q-p)\cdot x} \pmat{ -\gamma_\nu^\text{T} T^{b}{}^\text{T} \bar{\psi}^\text{T}_x  \\[1ex] \gamma_\nu T^b \psi_x }.
\end{align}
Here, we keep all the momentum and spacetime dependence in preparation for the systematic derivative expansion in the latter sections.

\subsection{Flow equation and vertex expansion}

The Wetterich equation \labelcref{eq:WetterichEq} can be deformed into
\begin{align}
    \partial_t \Gamma_k = \frac{1}{2} \tilde{\partial}_t \operatorname{STr} \log M,
    \label{eq:altWetterichEquation}
\end{align}
where we defined $M \equiv \Gamma_k^{(2)} + \mathcal{R}_k$.
The supermatrix $M$ reads
\begin{align}
    M = \pmat{ M_{BB} & M_{BF} \\ M_{FB} & M_{FF} } = \pmat{ M_{BB} & 0 \\ M_{FB} & 1 } \pmat{ 1 & M_{BB}^{-1} M_{BF} \\ 0 & M_{FF} - M_{FB} M_{BB}^{-1} M_{BF} }.
\end{align}
Then, the supertrace of the logarithm in \cref{eq:altWetterichEquation} is decomposed as
\begin{align}
    \operatorname{STr} \log M = \Tr \log M_{BB} - \Tr \log \big( M_{FF} - M_{FB} M_{BB}^{-1} M_{BF} \big).
    \label{eq:decomposedFlow}
\end{align}
Since the first term of \cref{eq:decomposedFlow} has no external fermionic field dependence, we shall drop this term in the latter analysis.
The fermionic field-dressed effective self-energy $- M_{FB} M_{BB}^{-1} M_{BF}$ reads
\begin{align}
    &- \left( M_{FB} \right)_\mu^a \left( M_{BB}^{-1} \right)_{\mu\nu}^{ab} \left( M_{BF} \right)_{\nu}^{b} \nonumber\\
    &= +g_s^2\int_{x,p_1,q_1,y} e^{-i (q - p_1)\cdot x} \pmat{ -\gamma_\mu^\text{T} T^{a,\text{T}} \bar{\psi}^\text{T}_x  \\ \gamma_\mu T^a \psi_x } \pmat{ \bar{\psi}_y \gamma_\nu T^b & - \psi^\text{T}_y T^{b,\text{T}} \gamma_\nu^\text{T} } e^{-i(q_1 - p)\cdot y}  (2\pi)^4 \delta^{(4)}(p_1 - q_1) \delta^{ab} D_{r,\mu\nu}(p_{1}) \nonumber\\
    &= +g_s^2\int_{x,p_1,y} e^{-i (q - p_1)\cdot x} e^{-i(p_1 - p)\cdot y}\pmat{ -\gamma_\mu^\text{T} T^{a,\text{T}} \bar{\psi}^\text{T}_x \bar{\psi}_y \gamma_\nu T^a & \gamma_\mu^\text{T} T^{a,\text{T}} \bar{\psi}^\text{T}_x \psi^\text{T}_y T^{b,\text{T}} \gamma_\nu^\text{T} \\ \\ \gamma_\mu T^a \psi_x \bar{\psi}_y \gamma_\nu T^b  & - \gamma_\mu T^a \psi_x \psi^\text{T}_y T^{b,\text{T}} \gamma_\nu^\text{T} }D_{r,\mu\nu}(p_{1}) \nonumber\\
    & \equiv \int_{p_1} \, \mathcal{V}^{\rm QCD}_{k,\mu\nu} D_{r,\mu\nu}(p_{1}),
    \label{eq:QCDInducedSelfEnergy}
\end{align}
where the superscript ``QCD'' denotes the QCD-induced contributions, and we defined
\begin{align}
    D_{r,\mu\nu}(p) \equiv \frac{1}{p^2(1+r^A_k)}\Pi_{\mu\nu}^{\xi}.
\end{align}
We denote the components of $\mathcal{V}^{\rm QCD}_{k,\mu\nu}$ as
\begin{align}
    \mathcal{V}^{\rm QCD}_{k,\mu\nu} = g_s^2 \int_{x,y} e^{-i (q - p_1)\cdot x} e^{-i(p_1 - p)\cdot y} \pmat{ V^{\rm QCD,NL_1}_{k,\mu\nu} & V^{\rm QCD,L_1}_{k,\mu\nu} \\[2ex]
    V^{\rm QCD,L_2}_{k,\mu\nu} & V^{\rm QCD,NL_2}_{k,\mu\nu} }_{x,y},
\end{align}
where the superscript ``L'' denotes ``ladder'' and ``NL'' denotes ``nonladder''.
See diagrammatic expressions in \cref{eq:feynmanDiagramsInTotal} below.

Then the flow equation \labelcref{eq:altWetterichEquation} is written as
\begin{align}
    \partial_t \Gamma_k = -\frac{1}{2} \tilde{\partial}_t \Tr \log \big( \mathcal{P}^{-1}_k + \mathcal{V}_k \big),
\end{align}
with
\begin{align}
    \mathcal{P}^{-1}_k \equiv \mathcal{K}_k + \mathcal{R}_k^\psi, \qquad \mathcal{V}_k \equiv \mathcal{V}_K^{4f,\rm LN} + \int_{p_1} \, \mathcal{V}^{\rm QCD}_{k,\mu\nu} D_{r,\mu\nu}(p_{1}).
\end{align}

We then perform a systematic vertex expansion with respect to $\mathcal{V}_k$, the trace of the logarithm reads
\begin{align}
    -\frac{1}{2} \Tr \log \big( \mathcal{P}^{-1}_k + \mathcal{V}_k \big) = -\frac{1}{2}  \Tr \log \big( \mathcal{P}^{-1}_k \big) - \frac{1}{2} \Tr \big[ \mathcal{P}_k \mathcal{V}_k \big] + \frac{1}{4} \Tr \big[ \mathcal{P}_k \mathcal{V}_k \mathcal{P}_k \mathcal{V}_k \big] + \cdots.
\end{align}
Given the definition of $\tilde{\partial}_t$
\begin{align}
    \tilde{\partial}_t \equiv \big( \partial_t \mathcal{R}_k \big)_{ab} \left(\frac{\delta}{\delta\mathcal{R}_k} \right)_{ba},
\end{align}
and the general structure of the regulator \labelcref{eq:generalRegulator}, we have
\begin{align}
    &\tilde{\partial}_t \mathcal{P}_k = - \mathcal{P}_k \mathcal{R}^\psi_k \mathcal{P}_k, \nonumber\\
    &\tilde{\partial}_t \mathcal{V}_k = \int_{p_1} \, \mathcal{V}^{\rm QCD}_{k,\mu\nu} \tilde{\partial}_t D_{r,\mu\nu}(p_{1}) = - \int_{p_1} \, \mathcal{V}^{\rm QCD}_{k,\mu\nu} D_{r,\mu\rho}(p_{1}) \left( \partial_t R^A_{k,\rho\sigma} \right) D_{r,\sigma\nu}(p_{1}).
    \label{eq:regulatorPartialDerivative}
\end{align}
The vertex expansion structure for the flow equation \labelcref{eq:altWetterichEquation} is then obtained after we perform the partial derivative $\tilde{\partial}_t$ using \labelcref{eq:regulatorPartialDerivative}.
This results in 
\begin{align}
    \partial_t \Gamma_k = & - \frac{1}{2} \Tr \left[ \mathcal{P}_k \, \partial_t \mathcal{R}^\psi_k \right] + \frac{1}{2} \Tr \left[ \mathcal{P}_k \, \mathcal{V}_k \, \mathcal{P}_k \, \partial_t \mathcal{R}^\psi_k \right] + \frac{1}{2} \Tr \left[ \mathcal{P}_k \, \mathcal{V}^{\rm QCD}_{k,\mu\nu} \, D_{r,\mu\rho} \, \partial_t R^A_{k,\rho\sigma} \, D_{r,\sigma\nu} \right] \nonumber\\ 
    & - \frac{1}{2} \Tr \left[ \mathcal{P}_k \, \mathcal{V}_k \, \mathcal{P}_k \, \mathcal{V}_k \, \mathcal{P}_k \, \partial_t \mathcal{R}^\psi_k \right] - \frac{1}{2} \Tr \left[ \mathcal{P}_k \, \mathcal{V}_k \, \mathcal{P}_k \, \mathcal{V}^{\rm QCD}_{k,\mu\nu} \, D_{r,\mu\rho} \, \partial_t R^A_{k,\rho\sigma} \, D_{r,\sigma\nu} \right] + \cdots.
    \label{eq:vertexExpansionFlow}
\end{align}
We here present the field-independent regulated fermion propagator
\begin{align}
    \mathcal{P}_k &= (2\pi)^4 \delta^{(4)}(p-q) \pmat{
    0 & \frac{1}{i \Slash{p} (1+r_k^\psi) + m e^{i \gamma_5 \theta / 2}} \\
    \frac{1}{i \Slash{p}^\text{T} (1+r_k^\psi) - m e^{i \gamma_5 \theta / 2}} & 0
    } \nonumber\\
    &= (2\pi)^4 \delta^{(4)}(p-q) \pmat{
    0 & \frac{-i \Slash{p} (1+r_k^\psi) + m e^{-i \gamma_5 \theta / 2}}{p^2 (1+r_k^\psi)^2 + m^2} \\
    \frac{-i \Slash{p}^\text{T} (1+r_k^\psi) - m e^{-i \gamma_5 \theta / 2}}{p^2 (1+r_k^\psi)^2 + m^2} & 0
    } \nonumber\\
    &\equiv (2\pi)^4 \delta^{(4)}(p-q) \pmat{
    0 & P_k \\
    P_k^\prime & 0
    }_p.
\end{align}

There is a clear correspondence between each term in \cref{eq:vertexExpansionFlow} and the Feynman diagrammatic expression as shown below.
\begin{subequations}
    \label{eq:feynmanDiagramsInTotal}
\begin{align}
    & + \frac{1}{2} \Tr \left[ \mathcal{P}_k \, \mathcal{V}_k \, \mathcal{P}_k \, \partial_t \mathcal{R}^\psi_k \right] \quad \sim \quad \adjincludegraphics[valign=c, width = 0.2\textwidth, raise=0.2\baselineskip, set vsize={1.3cm}{0cm}]{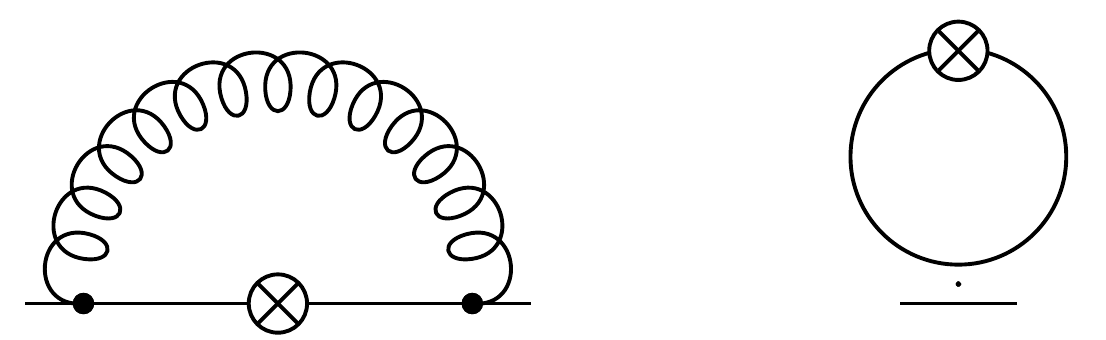}, \label{eq:feynmanDiagrams1}\\
    & + \frac{1}{2} \Tr \left[ \mathcal{P}_k \, \mathcal{V}^{\rm QCD}_{k,\mu\nu} \, D_{r,\mu\rho} \, \partial_t R^A_{k,\rho\sigma} \, D_{r,\sigma\nu} \right] \quad \sim \quad \adjincludegraphics[valign=c, width = 0.1\textwidth, raise=0.2\baselineskip, set vsize={1.3cm}{0cm}]{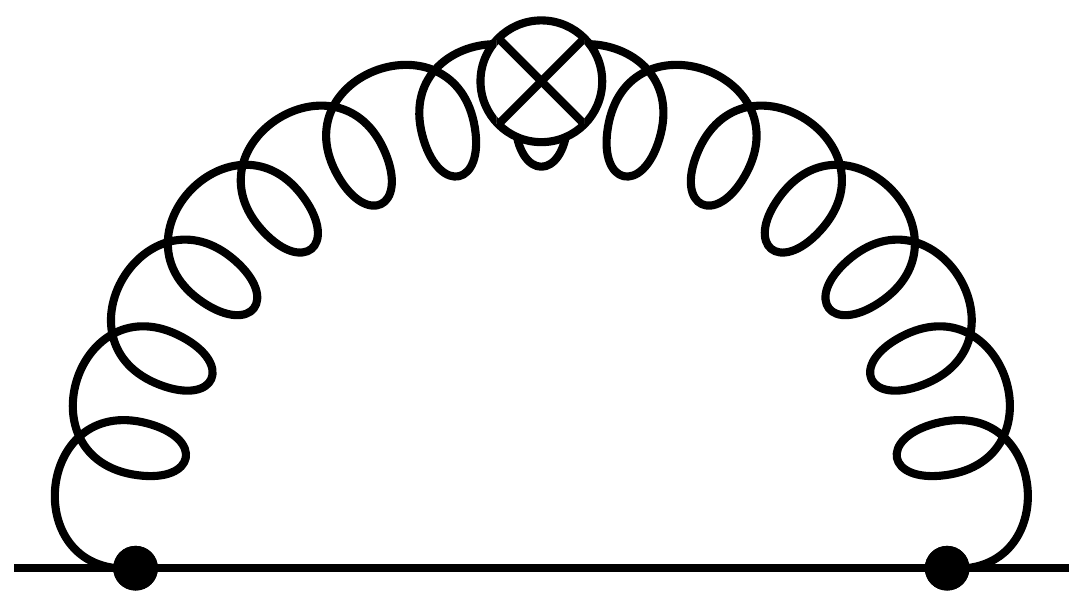}, 
    \label{eq:feynmanDiagrams2}\\
    & - \frac{1}{2} \Tr \left[ \mathcal{P}_k \, \mathcal{V}_k \, \mathcal{P}_k \, \mathcal{V}_k \, \mathcal{P}_k \, \partial_t \mathcal{R}^\psi_k \right] \quad \sim \quad \adjincludegraphics[valign=c, width = 0.5\textwidth, raise=0.2\baselineskip, set vsize={1.3cm}{0cm}]{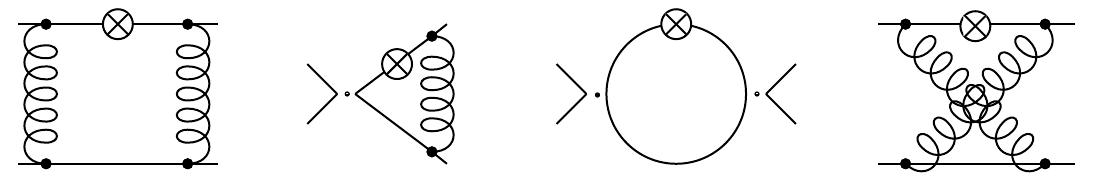},
       \label{eq:feynmanDiagrams3}\\
    & - \frac{1}{2} \Tr \left[ \mathcal{P}_k \, \mathcal{V}_k \, \mathcal{P}_k \, \mathcal{V}^{\rm QCD}_{k,\mu\nu} \, D_{r,\mu\rho} \, \partial_t R^A_{k,\rho\sigma} \, D_{r,\sigma\nu} \right] \quad \sim \quad \adjincludegraphics[valign=c, width = 0.35\textwidth, raise=0.2\baselineskip, set vsize={1.3cm}{0cm}]{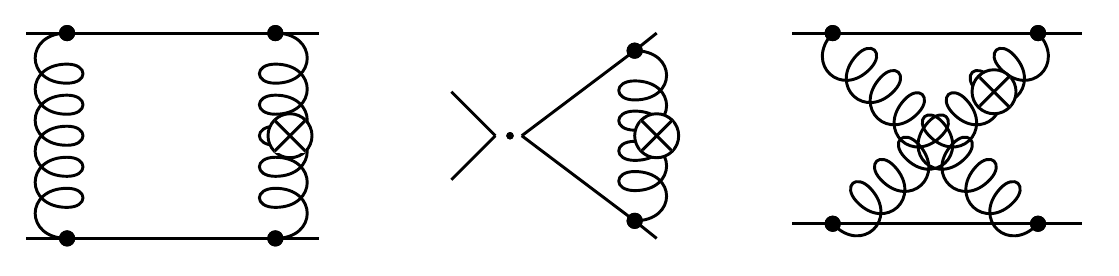},
        \label{eq:feynmanDiagrams4}
\end{align}
\end{subequations}
where the solid and curly lines stand for the fermion and gauge fields, respectively, and the cross circle in the propagators is the regulator insertion $\p_t R_k$.
In \cref{eq:feynmanDiagrams3,eq:feynmanDiagrams4}, the first term corresponds to the ladder diagram, while the last term is a nonladder (or cross ladder) one.

\subsection{Projection to Fierz subspace}
\label[appendix]{sec:ProjectionToFierzSubspace}

Before evaluating the diagrams in \cref{eq:feynmanDiagramsInTotal}, we shall review the way of projecting the flow onto the truncated subspace.
For the Dirac bilinears $\bar{\psi} \Gamma \psi$, we decompose them into 16 independent bases in the 4-dimensional spacetime
\begin{align}
    \Gamma \in \left\{ \textbf{1}_{\rm Dirac} , \, \gamma_5 , \, \gamma_\mu , \, \gamma_\mu \gamma_5 , \, \sigma_{\mu\nu} \right\} \equiv \left\{ \Gamma_I \right\},
\end{align}
which reads
\begin{align}
    \Gamma = \sum_I \, f_I \Gamma_I,
\end{align}
and the expansion coefficients are given by
\begin{align}
    f_I = \operatorname{tr} \left[ \Gamma \Gamma_I \right]/\operatorname{tr} \left[ \Gamma_I^2 \right].
    \label{eq:projectionOfBilinear}
\end{align}

In the evaluation of the four-point vertex flow, we encounter the following quadrilinear form
\begin{align}
    \bar{\psi}_1 \, M_1 \, \psi_2 \, \bar{\psi}_3 \, M_2 \, \psi_4.
\end{align}
In order to perform the projection in \cref{eq:projectionOfBilinear}, we perform the Fierz transformation for Dirac quadrilinears such that
\begin{align}
    \bar{\psi}_1 \, M_1 \, \psi_2 \, \bar{\psi}_3 \, M_2 \, \psi_4 = -\frac{1}{4} \sum_I \, \bar{\psi}_1 \left( M_1 \, \Gamma_I \, M_2  \right) \psi_4 \, \bar{\psi}_3 \left( \Gamma_I \right) \psi_2,
    \label{eq:fierzTransformation}
\end{align}
and then project the matrix $M_1 \, \Gamma_I \, M_2$ onto the basis $\left\{ \Gamma_I \right\}$.

\subsection{Mass terms}

Now, we are ready to calculate the flow equation for each coupling constant in the IR effective action \labelcref{eq:IRQCDaction}.
We first focus on the flow of the inverse two-point function, which is extracted from the first two lines of \cref{eq:feynmanDiagramsInTotal}, i.e.,
\begin{align}
    \partial_t \Gamma_k \ni \frac{1}{2} \Tr \left[ \mathcal{P}_k \, \mathcal{V}_k \, \mathcal{P}_k \, \partial_t \mathcal{R}^\psi_k \right] + \frac{1}{2} \Tr \left[ \mathcal{P}_k \, \mathcal{V}_{k,\mu\nu}^{\rm QCD} D_{r,\mu\nu} \, \mathcal{P}_k \, \partial_t \mathcal{R}^\psi_k \right] + \cdots.
    \label{eq:massTermLoops}
\end{align}
Also, the first term of \cref{eq:massTermLoops} can be decomposed into
\begin{align}
    \frac{1}{2} \Tr \left[ \mathcal{P}_k \, \mathcal{V}_k \, \mathcal{P}_k \, \partial_t \mathcal{R}^\psi_k \right] = \frac{1}{2} \Tr \left[ \mathcal{P}_k \, \mathcal{V}_k^{4f, \rm LN} \, \mathcal{P}_k \, \partial_t \mathcal{R}^\psi_k \right] + \frac{1}{2} \Tr \left[ \mathcal{P}_k \, \mathcal{V}_{k,\mu\nu}^{\rm QCD} D_{r,\mu\nu} \, \mathcal{P}_k \, \partial_t \mathcal{R}^\psi_k \right].
\end{align}
We shall evaluate these three terms individually in the following three subsections.

\subsubsection{Evaluation of \texorpdfstring{$\frac{1}{2} \Tr \left[ \mathcal{P}_k \, \mathcal{V}_k^{4f, \rm LN} \, \mathcal{P}_k \, \partial_t \mathcal{R}^\psi_k \right]$}{}}

The contribution from the four-fermion interaction reads
\begin{align}
    &\frac{1}{2} \Tr \left[ \mathcal{P}_k \, \mathcal{V}_k^{4f, \rm LN} \, \mathcal{P}_k \, \partial_t \mathcal{R}^\psi_k \right] \nonumber\\
    =& \frac{1}{2} \int_x \int_p \, \operatorname{tr} \left[ 
    \pmat{
    0 & P_k \\
    P_k^\prime & 0
    }_p
    \pmat{
     & V^{4f,\rm LN}_{12} \\
    V^{4f,\rm LN}_{21} & 
    }_x
    \pmat{
    0 & P_k \\
    P_k^\prime & 0
    }_p
    \pmat{
    0 & \partial_t r_k^\psi i \Slash{p}^\text{T} \\
    \partial_t r_k^\psi i \Slash{p} & 0
    }
    \right] \nonumber\\
    =& \frac{1}{2} \int_x \int_p \, \operatorname{tr} \left[ P_k \, V^{4f,\rm LN}_{21} \, P_k \, i \Slash{p} + P_k^\prime \, V^{4f,\rm LN}_{12} \, P_k^\prime \, i \Slash{p}^\text{T} \right] \partial_t r_k^\psi.
    \label{eq:beginingTwoPointFourFermionFermiReg}
\end{align}
Here we compute
\begin{align}
    &\operatorname{tr} \left[P_k \, V^{4f,\rm LN}_{21} \, P_k \, i \Slash{p} \right] \nonumber\\
    &\qquad = \operatorname{tr} \left\{ \frac{-i \Slash{p} (1+r_k^\psi) + m e^{-i \gamma_5 \theta / 2}}{p^2 (1+r_k^\psi)^2 + m^2} \, \left[ - \left( \bar{\psi} \psi \right)_x G_S \cdot \textbf{1}_{\rm total} - \left( \bar{\psi} i\gamma_5 \psi \right)_x G_P \cdot i\gamma_5 \right. \right. \nonumber\\
    &  \hspace{0.2\linewidth} - \left. \left. \left( \bar{\psi} i\gamma_5 \psi \right)_x C_4 \cdot \textbf{1}_{\rm total} - \left( \bar{\psi} \psi \right)_x C_4 \cdot i\gamma_5 \right] \, \frac{-i \Slash{p} (1+r_k^\psi) + m e^{-i \gamma_5 \theta / 2}}{p^2 (1+r_k^\psi)^2 + m^2} \, i \Slash{p} \right\}\nonumber\\
   &\qquad= \frac{- 8 N_c N_f m p^2 (1+r_k^\psi)}{\left[ p^2 (1+r_k^\psi)^2 + m^2 \right]^2} \biggl[ \biggl( C_4 \left( \bar{\psi} i\gamma_5 \psi \right)_x + G_S \left( \bar{\psi} \psi \right)_x \biggl) \cos{\frac{\theta}{2}} + \biggl( G_P \left( \bar{\psi} i\gamma_5 \psi \right)_x + C_4 \left( \bar{\psi} \psi \right)_x \biggl) \sin{\frac{\theta}{2}} \biggl],
    \label{eq:TwoPoint4fermiFermiReg}
\end{align}
\begin{align}
    &\operatorname{tr} \left[P_k^\prime \, V^{4f,\rm LN}_{12} \, P_k^\prime \, i \Slash{p}^\text{T} \right]
    = \operatorname{tr} \left[P_k \, V^{4f,\rm LN}_{21} \, P_k \, r_k^\psi i \Slash{p} \right]\nonumber\\
    &\qquad= \operatorname{tr} \left\{ \frac{-i \Slash{p}^\text{T} (1+r_k^\psi) - m e^{-i \gamma_5 \theta / 2}}{p^2 (1+r_k^\psi)^2 + m^2} \, \left[ + \left( \bar{\psi} \psi \right)_x G_S \cdot \textbf{1}_{\rm total} + \left( \bar{\psi} i\gamma_5 \psi \right)_x G_P \cdot i\gamma_5 \right. \right. \nonumber\\
    &  \hspace{0.2\linewidth} + \left. \left. \left( \bar{\psi} i\gamma_5 \psi \right)_x C_4 \cdot \textbf{1}_{\rm total} + \left( \bar{\psi} \psi \right)_x C_4 \cdot i\gamma_5 \right] \, \frac{-i \Slash{p}^\text{T} (1+r_k^\psi) - m e^{-i \gamma_5 \theta / 2}}{p^2 (1+r_k^\psi)^2 + m^2} \, i \Slash{p} \right\} \nonumber\\
    &\qquad= \frac{- 8 N_c N_f m p^2 (1+r_k^\psi)}{\left[ p^2 (1+r_k^\psi)^2 + m^2 \right]^2} \biggl[ \biggl( C_4 \left( \bar{\psi} i\gamma_5 \psi \right)_x + G_S \left( \bar{\psi} \psi \right)_x \biggl) \cos{\frac{\theta}{2}} + \biggl( G_P \left( \bar{\psi} i\gamma_5 \psi \right)_x + C_4 \left( \bar{\psi} \psi \right)_x \biggl) \sin{\frac{\theta}{2}} \biggl].
    \label{eq:antiTwoPoint4fermiFermiReg}
\end{align}
Thus, \cref{eq:beginingTwoPointFourFermionFermiReg} is given by
\begin{align}
    &\frac{1}{2} \Tr \left[ \mathcal{P}_k \, \mathcal{V}_k^{4f, \rm LN} \, \mathcal{P}_k \, \partial_t \mathcal{R}^\psi_k \right] \nonumber\\
    =& \int_x \int_p \, \frac{- 8\, N_c\, N_f\, m\, \partial_t r_k^\psi\, p^2 (1+r_k^\psi)}{\left[ p^2 (1+r_k^\psi)^2 + m^2 \right]^2}  \biggl[ \biggl( C_4 \left( \bar{\psi} i\gamma_5 \psi \right)_x + G_S \left( \bar{\psi} \psi \right)_x \biggl) \cos{\frac{\theta}{2}} + \biggl( G_P \left( \bar{\psi} i\gamma_5 \psi \right)_x + C_4 \left( \bar{\psi} \psi \right)_x \biggl) \sin{\frac{\theta}{2}} \biggl].
    \label{eq:twoPointFourFermionFermiReg}
\end{align}
Note here that this contribution is external-momentum independent.
We see that the anti-fermion loops, e.g., \labelcref{eq:antiTwoPoint4fermiFermiReg}, are always equivalent to the fermion loops, e.g., \labelcref{eq:TwoPoint4fermiFermiReg}, thanks to charge conjugation invariance.
Thus, hereafter, we will omit the detailed steps in the evaluation of the anti-fermion loops.

\subsubsection{Evaluation of \texorpdfstring{$\frac{1}{2} \Tr \left[ \mathcal{P}_k \, \mathcal{V}_{k,\mu\nu}^{\rm QCD} D_{r,\mu\nu} \, \mathcal{P}_k \, \partial_t \mathcal{R}^\psi_k \right]$}{}}

The QCD-induced loop reads
\begin{align}
    &\frac{1}{2} \Tr \left[ \mathcal{P}_k \, \mathcal{V}_{k,\mu\nu}^{\rm QCD} D_{r,\mu\nu} \, \mathcal{P}_k \, \partial_t \mathcal{R}^\psi_k \right] \nonumber\\
    =& \frac{g_s^2}{2} \int_p \int_{x,y,p_1} \, e^{-i (p-p_1) \cdot (x-y)} \nonumber\\
    &\hspace{0.1\textwidth} \times \operatorname{tr}\left[ 
    \pmat{
    0 & P_k \\
    P_k^\prime & 0
    }_p
    \pmat{ V^{\rm QCD,NL_1}_{k,\mu\nu} & V^{\rm QCD,L_1}_{k,\mu\nu} \\ V^{\rm QCD,L_2}_{k,\mu\nu} & V^{\rm QCD,NL_2}_{k,\mu\nu} }_{x,y}
    \pmat{
    0 & P_k \\
    P_k^\prime & 0
    }_p
    \pmat{
    0 & \partial_t r_k^\psi i \Slash{p}^\text{T} \\
    \partial_t r_k^\psi i \Slash{p} & 0
    }
    \right] D_{r,\mu\nu}(p_1) \nonumber\\
    =& \frac{g_s^2}{2} \int_p \int_{x,y,p_1} \, e^{-i (p-p_1) \cdot (x-y)} \operatorname{tr} \left[ P_k \, V^{\rm QCD,L_2}_{k,\mu\nu} \, P_k \, i \Slash{p} + P_k^\prime \, V^{\rm QCD,L_1}_{k,\mu\nu} \, P_k^\prime \, i \Slash{p}^\text{T} \right]  \partial_t r_k^\psi \, D_{r,\mu\nu}(p_1).
\end{align}
In order to project the flow onto the local interactions, we perform the derivative expansion with respect to the external momentum $p_{\rm ex} = p-p_1$ inside the loop, i.e., we replace the momentum argument of the gluon propagator by $D_{r,\mu\nu}(p_1) \rightarrow D_{r,\mu\nu}(p)$, which implies the lowest order of the derivative expansion.
Thus, the integral with respect to $p_1$ can be performed, leaving the local interaction
\begin{align}
    \int_{x,y,p_1} e^{i (p_1) \cdot (x-y)} = \int_{x,y} \delta^{(4)}(x-y).
    \label{eq:localInteractionDelta}
\end{align}

Then, the trace inside the fermion loop reads
\begin{align}
    &\operatorname{tr} \left[ P_k \, V^{\rm QCD,L_2}_{k,\mu\nu} \, P_k \, i \Slash{p} \right]  \partial_t r_k^\psi \, D_{r,\mu\nu}(p) \nonumber\\
    =& \operatorname{tr} \left[ \frac{-i \Slash{p} (1+r_k^\psi) + m e^{-i \gamma_5 \theta / 2}}{p^2 (1+r_k^\psi)^2 + m^2} \, \gamma_\mu T^a \psi_x \bar{\psi}_y \gamma_\nu T^a \, \frac{-i \Slash{p} (1+r_k^\psi) + m e^{-i \gamma_5 \theta / 2}}{p^2 (1+r_k^\psi)^2 + m^2} \, i \Slash{p} \right] \partial_t r_k^\psi \, \frac{1}{p^2(1+r_k^A)} \Pi^\xi_{\mu\nu} \nonumber\\
    =& - \bar{\psi}_y \left[ \gamma_\nu T^a \, \frac{-i \Slash{p} (1+r_k^\psi) + m e^{-i \gamma_5 \theta / 2}}{p^2 (1+r_k^\psi)^2 + m^2} \, i \Slash{p} \, \frac{-i \Slash{p} (1+r_k^\psi) + m e^{-i \gamma_5 \theta / 2}}{p^2 (1+r_k^\psi)^2 + m^2} \, \gamma_\mu T^a \right] \psi_x \, \frac{\partial_t r_k^\psi }{p^2(1+r_k^A)} \Pi^\xi_{\mu\nu} \nonumber\\
    \equiv& - \bar{\psi}_y \left[ \mathcal{O}_{1,\mu\nu} \right] \psi_x \, \frac{\partial_t r_k^\psi }{p^2(1+r_k^A)} \Pi^\xi_{\mu\nu}.
\end{align}
Following the discussion in \Cref{sec:ProjectionToFierzSubspace}, we project the above flow onto the tensorial structures $\textbf{1}_{\rm Dirac}$ and $\gamma_5$, which reads
\begin{align}
- \bar{\psi}_y \left[ \mathcal{O}_{1,\mu\nu} \right] \psi_x \, \frac{\partial_t r_k^\psi }{p^2(1+r_k^A)} \Pi^\xi_{\mu\nu} 
    \rightarrow - C_2 \bar{\psi}_y \biggl[ \textbf{1}_{\rm all} \cos{\frac{\theta}{2}} +  i \gamma_5 \sin{\frac{\theta}{2}} \biggl] \psi_x \frac{2 (3+\xi) m\, \partial_t r_k^\psi\, p^2 (1+r_k^\psi)}{\left[ p^2(1+r_k^\psi)^2 + m^2  \right]^2 \left[ p^2 (1+r_k^A) \right]}.
\end{align}
We have used the Casimir operatore $T^a T^a = (N_c^2 - 1)/(2 N_c) \mathbf{1}_{\rm color}$.
Similarly, we have
\begin{align}
\operatorname{tr} \left[P_k^\prime \, V^{\rm QCD,L_1}_{k,\mu\nu} \, P_k^\prime \, i \Slash{p}^\text{T} \right]  \partial_t r_k^\psi \, D_{r,\mu\nu}(p)
    \rightarrow - C_2 \bar{\psi}_x \biggl[ \textbf{1}_{\rm all} \cos{\frac{\theta}{2}} +  i \gamma_5 \sin{\frac{\theta}{2}} \biggl] \psi_y \frac{2 (3+\xi) m\, \partial_t r_k^\psi\, p^2 (1+r_k^\psi)}{\left[ p^2(1+r_k^\psi)^2 + m^2  \right]^2 \left[ p^2 (1+r_k^A) \right]}.
\end{align}
After performing the $y$-integral in \cref{eq:localInteractionDelta}, the loop induced by the QCD interaction at the leading order of the derivative expansion then becomes
\begin{align}
\frac{1}{2} \Tr \left[ \mathcal{P}_k \, \mathcal{V}_{k,\mu\nu}^{\rm QCD} D_{r,\mu\nu} \, \mathcal{P}_k \, \partial_t \mathcal{R}^\psi_k \right] 
    \rightarrow - g_s^2 C_2 \int_x \int_p \, \frac{2 (3+\xi) m\, \partial_t r_k^\psi\, p^2 (1+r_k^\psi)}{\left[ p^2(1+r_k^\psi)^2 + m^2  \right]^2 \left[ p^2 (1+r_k^A) \right]} \biggl[ \left( \bar{\psi} \psi \right)_x \cos{\frac{\theta}{2}} +  \left( \bar{\psi} i\gamma_5 \psi \right)_x\sin{\frac{\theta}{2}} \biggl].
    \label{eq:twoPointQCDFermiReg}
\end{align}

\subsubsection{Evaluation of \texorpdfstring{$\frac{1}{2} \Tr \left[ \mathcal{P}_k \, \mathcal{V}^{\rm QCD}_{k,\mu\nu} \, D_{r,\mu\rho} \, \partial_t R^A_{k,\rho\sigma} \, D_{r,\sigma\nu} \right]$}{}}

The second line of \cref{eq:feynmanDiagramsInTotal} is similar to the QCD-induced loop evaluated above, except for the position of the regulator insertion, which is in the gluon propagator
\begin{align}
    &\frac{1}{2} \Tr \left[ \mathcal{P}_k \, \mathcal{V}^{\rm QCD}_{k,\mu\nu} \, D_{r,\mu\rho} \, \partial_t R^A_{k,\rho\sigma} \, D_{r,\sigma\nu} \right] \nonumber\\
    =& \frac{g_s^2}{2} \int_p \int_{x,y,p_1} \, e^{-i (p-p_1)\cdot(x-y)} \operatorname{tr} \left[ 
    \pmat{
    0 & P_k \\
    P_k^\prime & 0
    }_p
    \pmat{ V^{\rm QCD,NL_1}_{k,\mu\nu} & V^{\rm QCD,L_1}_{k,\mu\nu} \\ V^{\rm QCD,L_2}_{k,\mu\nu} & V^{\rm QCD,NL_2}_{k,\mu\nu} }_{x,y}
     \right] \left( D_{r,\mu\rho} \, \partial_t R^A_{k,\rho\sigma} \, D_{r,\sigma\nu} \right)_{p_1} \nonumber\\
     =& \frac{g_s^2}{2} \int_p \int_{x,y,p_1} \, e^{-i (p-p_1)\cdot(x-y)} \operatorname{tr} \left[ P_k \, V^{\rm QCD,L_2}_{k,\mu\nu} + P_k^\prime \, V^{\rm QCD,L_1}_{k,\mu\nu} \right] \left( D_{r,\mu\rho} \, \partial_t R^A_{k,\rho\sigma} \, D_{r,\sigma\nu} \right)_{p_1}.
\end{align}
With the same treatment on the derivative expansion, the above trace can be performed as
\begin{align}
    &\operatorname{tr} \left[ P_k \, V^{\rm QCD,L_2}_{k,\mu\nu} + P_k^\prime \, V^{\rm QCD,L_1}_{k,\mu\nu} \right] \left( D_{r,\mu\rho} \, \partial_t R^A_{k,\rho\sigma} \, D_{r,\sigma\nu} \right)_{p} \nonumber\\
    &= \operatorname{tr} \left[ \frac{-i \Slash{p} (1+r_k^\psi) + m e^{-i \gamma_5 \theta / 2}}{p^2 (1+r_k^\psi)^2 + m^2} \, \gamma_\mu T^a \psi_x \bar{\psi}_y \gamma_\nu T^a \right] \left( \Pi_{r,\mu\rho}^\xi \, \Pi_{k,\rho\sigma}^{1/\xi} \, \Pi_{r,\sigma\nu}^\xi \right)_{p} \frac{\partial_t r_k^A \, p^2}{\left[ p^2 (1+r_k^A) \right]^2} \nonumber\\
    &= - \bar{\psi}_y \biggl[ \gamma_\nu T^a \, \frac{-i \Slash{p} (1+r_k^\psi) + m e^{-i \gamma_5 \theta / 2}}{p^2 (1+r_k^\psi)^2 + m^2} \, \gamma_\mu T^a \biggl] \psi_x \left( \Pi_{r,\mu\rho}^\xi \, \Pi_{k,\rho\sigma}^{1/\xi} \, \Pi_{r,\sigma\nu}^\xi \right)_{p} \frac{\partial_t r_k^A \, p^2}{\left[ p^2 (1+r_k^A) \right]^2} \nonumber\\
    &\rightarrow -C_2 \frac{(3+\xi)m \, \partial_t r_k^A \, p^2}{\left[ p^2 (1+r_k^\psi)^2 + m^2 \right] \left[ p^2 (1+r_k^A) \right]^2} \bar{\psi}_y \biggl[ \textbf{1}_{\rm all} \cos{\frac{\theta}{2}} +  i \gamma_5 \sin{\frac{\theta}{2}} \biggl] \psi_x,
\end{align}
and similarly
\begin{align}
    &\operatorname{tr} \left[ P_k^\prime \, V^{\rm QCD,L_1}_{k,\mu\nu} + P_k^\prime \, V^{\rm QCD,L_1}_{k,\mu\nu} \right] \left( D_{r,\mu\rho} \, \partial_t R^A_{k,\rho\sigma} \, D_{r,\sigma\nu} \right)_{p} \nonumber\\
    &\qquad
    \rightarrow  -C_2 \frac{(3+\xi)m \, \partial_t r_k^A \, p^2}{\left[ p^2 (1+r_k^\psi)^2 + m^2 \right] \left[ p^2 (1+r_k^A) \right]^2} \bar{\psi}_x \biggl[ \textbf{1}_{\rm all} \cos{\frac{\theta}{2}} +  i \gamma_5 \sin{\frac{\theta}{2}} \biggl] \psi_y.
\end{align}
After performing the $y$-integral as in \cref{eq:twoPointQCDFermiReg}, the loop becomes
\begin{align}
    &\frac{1}{2} \Tr \left[ \mathcal{P}_k \, \mathcal{V}^{\rm QCD}_{k,\mu\nu} \, D_{r,\mu\rho} \, \partial_t R^A_{k,\rho\sigma} \, D_{r,\sigma\nu} \right] \nonumber\\
    &\rightarrow - g_s^2 C_2 \int_x \int_p \frac{(3+\xi)m \, \partial_t r_k^A \, p^2}{\left[ p^2 (1+r_k^\psi)^2 + m^2 \right] \left[ p^2 (1+r_k^A) \right]^2} \biggl[ \left( \bar{\psi} \psi \right)_x \cos{\frac{\theta}{2}} +  \left( \bar{\psi} i\gamma_5 \psi \right)_x\sin{\frac{\theta}{2}} \biggl].
    \label{eq:twoPointQCDBoseReg}
\end{align}

\subsubsection{Warp-up for the flow of mass term}

In the current work, we apply the following Litim-type regulators as mentioned in \Cref{eq:LitimRegulatorFunctions}
\begin{align}
    r_k^\psi = \left( \sqrt{\frac{k^2}{p^2}} - 1 \right) \theta (k^2 - p^2), \quad r_k^A = \left( \frac{k^2}{p^2} - 1 \right) \theta (k^2 - p^2),
\end{align}
with the derivative with respect to $t$
\begin{align}
    &\partial_t r_k^\psi = \sqrt{\frac{k^2}{p^2}} \theta (k^2 - p^2) +  \left( \sqrt{\frac{k^2}{p^2}} - 1 \right) \delta (k^2 - p^2) \, 2k^2, \nonumber\\
    &\partial_t r_k^A = 2 \frac{k^2}{p^2} \theta (k^2 - p^2) + \left( \frac{k^2}{p^2} - 1 \right) \delta (k^2 - p^2) \, 2k^2.
    \label{eq:derOfLitimRegulator}
\end{align}
In the latter evaluations of the loop momentum integrals, those second terms appeared in \cref{eq:derOfLitimRegulator} would vanish according to the Morris's Lemma, thus we shall omit them below.

Combining \cref{eq:twoPointFourFermionFermiReg,eq:twoPointQCDFermiReg,eq:twoPointQCDBoseReg}, the flow of the inverse two-point function reads
\begin{align}
    &\frac{1}{2} \Tr \left[ \mathcal{P}_k \, \mathcal{V}_k \, \mathcal{P}_k \, \partial_t \mathcal{R}^\psi_k \right] + \frac{1}{2} \Tr \left[ \mathcal{P}_k \, \mathcal{V}^{\rm QCD}_{k,\mu\nu} \, D_{r,\mu\rho} \, \partial_t R^A_{k,\rho\sigma} \, D_{r,\sigma\nu} \right] \nonumber\\
    \rightarrow & \int_x \int_p \, \frac{- 8\, N_c\, N_f\, m\, \partial_t r_k^\psi\, p^2 (1+r_k^\psi)}{\left[ p^2 (1+r_k^\psi)^2 + m^2 \right]^2} \biggl[ \biggl( G_S \cos{\frac{\theta}{2}} + C_4 \sin{\frac{\theta}{2}} \biggl) \left( \bar{\psi} \psi \right)_x + \biggl( G_P \sin{\frac{\theta}{2}} + C_4 \cos{\frac{\theta}{2}} \biggl) \left( \bar{\psi} i\gamma_5 \psi \right)_x \biggl] \nonumber\\
    &\quad + \int_x \int_p \, \frac{- 2(3+\xi)\, g_s^2\, C_2\, m\, \partial_t r_k^\psi\, p^2 (1+r_k^\psi)}{\left[ p^2 (1+r_k^\psi)^2 + m^2 \right]^2 \left[ p^2 (1+r_k^A) \right]} \biggl[ \left( \bar{\psi} \psi \right)_x \cos{\frac{\theta}{2}} + \left( \bar{\psi} i\gamma_5 \psi \right)_x \sin{\frac{\theta}{2}} \biggl] \nonumber\\
    &\quad + \int_x \int_p \, \frac{- (3+\xi)\, g_s^2\, C_2\, m\, \partial_t r_k^A\, p^2}{\left[ p^2 (1+r_k^\psi)^2 + m^2 \right] \left[ p^2 (1+r_k^A) \right]^2} \biggl[ \left( \bar{\psi} \psi \right)_x \cos{\frac{\theta}{2}} + \left( \bar{\psi} i\gamma_5 \psi \right)_x \sin{\frac{\theta}{2}} \biggl] \nonumber\\
    =& \int_x \Biggl\{ - 8\, N_c\, N_f\, m \biggl[ \biggl( G_S \cos{\frac{\theta}{2}} + C_4 \sin{\frac{\theta}{2}} \biggl) \left( \bar{\psi} \psi \right)_x + \biggl( G_P \sin{\frac{\theta}{2}} + C_4 \cos{\frac{\theta}{2}} \biggl) \left( \bar{\psi} i\gamma_5 \psi \right)_x \biggl] \mathcal{M}_1 \nonumber\\
    &\quad - 2(3+\xi)\, g_s^2\, C_2\, m \biggl[ \left( \bar{\psi} \psi \right)_x \cos{\frac{\theta}{2}} + \left( \bar{\psi} i\gamma_5 \psi \right)_x \sin{\frac{\theta}{2}} \biggl] \left( \mathcal{M}_2 + \mathcal{M}_3 \right) \Biggl\},
\end{align}
where the threshold functions have been defined as
\begin{align}
    &\mathcal{M}_1 \equiv \int \frac{\df^4 p}{(2\pi)^4} \, \frac{ p^2 (1+r_k^\psi) \partial_t r_k^\psi}{\left[ p^2 (1+r_k^\psi)^2 + m^2 \right]^2} = \frac{1}{2(4\pi)^2} \frac{k^6}{\left(k^2+m^2 \right)^2}, \nonumber\\
    &\mathcal{M}_2 \equiv \int \frac{\df^4 p}{(2\pi)^4} \, \frac{ p^2 (1+r_k^\psi) \partial_t r_k^\psi}{\left[ p^2 (1+r_k^\psi)^2 + m^2 \right]^2 \left[ p^2 (1+r_k^A) \right]} = \frac{1}{2(4\pi)^2} \frac{k^4}{\left(k^2+m^2 \right)^2}, \nonumber\\
    &\mathcal{M}_3 \equiv \frac{1}{2}\int \frac{\df^4 p}{(2\pi)^4} \, \frac{ p^2  \partial_t r_k^A}{\left[ p^2 (1+r_k^\psi)^2 + m^2 \right] \left[ p^2 (1+r_k^A) \right]^2} = \frac{1}{2(4\pi)^2} \frac{k^2}{\left(k^2+m^2 \right)}.
\end{align}

Considering the ansatz of the IR effective action \labelcref{eq:IRQCDaction}, the left-hand side of the flow equation \labelcref{eq:vertexExpansionFlow} reads
\begin{align}
    \partial_t \Gamma_k = \int_x \Biggl\{ \left( \bar{\psi} \psi \right)_x \partial_t \left( m \cos{\frac{\theta}{2}} \right) + \left( \bar{\psi} i \gamma_5 \psi \right)_x \partial_t \left( m \sin{\frac{\theta}{2}} \right) \Biggl\} + \cdots,
\end{align}
and we read off the flow equations for the mass parameter and the $\theta$ parameter as
\begin{align}
    &\partial_t \left( m \cos{\frac{\theta}{2}} \right) = - 8\, N_c\, N_f\, m \biggl( G_S \cos{\frac{\theta}{2}} + C_4 \sin{\frac{\theta}{2}} \biggl) \mathcal{M}_1 - 2(3+\xi)\, g_s^2\, C_2\, m \cos{\frac{\theta}{2}} \left( \mathcal{M}_2 + \mathcal{M}_3 \right), \nonumber\\
    &\partial_t \left( m \sin{\frac{\theta}{2}} \right) = - 8\, N_c\, N_f\, m \biggl( G_P \sin{\frac{\theta}{2}} + C_4 \cos{\frac{\theta}{2}} \biggl) \mathcal{M}_1 - 2(3+\xi)\, g_s^2\, C_2\, m \sin{\frac{\theta}{2}} \left( \mathcal{M}_2 + \mathcal{M}_3 \right).
\end{align}
After linear transformation, we then find the flow equation for $m$ and $\theta$
\begin{align}
    \partial_t m &= \cos{\frac{\theta}{2}} \partial_t \left( m \cos{\frac{\theta}{2}} \right) + \sin{\frac{\theta}{2}} \partial_t \left( m \sin{\frac{\theta}{2}} \right) \nonumber\\
    &= - 4 N_c N_f \biggl[ (G_S + G_P) + 2 C_4 \sin{\theta} + (G_S - G_P)\cos{\theta} \biggl] m \mathcal{M}_1 - 2(3+\xi) C_2 g_s^2 m \left( \mathcal{M}_2 + \mathcal{M}_3 \right), \nonumber\\
    \partial_t \theta &= \frac{2}{m} \biggl[ - \sin{\frac{\theta}{2}} \partial_t \left( m \cos{\frac{\theta}{2}} \right) + \cos{\frac{\theta}{2}} \partial_t \left( m \sin{\frac{\theta}{2}} \right) \biggl] \nonumber\\
    &= 8 N_c N_f \biggl[ (G_S - G_P) \sin{\theta} - 2 C_4 \cos{\theta} \biggl] \mathcal{M}_1.
\end{align}
Defining dimensionless parameters
\begin{align}
    \tilde{m} = m k^{-1}, \quad \tilde{G}_S = G_S k^2, \quad \tilde{G}_P = G_P k^2, \quad \tilde{C}_4 = C_4 k^2,
    \label{eq:dimensionlessQuantities}
\end{align}
the flow equations for $\tilde{m}$ and $\theta$ are given by
\begin{align}
    \partial_t \tilde{m} &= - \tilde{m} - 4 N_c N_f \biggl[ (\tilde{G}_S + \tilde{G}_P) + 2 \tilde{C}_4 \sin{\theta} + (\tilde{G}_S - \tilde{G}_P)\cos{\theta} \biggl]\tilde{m} \tilde{\mathcal{M}}_{(2)} \nonumber\\
    &\quad\quad - 2(3+\xi) C_2 g_s^2 \tilde{m} \left( \tilde{\mathcal{M}}_{(2)} + \tilde{\mathcal{M}}_{(1)} \right), \\
    \partial_t \theta &= 8 N_c N_f \biggl[ (\tilde{G}_S - \tilde{G}_P) \sin{\theta} - 2 \tilde{C}_4 \cos{\theta} \biggl] \tilde{\mathcal{M}}_{(2)},
\end{align}
where
\begin{align}
    \tilde{\mathcal{M}}_{(1)} &= \mathcal{M}_3 = \frac{1}{2(4\pi)^2} \frac{1}{\left(1+\tilde{m}^2\right)}, &
    \tilde{\mathcal{M}}_{(2)} &= \frac{1}{k^2} \mathcal{M}_1 = \mathcal{M}_2 = \frac{1}{2(4\pi)^2} \frac{1}{\left(1+\tilde{m}^2\right)^2}.
    \label{eq:thresholdfunction1}
\end{align}

\subsection{Four-fermi couplings}

We derive the flow equations for the four-fermi couplings, which are extracted from the second lines of \cref{eq:feynmanDiagramsInTotal}, i.e., 
\begin{align}
    \partial_t \Gamma_k \ni - \frac{1}{2} \Tr \left[ \mathcal{P}_k \, \mathcal{V}_k \, \mathcal{P}_k \, \mathcal{V}_k \, \mathcal{P}_k \, \partial_t \mathcal{R}^\psi_k \right] - \frac{1}{2} \Tr \left[ \mathcal{P}_k \, \mathcal{V}_k \, \mathcal{P}_k \, \mathcal{V}^{\rm QCD}_{k,\mu\nu} \, D_{r,\mu\rho} \, \partial_t R^A_{k,\rho\sigma} \, D_{r,\sigma\nu} \right] + \cdots.
    \label{eq:fourPointLoops}
\end{align}
The first term of \cref{eq:fourPointLoops} with fermionic regulator can be decomposed into
\begin{align}
    - \frac{1}{2} \Tr \left[ \mathcal{P}_k \, \mathcal{V}_k \, \mathcal{P}_k \, \mathcal{V}_k \, \mathcal{P}_k \, \partial_t \mathcal{R}^\psi_k \right] = &-\frac{1}{2} \Tr \left[ \mathcal{P}_k \, \mathcal{V}_k^{4f,\rm LN} \, \mathcal{P}_k \, \mathcal{V}_k^{4f,\rm LN} \, \mathcal{P}_k \, \partial_t \mathcal{R}^\psi_k \right] \nonumber\\
    &- \frac{1}{2} \Tr \left[ \mathcal{P}_k \, \left( \mathcal{V}^{\rm QCD}_{k,\mu\nu} \, D_{r,\mu\nu} \right) \, \mathcal{P}_k \, \mathcal{V}_k^{4f,\rm LN} \, \mathcal{P}_k \, \partial_t \mathcal{R}^\psi_k \right] \nonumber\\
    &- \frac{1}{2} \Tr \left[ \mathcal{P}_k \, \mathcal{V}_k^{4f,\rm LN} \, \mathcal{P}_k \, \left( \mathcal{V}^{\rm QCD}_{k,\mu\nu} \, D_{r,\mu\nu} \right) \, \mathcal{P}_k \, \partial_t \mathcal{R}^\psi_k \right] \nonumber\\
    &- \frac{1}{2} \Tr \left[ \mathcal{P}_k \, \left( \mathcal{V}^{\rm QCD}_{k,\mu\nu} \, D_{r,\mu\nu} \right) \, \mathcal{P}_k \, \left( \mathcal{V}^{\rm QCD}_{k,\rho\sigma} \, D_{r,\rho\sigma} \right) \, \mathcal{P}_k \, \partial_t \mathcal{R}^\psi_k \right],
\end{align}
which correspond to the fermion vacuum polarization, the left triangle, the right triangle, and the ladder-type diagrams.
The second term of \cref{eq:fourPointLoops} with gluonic regulator can be decomposed into
\begin{align}
    &- \frac{1}{2} \Tr \left[ \mathcal{P}_k \, \mathcal{V}_k \, \mathcal{P}_k \, \mathcal{V}^{\rm QCD}_{k,\mu\nu} \, D_{r,\mu\rho} \, \partial_t R^A_{k,\rho\sigma} \, D_{r,\sigma\nu} \right] \nonumber\\
    &\quad= - \frac{1}{2} \Tr \left[ \mathcal{P}_k \, \mathcal{V}_k^{4f,\rm LN} \, \mathcal{P}_k \, \mathcal{V}^{\rm QCD}_{k,\mu\nu} \, D_{r,\mu\rho} \, \partial_t R^A_{k,\rho\sigma} \, D_{r,\sigma\nu} \right] \nonumber\\
    &\qquad - \frac{1}{2} \Tr \left[ \mathcal{P}_k \, \left( \mathcal{V}^{\rm QCD}_{k,\alpha\beta} \, D_{r,\alpha\beta} \right) \, \mathcal{P}_k \, \mathcal{V}^{\rm QCD}_{k,\mu\nu} \, D_{r,\mu\rho} \, \partial_t R^A_{k,\rho\sigma} \, D_{r,\sigma\nu} \right],
\end{align}
which correspond to the triangle and the ladder-type diagrams.
We shall evaluate these six terms individually in the following five subsections.

\subsubsection{Evaluation of \texorpdfstring{$-\frac{1}{2} \Tr \left[ \mathcal{P}_k \, \mathcal{V}_k^{4f,\rm LN} \, \mathcal{P}_k \, \mathcal{V}_k^{4f,\rm LN} \, \mathcal{P}_k \, \partial_t \mathcal{R}^\psi_k \right]$}{}}

The loop with the fermionic regulator insertion, which is proportional to the square of the four-fermion interactions, reads
\begin{align}
    &-\frac{1}{2} \Tr \left[ \mathcal{P}_k \, \mathcal{V}_k^{4f,\rm LN} \, \mathcal{P}_k \, \mathcal{V}_k^{4f,\rm LN} \, \mathcal{P}_k \, \partial_t \mathcal{R}^\psi_k \right] \nonumber\\
    &\qquad=  -\frac{1}{2} \int_{x,y} \int_{p,p_1} \, \operatorname{tr} \Biggl[ 
    \pmat{ 0 & P_k \\ P_k^\prime & 0 }_p
    \pmat{ 0 & V^{4f,\rm LN}_{12} \\ V^{4f,\rm LN}_{21} & 0 }_x
    \pmat{ 0 & P_k \\ P_k^\prime & 0 }_{p_1}
    \pmat{ 0 & V^{4f,\rm LN}_{12} \\ V^{4f,\rm LN}_{21} & 0 }_y
    \nonumber\\
    &\hspace{0.4\textwidth} \times \pmat{ 0 & P_k \\ P_k^\prime & 0 }_p 
    \pmat{ 0 & \partial_t r_k^\psi i \Slash{p}^\text{T} \\ \partial_t r_k^\psi i \Slash{p} & 0 }
    \Biggl] e^{-i (p-p_1) \cdot (x-y)} \nonumber\\
    &\qquad 
    = -\frac{1}{2} \int_{x,y} \int_{p,p_1} \, e^{-i (p-p_1) \cdot (x-y)} \operatorname{tr} \Big[
    P_k(p) \, V^{4f,\rm LN}_{21}(x) \, P_k(p_1) \, V^{4f,\rm LN}_{21}(y) P_k(p) \, i \Slash{p} \nonumber\\
    &\hspace{0.4\textwidth} +  P_k^\prime(p) \, V^{4f,\rm LN}_{12}(x) \, P_k^\prime(p_1) \, V^{4f,\rm LN}_{12}(y) P_k^\prime(p) \, i \Slash{p}^\text{T}
    \Big] \partial_t r_k^\psi.
    \label{eq:beginningFourPoint4FermionFermiReg}
\end{align}
We then perform the derivative expansion with respect to $(p-p_1)$ up to the leading order, i.e., replacing $p_1$ with $p$, and integrate out the momentum $p_1$ and one of the position $y$, resulting with the local poerators
\begin{align}
    &-\frac{1}{2} \Tr \left[ \mathcal{P}_k \, \mathcal{V}_k^{4f,\rm LN} \, \mathcal{P}_k \, \mathcal{V}_k^{4f,\rm LN} \, \mathcal{P}_k \, \partial_t \mathcal{R}^\psi_k \right] \nonumber\\
    &\quad\rightarrow -\frac{1}{2} \int_{x} \int_{p} \, \partial_t r_k^\psi  \operatorname{tr} \Big[
    P_k(p) \, V^{4f,\rm LN}_{21}(x) \, P_k(p) \, V^{4f,\rm LN}_{21}(x) P_k(p) \, i \Slash{p} + P_k^\prime(p) \, V^{4f,\rm LN}_{12}(x) \, P_k^\prime(p) \, V^{4f,\rm LN}_{12}(x) P_k^\prime(p) \, i \Slash{p}^T
    \Big]  \nonumber\\
    &\qquad
    =  - \int_{x} \int_{p} \, \partial_t r_k^\psi \operatorname{tr} \Big[
    P_k(p) \, V^{4f,\rm LN}_{21}(x) \, P_k(p) \, V^{4f,\rm LN}_{21}(x) P_k(p) \, i \Slash{p} \Big] \nonumber\\
    &\qquad
    =  \int_{x} \int_{p} \,  \partial_t r_k^\psi  \operatorname{tr} \Biggl[\frac{-i \Slash{p} (1+r_k^\psi) + m e^{-i \gamma_5 \theta / 2}}{p^2 (1+r_k^\psi)^2 + m^2} \, V^{4f,\rm LN}_{21} \, \frac{-i \Slash{p} (1+r_k^\psi) + m e^{-i \gamma_5 \theta / 2}}{p^2 (1+r_k^\psi)^2 + m^2} \, V^{4f,\rm LN}_{21} \, \frac{-i \Slash{p} (1+r_k^\psi) + m e^{-i \gamma_5 \theta / 2}}{p^2 (1+r_k^\psi)^2 + m^2} \, i \Slash{p} \Biggl] \nonumber\\
    \qquad
    &\qquad
    =  - \int_{x} \int_{p} \,  \frac{p^2 (1+r_k^\psi) \partial_t r_k^\psi}{\left[ p^2 (1+r_k^\psi)^2 + m^2 \right]^3} \nonumber\\
    &\qquad\quad \times \biggl[ 
    -4 N_c N_f  \Big( (G_S^2 + C_4^2) (p^2(1+r_k^\psi)^2 - m^2) - 2 m^2 (G_S^2 - C_4^2) \cos{\theta} - 4 m^2 G_S C_4 \sin{\theta} \Big) \left( \bar{\psi} \psi \right)_x^2 \nonumber\\
    &\qquad \qquad 
    - 4 N_c N_f \Big( (G_P^2 + C_4^2) (p^2(1+r_k^\psi)^2 - m^2) + 2 m^2 (G_P^2 - C_4^2) \cos{\theta} - 4 m^2 G_P C_4 \sin{\theta} \Big) \left( \bar{\psi} i\gamma_5 \psi \right)_x^2 \nonumber\\
    &\qquad \qquad 
    - 8 N_c N_f \Big( (G_S + G_P) C_4 (p^2(1+r_k^\psi)^2 - m^2) - 2 m^2 (G_S - G_P) C_4 \cos{\theta} - 2 m^2 (G_S G_P + C_4^2) \sin{\theta} \Big) \nonumber\\
    &\qquad\qquad\quad \times \left( \bar{\psi} \psi \right)_x \left( \bar{\psi} i\gamma_5 \psi \right)_x
    \biggl] \nonumber\\
    &\qquad
    =  \int_{x} \biggl[ 
    4 N_c N_f  \Big( (G_S^2 + C_4^2) (k^2 - m^2) - 2 m^2 (G_S^2 - C_4^2) \cos{\theta} - 4 m^2 G_S C_4 \sin{\theta} \Big) \left( \bar{\psi} \psi \right)_x^2 \nonumber\\
    &\qquad \qquad
    + 4 N_c N_f \Big( (G_P^2 + C_4^2) (k^2 - m^2) + 2 m^2 (G_P^2 - C_4^2) \cos{\theta} - 4 m^2 G_P C_4 \sin{\theta} \Big) \left( \bar{\psi} i\gamma_5 \psi \right)_x^2 \nonumber\\
    &\qquad \qquad 
    + 8 N_c N_f \Big( (G_S + G_P) C_4 (k^2 - m^2) - 2 m^2 (G_S - G_P) C_4 \cos{\theta} - 2 m^2 (G_S G_P + C_4^2) \sin{\theta} \Big) \nonumber\\
    &\qquad\qquad\quad \times \left( \bar{\psi} \psi \right)_x \left( \bar{\psi} i\gamma_5 \psi \right)_x
    \biggl] \mathcal{I}_1,
    \label{eq:fourPointFlow1}
\end{align}
where the threshold function reads
\begin{align}
    \mathcal{I}_1 \equiv \int \frac{\df^4 p}{(2\pi)^4} \, \frac{ p^2 (1+r_k^\psi) \partial_t r_k^\psi}{\left[ p^2 (1+r_k^\psi)^2 + m^2 \right]^3} = \frac{1}{2(4\pi)^2} \frac{k^6}{\left(k^2+m^2 \right)^3}.
\end{align}

\subsubsection{Evaluation of \texorpdfstring{$- \frac{1}{2} \Tr \left[ \mathcal{P}_k \, \left( \mathcal{V}^{\rm QCD}_{k,\mu\nu} \, D_{r,\mu\nu} \right) \, \mathcal{P}_k \, \mathcal{V}_k^{4f,\rm LN} \, \mathcal{P}_k \, \partial_t \mathcal{R}^\psi_k \right]$ and $- \frac{1}{2} \Tr \left[ \mathcal{P}_k \, \mathcal{V}_k^{4f,\rm LN} \, \mathcal{P}_k \, \left( \mathcal{V}^{\rm QCD}_{k,\mu\nu} \, D_{r,\mu\nu} \right) \, \mathcal{P}_k \, \partial_t \mathcal{R}^\psi_k \right]$}{}}

Since the left and the right triangle diagrams are identical to each other, we shall evaluate them in a single subsection.
The sum of the left and the right triangle diagram reads
\begin{align}
    & - \frac{1}{2} \Tr \left[ \mathcal{P}_k \, \left( \mathcal{V}^{\rm QCD}_{k,\mu\nu} \, D_{r,\mu\nu} \right) \, \mathcal{P}_k \, \mathcal{V}_k^{4f,\rm LN} \, \mathcal{P}_k \, \partial_t \mathcal{R}^\psi_k \right] - \frac{1}{2} \Tr \left[ \mathcal{P}_k \, \mathcal{V}_k^{4f,\rm LN} \, \mathcal{P}_k \, \left( \mathcal{V}^{\rm QCD}_{k,\mu\nu} \, D_{r,\mu\nu} \right) \, \mathcal{P}_k \, \partial_t \mathcal{R}^\psi_k \right] \nonumber\\
    &\quad
    = - \frac{g_s^2}{2} \int_{x,y,z} \int_{p,p_1,p_2} \, \operatorname{tr} \biggl[ 
    \pmat{ 0 & P_k \\ P_k^\prime & 0 }_p
    \pmat{ V^{\rm QCD,NL_1}_{k,\mu\nu} & V^{\rm QCD,L_1}_{k,\mu\nu} \\ V^{\rm QCD,L_2}_{k,\mu\nu} & V^{\rm QCD,NL_2}_{k,\mu\nu} }_{x,y}
    \pmat{ 0 & P_k \\ P_k^\prime & 0 }_{p_2} 
    \pmat{ 0 & V^{4f,\rm LN}_{12} \\ V^{4f,\rm LN}_{21} & 0 }_z
    \nonumber\\
    &\hspace{0.2\textwidth} \times 
    \pmat{ 0 & P_k \\ P_k^\prime & 0 }_p
    \pmat{ 0 & \partial_t r_k^\psi i \Slash{p}^\text{T} \\ \partial_t r_k^\psi i \Slash{p} & 0 }
    \biggl] D_{r,\mu\nu}(p_1) \, e^{-i (p-p_2)\cdot z} e^{-i (p_1-p)\cdot y} e^{-i (p_2 - p_1) \cdot x}
    \nonumber\\
    &\qquad- \frac{g_s^2}{2} \int_{x,y,z} \int_{p,p_1,p_2} \, \operatorname{tr} \biggl[ 
    \pmat{ 0 & P_k \\ P_k^\prime & 0 }_p
    \pmat{ 0 & V^{4f,\rm LN}_{12} \\ V^{4f,\rm LN}_{21} & 0 }_z
    \pmat{ 0 & P_k \\ P_k^\prime & 0 }_{p_2}
    \pmat{ V^{\rm QCD,NL_1}_{k,\mu\nu} & V^{\rm QCD,L_1}_{k,\mu\nu} \\ V^{\rm QCD,L_2}_{k,\mu\nu} & V^{\rm QCD,NL_2}_{k,\mu\nu} }_{x,y}
    \nonumber\\
    &\hspace{0.2\textwidth} \times 
    \pmat{ 0 & P_k \\ P_k^\prime & 0 }_p
    \pmat{ 0 & \partial_t r_k^\psi i \Slash{p}^\text{T} \\ \partial_t r_k^\psi i \Slash{p} & 0 }
    \biggl] D_{r,\mu\nu}(p_1) \, e^{-i (p_2 - p)\cdot z} e^{-i (p_1-p_2)\cdot y} e^{-i (p - p_1) \cdot x}.
\end{align}
Adopting the leading order of the derivative expansion, we replace $p_1$ and $p_2$ with $p$ in the loop integrand.
The following integrals give
\begin{align}
    \int_{p_1} \, e^{-i p_1 \cdot (y - x)} = \delta^{(4)}(y-x), \qquad \int_{p_2} \, e^{-i p_2 \cdot (z - y)} = \delta^{(4)}(z-y),
\end{align}
and the loop integral results with local operators after integrating out $y$ and $z$, i.e.,
\begin{align}
    & - \frac{1}{2} \Tr \left[ \mathcal{P}_k \, \left( \mathcal{V}^{\rm QCD}_{k,\mu\nu} \, D_{r,\mu\nu} \right) \, \mathcal{P}_k \, \mathcal{V}_k^{4f,\rm LN} \, \mathcal{P}_k \, \partial_t \mathcal{R}^\psi_k \right] 
    - \frac{1}{2} \Tr \left[ \mathcal{P}_k \, \mathcal{V}_k^{4f,\rm LN} \, \mathcal{P}_k \, \left( \mathcal{V}^{\rm QCD}_{k,\mu\nu} \, D_{r,\mu\nu} \right) \, \mathcal{P}_k \, \partial_t \mathcal{R}^\psi_k \right] 
    \nonumber\\
    &\quad
    \rightarrow -\frac{g_s^2}{2} \int_x \int_p \operatorname{tr} \Big[ 
    P_k(p) \, V^{\rm QCD,L_2}_{k,\mu\nu} \, P_k(p) \, V^{4f,\rm LN}_{21} \, P_k(p) \, i\Slash{p}
    +
    P_k^\prime(p) \, V^{\rm QCD,L_1}_{k,\mu\nu} \, P_k^\prime(p) \, V^{4f,\rm LN}_{12} \, P_k^\prime(p) \, i\Slash{p}^\text{T}
    \Big] \nonumber\\
    &\quad\quad\quad \times \partial_t r_k^\psi \, D_{r,\mu\nu}(p) \nonumber\\
    &\qquad
    -\frac{g_s^2}{2} \int_x \int_p \operatorname{tr} \Big[ 
    P_k(p) \, V^{4f,\rm LN}_{21} \, P_k(p) \, V^{\rm QCD,L_2}_{k,\mu\nu} \, P_k(p) \, i\Slash{p}
    +
    P_k^\prime(p) \, V^{4f,\rm LN}_{12} \, P_k^\prime(p) \, V^{\rm QCD,L_1}_{k,\mu\nu} \, P_k^\prime(p) \, i\Slash{p}^\text{T}
    \Big] \nonumber\\
    &\quad\quad\quad \times \partial_t r_k^\psi \, D_{r,\mu\nu}(p).
    \label{eq:algebracFlowOfTriangleFermiReg}
\end{align}
The terms in \cref{eq:algebracFlowOfTriangleFermiReg} are equivalent, so that it is enough to calculate the first term, which reads
\begin{align}
    &\operatorname{tr} \Big[ 
    P_k(p) \, V^{\rm QCD,L_2}_{k,\mu\nu} \, P_k(p) \, V^{4f,\rm LN}_{21} \, P_k(p) \, i\Slash{p}
    \Big] \partial_t r_k^\psi \, D_{r,\mu\nu}(p) \nonumber\\
    &= \operatorname{tr} \Biggl[
    \frac{-i \Slash{p} (1+r_k^\psi) + m e^{-i \gamma_5 \theta / 2}}{p^2 (1+r_k^\psi)^2 + m^2}
    \,
    \gamma_\mu T^a \psi_x \bar{\psi}_x \gamma_\nu T^a
    \,
    \frac{-i \Slash{p} (1+r_k^\psi) + m e^{-i \gamma_5 \theta / 2}}{p^2 (1+r_k^\psi)^2 + m^2}
    \,
    V^{4f,\rm LN}_{21} 
    \nonumber\\
    &\hspace{0.4\textwidth} \times
    \frac{-i \Slash{p} (1+r_k^\psi) + m e^{-i \gamma_5 \theta / 2}}{p^2 (1+r_k^\psi)^2 + m^2} 
    \, 
    i \Slash{p} 
    \Biggl]
    \partial_t r_k^\psi \, D_{r,\mu\nu}(p) \nonumber\\
    &= \bar{\psi}_x \Biggl[ \gamma_\nu T^a
    \,
    \frac{-i \Slash{p} (1+r_k^\psi) + m e^{-i \gamma_5 \theta / 2}}{p^2 (1+r_k^\psi)^2 + m^2}
    \,
    V^{4f,\rm LN}_{21} 
    \,
    \frac{-i \Slash{p} (1+r_k^\psi) + m e^{-i \gamma_5 \theta / 2}}{p^2 (1+r_k^\psi)^2 + m^2} 
    \, 
    i \Slash{p}
    \nonumber\\
    &\hspace{0.4\textwidth} \times
    \frac{-i \Slash{p} (1+r_k^\psi) + m e^{-i \gamma_5 \theta / 2}}{p^2 (1+r_k^\psi)^2 + m^2}
    \,
    \gamma_\mu T^a \Biggl] \psi_x \, \partial_t r_k^\psi \, D_{r,\mu\nu}(p).
    \label{eq:OneOfTheTraceInTriangleFermiReg}
\end{align}
We project the above formula onto the tensorial structures $\textbf{1}_{\rm Dirac}$ and $\gamma_5$ to obtain
\begin{align}
    \labelcref{eq:OneOfTheTraceInTriangleFermiReg} \rightarrow & - C_2 (3+\xi) \frac{p^2 (1+r_k^\psi)\partial_t r_k^\psi}{\left[ p^2(1+r_k^\psi)^2 + m^2 \right]^3 \left[ p^2 (1+r_k^A) \right]} \nonumber\\
    &\quad \times \Biggl\{ 
    \left( \bar{\psi}\psi \right)_x^2 \Big[ G_S (p^2(1+r_k^\psi)^2 - m^2) - 2 m^2 G_S \cos{\theta} - 2m^2 C_4 \sin{\theta} \Big] \nonumber\\
    &\quad\quad + \left( \bar{\psi} i\gamma_5 \psi \right)_x^2 \Big[ G_P (p^2(1+r_k^\psi)^2 - m^2) + 2 m^2 G_P \cos{\theta} - 2m^2 C_4 \sin{\theta} \Big] \nonumber\\
    &\quad\quad + \left( \bar{\psi}\psi \right)_x \left( \bar{\psi} i\gamma_5 \psi \right)_x \Big[ 2 C_4 (p^2(1+r_k^\psi)^2 - m^2) - 2 m^2 (G_S + G_P) \sin{\theta} \Big]
    \Biggl\}.
\end{align}
Collecting all terms in \cref{eq:algebracFlowOfTriangleFermiReg}, we get
\begin{align}
    & - \frac{1}{2} \Tr \left[ \mathcal{P}_k \, \left( \mathcal{V}^{\rm QCD}_{k,\mu\nu} \, D_{r,\mu\nu} \right) \, \mathcal{P}_k \, \mathcal{V}_k^{4f,\rm LN} \, \mathcal{P}_k \, \partial_t \mathcal{R}^\psi_k \right] - \frac{1}{2} \Tr \left[ \mathcal{P}_k \, \mathcal{V}_k^{4f,\rm LN} \, \mathcal{P}_k \, \left( \mathcal{V}^{\rm QCD}_{k,\mu\nu} \, D_{r,\mu\nu} \right) \, \mathcal{P}_k \, \partial_t \mathcal{R}^\psi_k \right] \nonumber\\
    &\rightarrow  \frac{4}{2} g_s^2 C_2 (3+\xi) \int_x \int_p \frac{p^2 (1+r_k^\psi)\partial_t r_k^\psi}{\left[ p^2(1+r_k^\psi)^2 + m^2 \right]^3 \left[ p^2 (1+r_k^A) \right]} \nonumber\\
    &\hspace{0.2\textwidth} \times \Biggl\{ 
    \left( \bar{\psi}\psi \right)_x^2 \Big[ G_S (p^2(1+r_k^\psi)^2 - m^2) - 2 m^2 G_S \cos{\theta} - 2m^2 C_4 \sin{\theta} \Big] \nonumber\\
    &\hspace{0.2\textwidth}\quad + \left( \bar{\psi} i\gamma_5 \psi \right)_x^2 \Big[ G_P (p^2(1+r_k^\psi)^2 - m^2) + 2 m^2 G_P \cos{\theta} - 2m^2 C_4 \sin{\theta} \Big] \nonumber\\
    &\hspace{0.2\textwidth}\quad + \left( \bar{\psi}\psi \right)_x \left( \bar{\psi} i\gamma_5 \psi \right)_x \Big[ 2 C_4 (p^2(1+r_k^\psi)^2 - m^2) - 2 m^2 (G_S + G_P) \sin{\theta} \Big]
    \Biggl\} \nonumber\\
    &=  \int_x \, 2 g_s^2 C_2 (3+\xi) \Biggl\{ 
    \left( \bar{\psi}\psi \right)_x^2 \Big[ G_S (k^2 - m^2) - 2 m^2 G_S \cos{\theta} - 2m^2 C_4 \sin{\theta} \Big] \nonumber\\
    &\hspace{0.2\textwidth}\quad + \left( \bar{\psi} i\gamma_5 \psi \right)_x^2 \Big[ G_P (k^2 - m^2) + 2 m^2 G_P \cos{\theta} - 2m^2 C_4 \sin{\theta} \Big] \nonumber\\
    &\hspace{0.2\textwidth}\quad + \left( \bar{\psi}\psi \right)_x \left( \bar{\psi} i\gamma_5 \psi \right)_x \Big[ 2 C_4 (k^2 - m^2) - 2 m^2 (G_S + G_P) \sin{\theta} \Big]
    \Biggl\} \times \mathcal{I}_2,
    \label{eq:fourPointFlow2}
\end{align}
where the threshold function was defined as
\begin{align}
    \mathcal{I}_2 \equiv \int \frac{\df^4 p}{(2\pi)^4} \, \frac{p^2 (1+r_k^\psi)\partial_t r_k^\psi}{\left[ p^2(1+r_k^\psi)^2 + m^2 \right]^3 \left[ p^2 (1+r_k^A) \right]} = \frac{1}{2(4\pi)^2} \frac{k^4}{\left( k^2 + m^2 \right)^3}.
\end{align}

\subsubsection{Evaluation of \texorpdfstring{$- \frac{1}{2} \Tr \left[ \mathcal{P}_k \, \left( \mathcal{V}^{\rm QCD}_{k,\mu\nu} \, D_{r,\mu\nu} \right) \, \mathcal{P}_k \, \left( \mathcal{V}^{\rm QCD}_{k,\rho\sigma} \, D_{r,\rho\sigma} \right) \, \mathcal{P}_k \, \partial_t \mathcal{R}^\psi_k \right]$}{}}

The flow of the four-point function, which is proportional to $g_s^4$ with the fermionic regulator insertion $\p_t \mathcal R_k^\psi$, is given by
\begin{align}
    &- \frac{1}{2} \Tr \left[ \mathcal{P}_k \, \left( \mathcal{V}^{\rm QCD}_{k,\mu\nu} \, D_{r,\mu\nu} \right) \, \mathcal{P}_k \, \left( \mathcal{V}^{\rm QCD}_{k,\rho\sigma} \, D_{r,\rho\sigma} \right) \, \mathcal{P}_k \, \partial_t \mathcal{R}^\psi_k \right] \nonumber\\
    &\quad
    =  -\frac{g_s^4}{2} \int_{x,y,z,w} \int_{p,p_1,p_2,p_3} \, \operatorname{tr} \biggl[ 
    \pmat{ 0 & P_k \\ P_k^\prime & 0 }_p
    \pmat{ V^{\rm QCD,NL_1}_{k,\mu\nu} & V^{\rm QCD,L_1}_{k,\mu\nu} \\ V^{\rm QCD,L_2}_{k,\mu\nu} & V^{\rm QCD,NL_2}_{k,\mu\nu} }_{x,y}
    \pmat{ 0 & P_k \\ P_k^\prime & 0 }_{p_2}
    \nonumber\\
    &\hspace{0.1\textwidth} \times
    \pmat{ V^{\rm QCD,NL_1}_{k,\rho\sigma} & V^{\rm QCD,L_1}_{k,\rho\sigma} \\ V^{\rm QCD,L_2}_{k,\rho\sigma} & V^{\rm QCD,NL_2}_{k,\rho\sigma} }_{z,w}
    \pmat{ 0 & P_k \\ P_k^\prime & 0 }_{p}
    \pmat{ 0 & \partial_t r_k^\psi i \Slash{p}^\text{T} \\ \partial_t r_k^\psi i \Slash{p} & 0 }
    \biggl] D_{r,\mu\nu}(p_1) \, D_{r,\rho\sigma}(p_3) \nonumber\\
    &\hspace{0.1\textwidth} \times 
    e^{-i (p_3 - p)\cdot w} e^{-i (p_2 - p_3)\cdot z} e^{-i (p_1-p_2)\cdot y} e^{-i (p - p_1) \cdot x}.
\end{align}
Again, with the same strategy of the derivative expansion, we replace $p_1$, $p_2$, and $p_3$ by $p$, and integrate them out, leaving with local operators
\begin{align}
    &- \frac{1}{2} \Tr \left[ \mathcal{P}_k \, \left( \mathcal{V}^{\rm QCD}_{k,\mu\nu} \, D_{r,\mu\nu} \right) \, \mathcal{P}_k \, \left( \mathcal{V}^{\rm QCD}_{k,\rho\sigma} \, D_{r,\rho\sigma} \right) \, \mathcal{P}_k \, \partial_t \mathcal{R}^\psi_k \right] \nonumber\\
    &\quad
    \rightarrow  - \frac{1}{2} \int_x \int_p \, \operatorname{tr} \Big[ 
    P_k(p) \, V^{\rm QCD,L_2}_{k,\mu\nu} \, P_k(p) \, V^{\rm QCD,L_2}_{k,\rho\sigma} \, P_k(p) \, i \Slash{p}
    + P_k(p) \, V^{\rm QCD,NL_2}_{k,\mu\nu} \, P_k^\prime(p) \, V^{\rm QCD,NL_1}_{k,\rho\sigma} \, P_k(p) \, i \Slash{p}
    \nonumber\\
    &\hspace{0.1\textwidth}
    +P_k^\prime(p) \, V^{\rm QCD,L_1}_{k,\mu\nu} \, P_k^\prime(p) \, V^{\rm QCD,L_1}_{k,\rho\sigma} \, P_k^\prime(p) \, i \Slash{p}^\text{T}
    + P_k^\prime(p) \, V^{\rm QCD,NL_1}_{k,\mu\nu} \, P_k(p) \, V^{\rm QCD,NL_2}_{k,\rho\sigma} \, P_k^\prime(p) \, i \Slash{p}^\text{T}
    \Big] \nonumber\\
    &\hspace{0.15\textwidth} \times
    \partial_t r_k^\psi \, D_{r,\mu\nu}(p) \, D_{r,\rho\sigma}(p).
    \label{eq:ladderNnonladderLPA}
\end{align}
Here we see there are four terms in the square bracket in \cref{eq:ladderNnonladderLPA}.
The first and the third terms correspond to the so-called ``ladder'' type of diagrams, while the second and the fourth terms are ``nonladder'' type ones.
Hence, the second and the third lines in \cref{eq:ladderNnonladderLPA} are equivalent to each other, so that we shall demonstrate only the second line.

The first term in \cref{eq:ladderNnonladderLPA}, which corresponds to the ``ladder'' part, reads
\begin{align}
    &\operatorname{tr} \Big[ 
    P_k(p) \, V^{\rm QCD,L_2}_{k,\mu\nu} \, P_k(p) \, V^{\rm QCD,L_2}_{k,\rho\sigma} \, P_k(p) \, i \Slash{p}
    \Big] \partial_t r_k^\psi \, D_{r,\mu\nu}(p) \, D_{r,\rho\sigma}(p)
    \nonumber\\
    &\quad
    = \operatorname{tr} \Big[ 
    \frac{-i \Slash{p} (1+r_k^\psi) + m e^{-i \gamma_5 \theta / 2}}{p^2 (1+r_k^\psi)^2 + m^2}
    \,
    \gamma_\mu T^a \psi_x \bar{\psi}_x \gamma_\nu T^a
    \,
    \frac{-i \Slash{p} (1+r_k^\psi) + m e^{-i \gamma_5 \theta / 2}}{p^2 (1+r_k^\psi)^2 + m^2}
    \,
    \gamma_\rho T^b \psi_x \bar{\psi}_x \gamma_\sigma T^b
    \nonumber\\
    &\hspace{0.4\textwidth} \times
    \frac{-i \Slash{p} (1+r_k^\psi) + m e^{-i \gamma_5 \theta / 2}}{p^2 (1+r_k^\psi)^2 + m^2}
    \,
    i\Slash{p}
    \Big] \partial_t r_k^\psi \, D_{r,\mu\nu}(p) \, D_{r,\rho\sigma}(p)
    \nonumber\\
    &= - \bar{\psi}_x \Big[ \gamma_\sigma T^b
    \,
    \frac{-i \Slash{p} (1+r_k^\psi) + m e^{-i \gamma_5 \theta / 2}}{p^2 (1+r_k^\psi)^2 + m^2}
    \,
    i\Slash{p}
    \,
    \frac{-i \Slash{p} (1+r_k^\psi) + m e^{-i \gamma_5 \theta / 2}}{p^2 (1+r_k^\psi)^2 + m^2}
    \,
    \gamma_\mu T^a \Big] \psi_x
    \nonumber\\
    & \times    
    \bar{\psi}_x \Big[ \gamma_\nu T^a
    \,
    \frac{-i \Slash{p} (1+r_k^\psi) + m e^{-i \gamma_5 \theta / 2}}{p^2 (1+r_k^\psi)^2 + m^2}
    \,
    \gamma_\rho T^b \Big] \psi_x \, \partial_t r_k^\psi \, D_{r,\mu\nu}(p) \, D_{r,\rho\sigma}(p)
    \nonumber\\
    &\equiv  - \bar{\psi}_x \Big[ 
    \mathcal{O}_{2,\sigma\mu}^{ba}
    \Big] \psi_x
    \,   
    \bar{\psi}_x \Big[ 
    \mathcal{O}_{3,\nu\rho}^{ab}
    \Big] \psi_x \, \partial_t r_k^\psi \, D_{r,\mu\nu}(p) \, D_{r,\rho\sigma}(p).
    \label{eq:intermediateMark1}
\end{align}
Following the strategy discussed in \Cref{sec:ProjectionToFierzSubspace}, we perform the Fierz transformation as
\begin{align}
    \bar{\psi}_x \Big[ \mathcal{O}_{2,\sigma\mu}^{ba} \Big] \psi_x \, \bar{\psi}_x \Big[ \mathcal{O}_{3,\nu\rho}^{ab} \Big] \psi_x
    &=
    -\frac{1}{4N_c} \sum_I \, \bar{\psi}_x \left( \mathcal{O}_{2,\sigma\mu}^{ba} \, \Gamma_I \otimes \textbf{1}_{\rm color} \, \mathcal{O}_{3,\nu\rho}^{ab} \right) \psi_x \, \bar{\psi}_x \left( \Gamma_I \otimes \textbf{1}_{\rm color} \right) \psi_x \nonumber\\
    &= -\frac{1}{4N_c} \sum_{I,J} \, f_{IJ} \, \bar{\psi}_x \left( \Gamma_J \right) \psi_x \, \bar{\psi}_x \left( \Gamma_I \right) \psi_x,
\end{align}
where
\begin{align}
    f_{IJ} = \operatorname{tr} \Big[ \mathcal{O}_{2,\sigma\mu}^{ba} \, \Gamma_I \otimes \textbf{1}_{\rm color} \, \mathcal{O}_{3,\nu\rho}^{ab} \, \Gamma_J \otimes \textbf{1}_{\rm color} \Big] / (4 N_c).
\end{align}
Since we are focusing on only the $\Gamma_0 = \textbf{1}_{\rm Dirac}$ and the $\Gamma_5 = \gamma_5$ channels, thus $I,J = 0,5$, we have
\begin{align}
    \labelcref{eq:intermediateMark1} \ni& (-1)^2 (3+\xi)^2 \left( \frac{C_2}{4N_c} \right)^2 4 N_c 
    \nonumber\\
    &\quad \times
    \Biggl\{ 
    \left( \bar{\psi} \psi \right)_x^2 \Big[ - p^2 (1+r_k^\psi)^2 + m^2 + 2 m^2 \cos{\theta} \Big]
    + \left( \bar{\psi} i\gamma_5 \psi \right)_x^2 \Big[ - p^2 (1+r_k^\psi)^2 + m^2 - 2 m^2 \cos{\theta} \Big]
    \nonumber\\
    &\quad\quad
    + \left( \bar{\psi} \psi \right)_x \left( \bar{\psi} i\gamma_5 \psi \right)_x \Big[ 4 m^2 \sin{\theta} \Big]
    \Biggl\} \frac{p^2 (1+r_k^\psi) \partial_t r_k^\psi}{\left[ p^2 (1+r_k^\psi)^2 + m^2 \right]^3 \left[ p^2 (1+r_k^A) \right]^2}.
\end{align}
After collecting the anti-fermionic loop, the flow from the ``ladder'' diagrams reads
\begin{align}
    &- \frac{1}{2} \Tr \left[ \mathcal{P}_k \, \left( \mathcal{V}^{\rm QCD}_{k,\mu\nu} \, D_{r,\mu\nu} \right) \, \mathcal{P}_k \, \left( \mathcal{V}^{\rm QCD}_{k,\rho\sigma} \, D_{r,\rho\sigma} \right) \, \mathcal{P}_k \, \partial_t \mathcal{R}^\psi_k \right] \Biggl|_{\rm ladder}
    \nonumber\\
    &\quad
    \rightarrow  \int_x (3+\xi)^2 \left( \frac{g_s^2 C_2}{4N_c} \right)^2 4 N_c 
    \Biggl\{ 
    \left( \bar{\psi} \psi \right)_x^2 \Big[ k^2 - m^2 - 2 m^2 \cos{\theta} \Big]
    + \left( \bar{\psi} i\gamma_5 \psi \right)_x^2 \Big[ k^2 - m^2 + 2 m^2 \cos{\theta} \Big]
    \nonumber\\
    &\qquad\qquad
    + \left( \bar{\psi} \psi \right)_x \left( \bar{\psi} i\gamma_5 \psi \right)_x \Big[ - 4 m^2 \sin{\theta} \Big]
    \Biggl\} \, \mathcal{I}_3,
    \label{eq:fourPointFlow3}
\end{align}
where the threshold function has been defined as
\begin{align}
    \mathcal{I}_3 \equiv \int \frac{\df^4 p}{(2\pi)^4} \frac{p^2 (1+r_k^\psi) \partial_t r_k^\psi}{\left[ p^2 (1+r_k^\psi)^2 + m^2 \right]^3 \left[ p^2 (1+r_k^A) \right]^2} = \frac{1}{2(4\pi)^2} \frac{k^2}{\left( k^2+m^2 \right)^3}.
\end{align}

Similarly, the ``nonladder'' part reads
\begin{align}
    &\operatorname{tr} \Big[ P_k(p) \, V^{\rm QCD,NL_2}_{k,\mu\nu} \, P_k^\prime(p) \, V^{\rm QCD,NL_1}_{k,\rho\sigma} \, P_k(p) \, i \Slash{p} \Big] 
    \nonumber\\
    &\quad
    = \operatorname{tr} \Big[ 
    \frac{-i \Slash{p} (1+r_k^\psi) + m e^{-i \gamma_5 \theta / 2}}{p^2 (1+r_k^\psi)^2 + m^2}
    \,
    \gamma_\mu T^a \psi_x \psi^\text{T}_x \gamma_\nu^\text{T} T^{\text{T},a}
    \,
    \frac{-i \Slash{p}^\text{T} (1+r_k^\psi) - m e^{-i \gamma_5 \theta / 2}}{p^2 (1+r_k^\psi)^2 + m^2}
    \,
    \gamma_\rho^\text{T} T^{\text{T},b} \bar{\psi}^\text{T}_x \bar{\psi}_x \gamma_\sigma T^b
    \nonumber\\
    &\hspace{0.4\textwidth} \times
    \frac{-i \Slash{p} (1+r_k^\psi) + m e^{-i \gamma_5 \theta / 2}}{p^2 (1+r_k^\psi)^2 + m^2}
    \,
    i\Slash{p}
    \Big] \partial_t r_k^\psi \, D_{r,\mu\nu}(p) \, D_{r,\rho\sigma}(p)
    \nonumber\\
    &\quad
    = (-1)^2 \, \bar{\psi}_x \Big[
    \gamma_\rho T^b 
    \, 
    \frac{-i \Slash{p}^\text{T} (1+r_k^\psi) - m e^{-i \gamma_5 \theta / 2}}{p^2 (1+r_k^\psi)^2 + m^2}   
    \,
    \gamma_\nu T^a
    \Big] \psi_x
    \nonumber\\
    &\times \bar{\psi}_x \Big[ 
    \gamma_\sigma T^b
    \,
    \frac{-i \Slash{p} (1+r_k^\psi) + m e^{-i \gamma_5 \theta / 2}}{p^2 (1+r_k^\psi)^2 + m^2}
    \,
    i\Slash{p}
    \,
    \frac{-i \Slash{p} (1+r_k^\psi) + m e^{-i \gamma_5 \theta / 2}}{p^2 (1+r_k^\psi)^2 + m^2}
    \,
    \gamma_\mu T^a 
    \Big] \psi_x \, \partial_t r_k^\psi \, D_{r,\mu\nu}(p) \, D_{r,\rho\sigma}(p)
    \nonumber\\
    &\quad
    \ni  (-1)^3 \left( \frac{C_2}{4N_c} \right) \frac{1}{2 N_c}
    \nonumber\\
    &\qquad \times
    \Biggl\{ 
    \left( \bar{\psi} \psi \right)_x^2 \Big[ - \zeta^- p^2 (1+r_k^\psi)^2 + \zeta^- m^2 - 2 \zeta^+ m^2 \cos{\theta} \Big]
    \nonumber\\
    &\quad\quad
    + \left( \bar{\psi} i\gamma_5 \psi \right)_x^2 \Big[ - \zeta^- p^2 (1+r_k^\psi)^2 + \zeta^- m^2 + 2 \zeta^+ m^2 \cos{\theta} \Big]
    \nonumber\\
    &\quad\quad
    + \left( \bar{\psi} \psi \right)_x \left( \bar{\psi} i\gamma_5 \psi \right)_x \Big[ - 4 \zeta^+ m^2 \sin{\theta} \Big]
    \Biggl\} \frac{p^2 (1+r_k^\psi) \partial_t r_k^\psi}{\left[ p^2 (1+r_k^\psi)^2 + m^2 \right]^3 \left[ p^2 (1+r_k^A) \right]^2}.
\end{align}
where $\zeta^{\pm} = 3 \pm \xi (6 \mp \xi)$.
After collecting the anti-fermionic loop, the flow from the ``nonladder'' diagrams reads
\begin{align}
    &- \frac{1}{2} \Tr \left[ \mathcal{P}_k \, \left( \mathcal{V}^{\rm QCD}_{k,\mu\nu} \, D_{r,\mu\nu} \right) \, \mathcal{P}_k \, \left( \mathcal{V}^{\rm QCD}_{k,\rho\sigma} \, D_{r,\rho\sigma} \right) \, \mathcal{P}_k \, \partial_t \mathcal{R}^\psi_k \right] \Biggl|_{\rm nonladder}
    \nonumber\\
    &\quad
    \rightarrow  \int_x \left( \frac{g_s^4 C_2}{4N_c} \right) \frac{1}{2 N_c} 
    \Biggl\{ 
    \left( \bar{\psi} \psi \right)_x^2 \Big[ - \zeta^- \big( k^2 - m^2 \big) - 2 \zeta^+ m^2 \cos{\theta} \Big]
    + \left( \bar{\psi} i\gamma_5 \psi \right)_x^2 \Big[ - \zeta^- \big( k^2 - m^2 \big) + 2 \zeta^+ m^2 \cos{\theta} \Big]
    \nonumber\\
    &\qquad\qquad
    + \left( \bar{\psi} \psi \right)_x \left( \bar{\psi} i\gamma_5 \psi \right)_x \Big[ - 4 \zeta^+ m^2 \sin{\theta} \Big]
    \Biggl\} \, \mathcal{I}_3.
    \label{eq:fourPointFlow4}
\end{align}

\subsubsection{Evaluation of \texorpdfstring{$- \frac{1}{2} \Tr \left[ \mathcal{P}_k \, \mathcal{V}_k^{4f,\rm LN} \, \mathcal{P}_k \, \mathcal{V}^{\rm QCD}_{k,\mu\nu} \, D_{r,\mu\rho} \, \partial_t R^A_{k,\rho\sigma} \, D_{r,\sigma\nu} \right]$}{}}

The triangle diagram with gluonic regulator insertion $\p_t R_k^A$ reads
\begin{align}
    &- \frac{1}{2} \Tr \left[ \mathcal{P}_k \, \mathcal{V}_k^{4f,\rm LN} \, \mathcal{P}_k \, \mathcal{V}^{\rm QCD}_{k,\mu\nu} \, D_{r,\mu\rho} \, \partial_t R^A_{k,\rho\sigma} \, D_{r,\sigma\nu} \right]
    \nonumber\\
    &\quad
    \rightarrow  - \frac{g_s^2}{2} \int_{x} \int_{p} \, \operatorname{tr} \biggl[ 
    \pmat{ 0 & P_k \\ P_k^\prime & 0 }_p
    \pmat{ 0 & V^{4f,\rm LN}_{12} \\ V^{4f,\rm LN}_{21} & 0 }_x
    \pmat{ 0 & P_k \\ P_k^\prime & 0 }_{p}
    \pmat{ V^{\rm QCD,NL_1}_{k,\mu\nu} & V^{\rm QCD,L_1}_{k,\mu\nu} \\ V^{\rm QCD,L_2}_{k,\mu\nu} & V^{\rm QCD,NL_2}_{k,\mu\nu} }_{x}
    \biggl]  
  \left( D_{r,\mu\rho} \, \partial_t R^A_{k,\rho\sigma} \, D_{r,\sigma\nu} \right)_p 
    \nonumber\\
    &=  - \frac{g_s^2}{2} \int_{x} \int_{p} \, \operatorname{tr} \Big[ 
    P_k(p) \, V^{4f,\rm LN}_{21} \, P_k(p) \, V^{\rm QCD,L_2}_{k,\mu\nu}
    +
    P^\prime_k(p) \, V^{4f,\rm LN}_{12} \, P^\prime_k(p) \, V^{\rm QCD,L_1}_{k,\mu\nu}
    \Big]\left( D_{r,\mu\rho} \, \partial_t R^A_{k,\rho\sigma} \, D_{r,\sigma\nu} \right)_p .
\end{align}
The first trace reads
\begin{align}
    & \operatorname{tr} \Big[ 
    P_k(p) \, V^{4f,\rm LN}_{21} \, P_k(p) \, V^{\rm QCD,L_2}_{k,\mu\nu}
    \Big] 
    \left( D_{r,\mu\rho} \, \partial_t R^A_{k,\rho\sigma} \, D_{r,\sigma\nu} \right)_p
    \nonumber\\
    &\quad
    = \operatorname{tr} \Big[ 
    \frac{-i \Slash{p} (1+r_k^\psi) + m e^{-i \gamma_5 \theta / 2}}{p^2 (1+r_k^\psi)^2 + m^2}
    \, 
    V^{4f,\rm LN}_{21}
    \, 
    \frac{-i \Slash{p} (1+r_k^\psi) + m e^{-i \gamma_5 \theta / 2}}{p^2 (1+r_k^\psi)^2 + m^2}
    \, 
    \gamma_\mu T^a \psi_x \bar{\psi}_x \gamma_\nu T^a
    \Big] 
  \left( D_{r,\mu\rho} \, \partial_t R^A_{k,\rho\sigma} \, D_{r,\sigma\nu} \right)_p
    \nonumber\\
    &\quad
    = - C_2 \bar{\psi}_x \Big[ 
    \gamma_\nu 
    \,
    \frac{-i \Slash{p} (1+r_k^\psi) + m e^{-i \gamma_5 \theta / 2}}{p^2 (1+r_k^\psi)^2 + m^2}
    \, 
    V^{4f,\rm LN}_{21}
    \, 
    \frac{-i \Slash{p} (1+r_k^\psi) + m e^{-i \gamma_5 \theta / 2}}{p^2 (1+r_k^\psi)^2 + m^2}
    \, 
    \gamma_\mu
    \Big] \psi_x
    \left( D_{r,\mu\rho} \, \partial_t R^A_{k,\rho\sigma} \, D_{r,\sigma\nu} \right)_p.
    \label{eq:OneOfTheTraceInTriangleBoseReg}
\end{align}
We project the above formula onto the tensorial structures $\textbf{1}_{\rm Dirac}$ and $\gamma_5$, which results in
\begin{align}
    \labelcref{eq:OneOfTheTraceInTriangleBoseReg} \rightarrow & - C_2 (3+\xi) \frac{p^2 \partial_t r_k^A}{\left[ p^2(1+r_k^\psi)^2 + m^2 \right]^2 \left[ p^2 (1+r_k^A) \right]^2} \nonumber\\
    &\quad \times \Biggl\{ 
    \left( \bar{\psi}\psi \right)_x^2 \Big[ G_S (p^2(1+r_k^\psi)^2 - m^2 \cos{\theta}) - m^2 C_4 \sin{\theta} \Big] \nonumber\\
    &\quad\quad + \left( \bar{\psi} i\gamma_5 \psi \right)_x^2 \Big[ G_P (p^2(1+r_k^\psi)^2 + m^2 \cos{\theta}) - m^2 C_4 \sin{\theta} \Big] \nonumber\\
    &\quad\quad + \left( \bar{\psi}\psi \right)_x \left( \bar{\psi} i\gamma_5 \psi \right)_x \Big[ 2 C_4 p^2(1+r_k^\psi)^2 - m^2 (G_S + G_P) \sin{\theta} \Big]
    \Biggl\}.
\end{align}
Then, the triangle loop with gluonic regulator insertion reads
\begin{align}
    &- \frac{1}{2} \Tr \left[ \mathcal{P}_k \, \mathcal{V}_k^{4f,\rm LN} \, \mathcal{P}_k \, \mathcal{V}^{\rm QCD}_{k,\mu\nu} \, D_{r,\mu\rho} \, \partial_t R^A_{k,\rho\sigma} \, D_{r,\sigma\nu} \right] 
    \nonumber\\
    &\quad
    \rightarrow  \int_x 2 g_s^2 C_2 (3+\xi) \mathcal{I}_4 \Biggl\{ 
    \left( \bar{\psi}\psi \right)_x^2 \Big[ G_S (k^2 - m^2 \cos{\theta}) - m^2 C_4 \sin{\theta} \Big] 
    \nonumber\\
    &\qquad
    + \left( \bar{\psi} i\gamma_5 \psi \right)_x^2 \Big[ G_P (k^2 + m^2 \cos{\theta}) - m^2 C_4 \sin{\theta} \Big] + \left( \bar{\psi}\psi \right)_x \left( \bar{\psi} i\gamma_5 \psi \right)_x \Big[ 2 C_4 k^2 - m^2 (G_S + G_P) \sin{\theta} \Big]
    \Biggl\} ,
    \label{eq:fourPointFlow5}
\end{align}
where we have defined the threshold function as
\begin{align}
    \mathcal{I}_4 \equiv \frac{1}{2}\int \frac{\df^4 p}{(2\pi)^4}  \frac{p^2 \partial_t r_k^A}{\left[ p^2(1+r_k^\psi)^2 + m^2 \right]^2 \left[ p^2 (1+r_k^A) \right]^2} = \frac{1}{2(4\pi)^2} \frac{k^2}{\left( k^2 + m^2 \right)^2}.
\end{align}

\subsubsection{Evauation of \texorpdfstring{$- \frac{1}{2} \Tr \left[ \mathcal{P}_k \, \left( \mathcal{V}^{\rm QCD}_{k,\alpha\beta} \, D_{r,\alpha\beta} \right) \, \mathcal{P}_k \, \mathcal{V}^{\rm QCD}_{k,\mu\nu} \, D_{r,\mu\rho} \, \partial_t R^A_{k,\rho\sigma} \, D_{r,\sigma\nu} \right]$}{}}

The ``ladder'' and ``nonladder'' diagram with gluonic regulator insertion reads
\begin{align}
    &- \frac{1}{2} \Tr \left[ \mathcal{P}_k \, \left( \mathcal{V}^{\rm QCD}_{k,\alpha\beta} \, D_{r,\alpha\beta} \right) \, \mathcal{P}_k \, \mathcal{V}^{\rm QCD}_{k,\mu\nu} \, D_{r,\mu\rho} \, \partial_t R^A_{k,\rho\sigma} \, D_{r,\sigma\nu} \right]
    \nonumber\\
    &\quad
    \rightarrow  -\frac{g_s^4}{2} \int_{x} \int_{p} \, \operatorname{tr} \biggl[ 
    \pmat{ 0 & P_k \\ P_k^\prime & 0 }_p
    \pmat{ V^{\rm QCD,NL_1}_{k,\alpha\beta} & V^{\rm QCD,L_1}_{k,\alpha\beta} \\ V^{\rm QCD,L_2}_{k,\alpha\beta} & V^{\rm QCD,NL_2}_{k,\alpha\beta} }_{x}
    \pmat{ 0 & P_k \\ P_k^\prime & 0 }_{p} 
    \pmat{ V^{\rm QCD,NL_1}_{k,\mu\nu} & V^{\rm QCD,L_1}_{k,\mu\nu} \\ V^{\rm QCD,L_2}_{k,\mu\nu} & V^{\rm QCD,NL_2}_{k,\mu\nu} }_{x}
    \biggl] 
    \nonumber\\
    &\hspace{0.1\textwidth} \times D_{r,\alpha\beta}(p)
    \left( D_{r,\mu\rho} \, \partial_t R^A_{k,\rho\sigma} \, D_{r,\sigma\nu} \right)_p
    \nonumber\\
    &\quad
    = -\frac{g_s^4}{2} \int_{x} \int_{p} \, \operatorname{tr} \Big[ 
    P_k(p) \, V^{\rm QCD,L_2}_{k,\alpha\beta} \, P_k(p) \, V^{\rm QCD,L_2}_{k,\mu\nu}
    + P_k(p) \, V^{\rm QCD,NL_2}_{k,\alpha\beta} \, P_k^\prime(p) \, V^{\rm QCD,NL_1}_{k,\mu\nu}
    \nonumber\\
    &\qquad
    + P_k^\prime(p) \, V^{\rm QCD,L_1}_{k,\alpha\beta} \, P_k^\prime(p) \, V^{\rm QCD,L_1}_{k,\mu\nu}
    + P_k^\prime(p) \, V^{\rm QCD,NL_1}_{k,\alpha\beta} \, P_k(p) \, V^{\rm QCD,NL_2}_{k,\mu\nu}
    \Big] D_{r,\alpha\beta}(p)
    \left( D_{r,\mu\rho} \, \partial_t R^A_{k,\rho\sigma} \, D_{r,\sigma\nu} \right)_p.
\end{align}

The trace of the ladder diagram is evaluated as
\begin{align}
    &\operatorname{tr} \Big[ 
    P_k(p) \, V^{\rm QCD,L_2}_{k,\alpha\beta} \, P_k(p) \, V^{\rm QCD,L_2}_{k,\mu\nu}
    \Big] 
    D_{r,\alpha\beta}(p) \left( D_{r,\mu\rho} \, \partial_t R^A_{k,\rho\sigma} \, D_{r,\sigma\nu} \right)_p 
    \nonumber\\
    &
    = \operatorname{tr} \Big[ 
    \frac{-i \Slash{p} (1+r_k^\psi) + m e^{-i \gamma_5 \theta / 2}}{p^2 (1+r_k^\psi)^2 + m^2}
    \,
    \gamma_\mu T^a \psi_x \bar{\psi}_x \gamma_\nu T^a
    \,
    \frac{-i \Slash{p} (1+r_k^\psi) + m e^{-i \gamma_5 \theta / 2}}{p^2 (1+r_k^\psi)^2 + m^2}
    \,
    \gamma_\rho T^b \psi_x \bar{\psi}_x \gamma_\sigma T^b
    \Big] 
    D_{r,\alpha\beta}(p) \left( D_{r,\mu\rho} \, \partial_t R^A_{k,\rho\sigma} \, D_{r,\sigma\nu} \right)_p
    \nonumber\\
    &
    = - \bar{\psi}_x \Big[ \gamma_\nu T^a
    \,
    \frac{-i \Slash{p} (1+r_k^\psi) + m e^{-i \gamma_5 \theta / 2}}{p^2 (1+r_k^\psi)^2 + m^2}
    \,
    \gamma_\rho T^b \Big] \psi_x
    \,
    \bar{\psi}_x \Big[ \gamma_\sigma T^b
    \,
    \frac{-i \Slash{p} (1+r_k^\psi) + m e^{-i \gamma_5 \theta / 2}}{p^2 (1+r_k^\psi)^2 + m^2}
    \,
    \gamma_\mu T^a \Big] \psi_x
    D_{r,\alpha\beta}(p) \left( D_{r,\mu\rho} \, \partial_t R^A_{k,\rho\sigma} \, D_{r,\sigma\nu} \right)_p
    \nonumber\\
    &
    \ni  (-1)^2 (3+\xi)^2 \left( \frac{C_2}{4N_c} \right)^2 4 N_c 
    \Biggl\{ 
    \left( \bar{\psi} \psi \right)_x^2 \Big[ - p^2 (1+r_k^\psi)^2 + m^2 \cos{\theta} \Big]
    + \left( \bar{\psi} i\gamma_5 \psi \right)_x^2 \Big[ - p^2 (1+r_k^\psi)^2 - m^2 \cos{\theta} \Big]
    \nonumber\\
    &\quad\quad
    + \left( \bar{\psi} \psi \right)_x \left( \bar{\psi} i\gamma_5 \psi \right)_x \Big[ 2 m^2 \sin{\theta} \Big]
    \Biggl\} \frac{p^2 \partial_t r_k^A}{\left[ p^2 (1+r_k^\psi)^2 + m^2 \right]^2 \left[ p^2 (1+r_k^A) \right]^3}.
\end{align}
After collecting the anti-fermionic loop, the ``ladder'' part reads
\begin{align}
    & - \frac{1}{2} \Tr \left[ \mathcal{P}_k \, \left( \mathcal{V}^{\rm QCD}_{k,\alpha\beta} \, D_{r,\alpha\beta} \right) \, \mathcal{P}_k \, \mathcal{V}^{\rm QCD}_{k,\mu\nu} \, D_{r,\mu\rho} \, \partial_t R^A_{k,\rho\sigma} \, D_{r,\sigma\nu} \right] \Biggl|_{\rm ladder} 
    \nonumber\\
    &\quad
    \rightarrow  \int_x (3+\xi)^2 \left( \frac{g_s^2 C_2}{4N_c} \right)^2 4 N_c \cdot 2
    \nonumber\\
    &\qquad \times
    \Biggl\{ 
    \left( \bar{\psi} \psi \right)_x^2 \Big[ k^2 - m^2 \cos{\theta} \Big]
    + \left( \bar{\psi} i\gamma_5 \psi \right)_x^2 \Big[ k^2 + m^2 \cos{\theta} \Big]
    + \left( \bar{\psi} \psi \right)_x \left( \bar{\psi} i\gamma_5 \psi \right)_x \Big[ - 2 m^2 \sin{\theta} \Big]
    \Biggl\} \, \mathcal{I}_5,
    \label{eq:fourPointFlow6}
\end{align}
where we have defined the threshold function
\begin{align}
    \mathcal{I}_5 \equiv \frac{1}{2} \int \frac{\df^4 p}{(2\pi)^4} \frac{p^2 \partial_t r_k^A}{\left[ p^2 (1+r_k^\psi)^2 + m^2 \right]^2 \left[ p^2 (1+r_k^A) \right]^3} = \frac{1}{2(4\pi)^2} \frac{1}{\left( k^2+m^2 \right)^2}.
\end{align}

Similarly, the trace of the ladder diagram is evaluated as
\begin{align}
    &\operatorname{tr} \Big[ 
    P_k(p) \, V^{\rm QCD,NL_2}_{k,\alpha\beta} \, P_k^\prime(p) \, V^{\rm QCD,NL_1}_{k,\mu\nu}
    \Big] 
    D_{r,\alpha\beta}(p) \left( D_{r,\mu\rho} \, \partial_t R^A_{k,\rho\sigma} \, D_{r,\sigma\nu} \right)_p 
    \nonumber\\
    &\quad
    = \operatorname{tr} \Big[ 
    \frac{-i \Slash{p} (1+r_k^\psi) + m e^{-i \gamma_5 \theta / 2}}{p^2 (1+r_k^\psi)^2 + m^2}
    \,
    \gamma_\mu T^a \psi_x \psi^\text{T}_x \gamma_\nu^\text{T} T^{\text{T},a}
    \,
    \frac{-i \Slash{p}^\text{T} (1+r_k^\psi) - m e^{-i \gamma_5 \theta / 2}}{p^2 (1+r_k^\psi)^2 + m^2}
    \,
    \gamma_\rho^\text{T} T^{\text{T},b} \bar{\psi}^\text{T}_x \bar{\psi}_x \gamma_\sigma T^b
    \nonumber\\
    &\qquad \times 
    D_{r,\alpha\beta}(p) \left( D_{r,\mu\rho} \, \partial_t R^A_{k,\rho\sigma} \, D_{r,\sigma\nu} \right)_p
    \nonumber\\
    &\quad
    = (-1)^2 \, \bar{\psi}_x \Big[
    \gamma_\rho T^b 
    \, 
    \frac{-i \Slash{p}^\text{T} (1+r_k^\psi) - m e^{-i \gamma_5 \theta / 2}}{p^2 (1+r_k^\psi)^2 + m^2}   
    \,
    \gamma_\nu T^a
    \Big] \psi_x
    \,
    \bar{\psi}_x \Big[ 
    \gamma_\sigma T^b
    \,
    \frac{-i \Slash{p} (1+r_k^\psi) + m e^{-i \gamma_5 \theta / 2}}{p^2 (1+r_k^\psi)^2 + m^2}
    \,
    \gamma_\mu T^a 
    \Big] \psi_x
    \nonumber\\
    &\qquad \times 
    D_{r,\alpha\beta}(p) \left( D_{r,\mu\rho} \, \partial_t R^A_{k,\rho\sigma} \, D_{r,\sigma\nu} \right)_p
    \nonumber\\
    &\quad
    \ni (-1)^3  \left( \frac{C_2}{4N_c} \right) \frac{1}{2 N_c}
    \nonumber\\
    &\qquad \times
    \Biggl\{ 
    \left( \bar{\psi} \psi \right)_x^2 \Big[ - \zeta^- p^2 (1+r_k^\psi)^2 - \zeta^+ m^2 \cos{\theta} \Big]
    + \left( \bar{\psi} i\gamma_5 \psi \right)_x^2 \Big[ - \zeta^- p^2 (1+r_k^\psi)^2 + \zeta^+ m^2 \cos{\theta} \Big]
    \nonumber\\
    &\qquad\qquad
    + \left( \bar{\psi} \psi \right)_x \left( \bar{\psi} i\gamma_5 \psi \right)_x \Big[ - 2 \zeta^+ m^2 \sin{\theta} \Big]
    \Biggl\} \frac{p^2 \partial_t r_k^A}{\left[ p^2 (1+r_k^\psi)^2 + m^2 \right]^2 \left[ p^2 (1+r_k^A) \right]^3}.
\end{align}
After collecting the anti-fermionic loop, the ``nonladder'' part reads
\begin{align}
    & - \frac{1}{2} \Tr \left[ \mathcal{P}_k \, \left( \mathcal{V}^{\rm QCD}_{k,\alpha\beta} \, D_{r,\alpha\beta} \right) \, \mathcal{P}_k \, \mathcal{V}^{\rm QCD}_{k,\mu\nu} \, D_{r,\mu\rho} \, \partial_t R^A_{k,\rho\sigma} \, D_{r,\sigma\nu} \right] \Biggl|_{\rm nonladder} 
    \nonumber\\
    &\quad
    \rightarrow  \int_x \left( \frac{g_s^4 C_2}{4N_c} \right) \frac{1}{2 N_c} \cdot 2
    \nonumber\\
    &\qquad \times
    \Biggl\{ 
    \left( \bar{\psi} \psi \right)_x^2 \Big[ - \zeta^- k^2 - \zeta^+ m^2 \cos{\theta} \Big]
    + \left( \bar{\psi} i\gamma_5 \psi \right)_x^2 \Big[ - \zeta^- k^2 + \zeta^+ m^2 \cos{\theta} \Big]
    + \left( \bar{\psi} \psi \right)_x \left( \bar{\psi} i\gamma_5 \psi \right)_x \Big[ - 2 \zeta^+ m^2 \sin{\theta} \Big]
    \Biggl\} \, \mathcal{I}_5.
    \label{eq:fourPointFlow7}
\end{align}

\subsubsection{Flow equation for four-fermion couplings}

Considering the ansatz of the effective average action \labelcref{eq:IRQCDaction}, the left-hand side of the flow equation \labelcref{eq:vertexExpansionFlow} is given by
\begin{align}
    \partial_t \Gamma_k = \int_x \Biggl\{ \left( \bar{\psi} \psi \right)_x^2 \partial_t \left( - \frac{G_S}{2} \right) + \left( \bar{\psi} i \gamma_5 \psi \right)_x^2 \partial_t \left( - \frac{G_P}{2} \right) + \left( \bar{\psi} \psi \right)_x \left( \bar{\psi} i \gamma_5 \psi \right)_x \partial_t \left( -C_4 \right) \Biggl\} + \cdots.
\end{align}
We read off the flow equations for $G_S$, $G_P$, and $C_4$ by comparing the coefficient with \cref{eq:fourPointFlow1,eq:fourPointFlow2,eq:fourPointFlow3,eq:fourPointFlow4,eq:fourPointFlow5,eq:fourPointFlow6,eq:fourPointFlow7}.
We find the flow equations as
\begin{align}
    \partial_t G_S =& - 8 N_c  N_f \Big( (G_S^2 + C_4^2) (k^2 - m^2) - 2 m^2 (G_S^2 - C_4^2) \cos{\theta} - 4 m^2 G_S C_4 \sin{\theta} \Big) \mathcal{I}_1 \nonumber\\
    & - 4 g_s^2 C_2 (3+\xi) \Big( G_S (k^2 - m^2) - 2 m^2 G_S \cos{\theta} - 2 m^2 C_4 \sin{\theta} \Big) \mathcal{I}_2 \nonumber\\
    & - 8 N_c (3+\xi)^2 \left( \frac{g_s^2 C_2}{4 N_c} \right)^2 \Big( k^2 - m^2 - 2 m^2 \cos{\theta} \Big) \mathcal{I}_3 \nonumber\\
    & + \frac{1}{N_c} \left( \frac{g_s^4 C_2}{4 N_c} \right) \Big( \zeta^- (k^2 - m^2) + 2 \zeta^+ m^2 \cos{\theta} \Big) \mathcal{I}_3 \nonumber\\
    & - 4 g_s^2 C_2 (3+\xi) \Big( G_S (k^2 - m^2\cos{\theta}) - m^2 C_4 \sin{\theta} \Big) \mathcal{I}_4 \nonumber\\
    &- 16 N_c (3+\xi)^2 \left( \frac{g_s^2 C_2}{4 N_c} \right)^2 \Big( k^2 - m^2 \cos{\theta} \Big) \mathcal{I}_5 \nonumber\\
    & + \frac{2}{N_c} \left( \frac{g_s^4 C_2}{4 N_c} \right) \Big( \zeta^- k^2 + \zeta^+ m^2 \cos{\theta} \Big) \mathcal{I}_5 ,
    \\
    \nonumber\\
    \partial_t G_P =& - 8 N_c  N_f \Big( (G_P^2 + C_4^2) (k^2 - m^2) + 2 m^2 (G_P^2 - C_4^2) \cos{\theta} - 4 m^2 G_P C_4 \sin{\theta} \Big) \mathcal{I}_1 \nonumber\\
    & - 4 g_s^2 C_2 (3+\xi) \Big( G_P (k^2 - m^2) + 2 m^2 G_P \cos{\theta} - 2 m^2 C_4 \sin{\theta} \Big) \mathcal{I}_2 \nonumber\\
    & - 8 N_c (3+\xi)^2 \left( \frac{g_s^2 C_2}{4 N_c} \right)^2 \Big( k^2 - m^2 + 2 m^2 \cos{\theta} \Big) \mathcal{I}_3 \nonumber\\
    & + \frac{1}{N_c} \left( \frac{g_s^4 C_2}{4 N_c} \right) \Big( \zeta^- (k^2 - m^2) - 2 \zeta^+ m^2 \cos{\theta} \Big) \mathcal{I}_3 \nonumber\\
    & - 4 g_s^2 C_2 (3+\xi) \Big( G_P (k^2 + m^2\cos{\theta}) - m^2 C_4 \sin{\theta} \Big) \mathcal{I}_4 \nonumber\\
    &- 16 N_c (3+\xi)^2 \left( \frac{g_s^2 C_2}{4 N_c} \right)^2 \Big( k^2 + m^2 \cos{\theta} \Big) \mathcal{I}_5, \nonumber\\
    & + \frac{2}{N_c} \left( \frac{g_s^4 C_2}{4 N_c} \right) \Big( \zeta^- k^2 - \zeta^+ m^2 \cos{\theta} \Big) \mathcal{I}_5,
    \\
    \nonumber\\
    \partial_t C_4 =& - 8 N_c  N_f \Big( (G_S + G_P) C_4 (k^2 - m^2) - 2 m^2 (G_S - G_P) C_4 \cos{\theta} - 2 m^2 (G_S G_P + C_4^2) \sin{\theta} \Big) \mathcal{I}_1 \nonumber\\
    & - 4 g_s^2 C_2 (3+\xi) \Big( C_4 (k^2 - m^2) - m^2 (G_S+G_P) \sin{\theta} \Big) \mathcal{I}_2 \nonumber\\
    & + 8 N_c (3+\xi)^2 \left( \frac{g_s^2 C_2}{4 N_c} \right)^2 \Big( 2 m^2 \sin{\theta} \Big) \mathcal{I}_3 \nonumber\\
    & - \frac{1}{N_c} \left( \frac{g_s^4 C_2}{4 N_c} \right) \Big( 2 \zeta^+ m^2 \sin{\theta} \Big) \mathcal{I}_3 \nonumber\\
    & - 4 g_s^2 C_2 (3+\xi) \Big( C_4 k^2 - m^2 (G_S+G_P) \sin{\theta} \Big) \mathcal{I}_4 \nonumber\\
    & + 16 N_c (3+\xi)^2 \left( \frac{g_s^2 C_2}{4 N_c} \right)^2 \Big( m^2 \sin{\theta} \Big) \mathcal{I}_5, \nonumber\\
    & - \frac{2}{N_c} \left( \frac{g_s^4 C_2}{4 N_c} \right) \Big( \zeta^+ m^2 \sin{\theta} \Big) \mathcal{I}_5.
\end{align}
For the dimensionless quantities as defined in \cref{eq:dimensionlessQuantities}, these equations become
\begin{align}
    \partial_t \tilde G_S =& 2 \tilde G_S - 8 N_c  N_f \Big( (\tilde G_S^2 + \tilde C_4^2) (1 - \tilde m^2) - 2 \tilde m^2 (\tilde G_S^2 - \tilde C_4^2) \cos{\theta} - 4 \tilde m^2 \tilde G_S \tilde C_4 \sin{\theta} \Big) \tilde{\mathcal{I}}_{(3)} \nonumber\\
    & - 4 g_s^2 C_2 (3+\xi) \Big( \tilde G_S (1 - \tilde m^2) - 2 \tilde m^2 \tilde G_S \cos{\theta} - 2 \tilde m^2 \tilde C_4 \sin{\theta} \Big) \tilde{\mathcal{I}}_{(3)} \nonumber\\
    & - 8 N_c (3+\xi)^2 \left( \frac{g_s^2 C_2}{4 N_c} \right)^2 \Big( 1 - \tilde m^2 - 2 \tilde m^2 \cos{\theta} \Big) \tilde{\mathcal{I}}_{(3)} \nonumber\\
    & + \frac{1}{N_c} \left( \frac{g_s^4 C_2}{4 N_c} \right) \Big( \zeta^- (1 - \tilde m^2) + 2 \zeta^+ \tilde m^2 \cos{\theta} \Big) \tilde{\mathcal{I}}_{(3)} \nonumber\\
    & - 4 g_s^2 C_2 (3+\xi) \Big( \tilde G_S (1 - \tilde m^2\cos{\theta}) - \tilde m^2 \tilde C_4 \sin{\theta} \Big) \tilde{\mathcal{I}}_{(2)} \nonumber\\
    &- 16 N_c (3+\xi)^2 \left( \frac{g_s^2 C_2}{4 N_c} \right)^2 \Big( 1 - \tilde m^2 \cos{\theta} \Big) \tilde{\mathcal{I}}_{(2)} \nonumber\\
    & + \frac{2}{N_c} \left( \frac{g_s^4 C_2}{4 N_c} \right) \Big( \zeta^- + \zeta^+ \tilde m^2 \cos{\theta} \Big) \tilde{\mathcal{I}}_{(2)} ,
    \\
    \nonumber\\
    \partial_t \tilde G_P =& 2 \tilde G_P - 8 N_c  N_f \Big( (\tilde G_P^2 + \tilde C_4^2) (1 - \tilde m^2) + 2 \tilde m^2 (\tilde G_P^2 - \tilde C_4^2) \cos{\theta} - 4 \tilde m^2 \tilde G_P \tilde C_4 \sin{\theta} \Big) \tilde{\mathcal{I}}_{(3)} \nonumber\\
    & - 4 g_s^2 C_2 (3+\xi) \Big( \tilde G_P (1 - \tilde m^2) + 2 \tilde m^2 \tilde G_P \cos{\theta} - 2 \tilde m^2 \tilde C_4 \sin{\theta} \Big) \tilde{\mathcal{I}}_{(3)} \nonumber\\
    & - 8 N_c (3+\xi)^2 \left( \frac{g_s^2 C_2}{4 N_c} \right)^2 \Big( 1 - \tilde m^2 + 2 \tilde m^2 \cos{\theta} \Big) \tilde{\mathcal{I}}_{(3)} \nonumber\\
    & + \frac{1}{N_c} \left( \frac{g_s^4 C_2}{4 N_c} \right) \Big( \zeta^- (1 - \tilde m^2) - 2 \zeta^+ \tilde m^2 \cos{\theta} \Big) \tilde{\mathcal{I}}_{(3)} \nonumber\\
    & - 4 g_s^2 C_2 (3+\xi) \Big( \tilde G_P (1 + \tilde m^2\cos{\theta}) - \tilde m^2 \tilde C_4 \sin{\theta} \Big) \tilde{\mathcal{I}}_{(2)} \nonumber\\
    &- 16 N_c (3+\xi)^2 \left( \frac{g_s^2 C_2}{4 N_c} \right)^2 \Big( 1 + \tilde m^2 \cos{\theta} \Big) \tilde{\mathcal{I}}_{(2)}, \nonumber\\
    & + \frac{2}{N_c} \left( \frac{g_s^4 C_2}{4 N_c} \right) \Big( \zeta^- - \zeta^+ \tilde m^2 \cos{\theta} \Big) \tilde{\mathcal{I}}_{(2)},
    \\
    \nonumber\\
    \partial_t \tilde C_4 =& 2 \tilde C_4 - 8 N_c  N_f \Big( (\tilde G_S + \tilde G_P) \tilde C_4 (1 - \tilde m^2) - 2 m^2 (\tilde G_S - \tilde G_P) \tilde C_4 \cos{\theta} - 2 \tilde m^2 (\tilde G_S \tilde G_P + \tilde C_4^2) \sin{\theta} \Big) \tilde{\mathcal{I}}_{(3)} \nonumber\\
    & - 4 g_s^2 C_2 (3+\xi) \Big( \tilde C_4 (1 - \tilde m^2) - \tilde m^2 (\tilde G_S+\tilde G_P) \sin{\theta} \Big) \tilde{\mathcal{I}}_{(3)} \nonumber\\
    & + 8 N_c (3+\xi)^2 \left( \frac{g_s^2 C_2}{4 N_c} \right)^2 \Big( 2 \tilde m^2 \sin{\theta} \Big) \tilde{\mathcal{I}}_{(3)} \nonumber\\
    & - \frac{1}{N_c} \left( \frac{g_s^4 C_2}{4 N_c} \right) \Big( 2 \zeta^+ \tilde m^2 \sin{\theta} \Big) \tilde{\mathcal{I}}_{(3)} \nonumber\\
    & - 4 g_s^2 C_2 (3+\xi) \Big( \tilde C_4 - \tilde m^2 (\tilde G_S+\tilde G_P) \sin{\theta} \Big) \tilde{\mathcal{I}}_{(2)} \nonumber\\
    & + 16 N_c (3+\xi)^2 \left( \frac{g_s^2 C_2}{4 N_c} \right)^2 \Big( \tilde m^2 \sin{\theta} \Big) \tilde{\mathcal{I}}_{(2)}, \nonumber\\
    & - \frac{2}{N_c} \left( \frac{g_s^4 C_2}{4 N_c} \right) \Big( \zeta^+ \tilde m^2 \sin{\theta} \Big) \tilde{\mathcal{I}}_{(2)},
\end{align}
where the dimensionless threshold functions are
\begin{align}
    &\tilde{\mathcal{I}}_{(2)} = k^2 \mathcal{I}_4 = k^4 \mathcal{I}_5 = \frac{1}{2(4\pi)^2}\frac{1}{\left( 1 + \tilde m^2 \right)^2},
    \nonumber\\
    &\tilde{\mathcal{I}}_{(3)} = \mathcal{I}_1 = k^2 \mathcal{I}_2 = k^4 \mathcal{I}_3 = \frac{1}{2(4\pi)^2}\frac{1}{\left( 1 + \tilde m^2 \right)^3}.
    \label{eq:thresholdfunction2}
\end{align}

In \cref{eq:flowequationsinmaintext}, we introduce following coefficient functions
\begin{subequations}
    \label{eq:coefficientfunctions}
\begin{align}
    \mathcal A_S(\tilde m, \theta) = &-4 C_2 (3+\xi) \Big( (1 -  (1+2\cos\theta)\tilde m^2)\tilde{\mathcal{I}}_{(3)} + (1 - \tilde m^2\cos{\theta})\tilde{\mathcal{I}}_{(2)} \Big) ,\\
    \mathcal B_S(\tilde m, \theta) = &+ 4 C_2 (3+\xi) \tilde m^2 \sin{\theta} \Big(  2 \tilde{\mathcal{I}}_{(3)} + \tilde{\mathcal{I}}_{(2)} \Big),\\
    \mathcal C_S(\tilde m, \theta) = &- 8 N_c (3+\xi)^2 \left( \frac{C_2}{4 N_c} \right)^2 \Big( 1 - \tilde m^2 - 2 \tilde m^2 \cos{\theta} \Big) \tilde{\mathcal{I}}_{(3)} + \frac{1}{N_c} \left( \frac{C_2}{4 N_c} \right) \Big( \zeta^- (1 - \tilde m^2) + 2 \zeta^+ \tilde m^2 \cos{\theta} \Big) \tilde{\mathcal{I}}_{(3)}\nonumber\\
    &\quad - 16 N_c (3+\xi)^2 \left( \frac{C_2}{4 N_c} \right)^2 \Big( 1 - \tilde m^2 \cos{\theta} \Big) \tilde{\mathcal{I}}_{(2)} + \frac{2}{N_c} \left( \frac{C_2}{4 N_c} \right) \Big( \zeta^- + \zeta^+ \tilde m^2 \cos{\theta} \Big) \tilde{\mathcal{I}}_{(2)},
    \\
    \mathcal A_P(\tilde m, \theta) = &-4 C_2 (3+\xi) \Big( (1 -  (1-2\cos\theta)\tilde m^2)\tilde{\mathcal{I}}_{(3)} + (1 + \tilde m^2\cos{\theta})\tilde{\mathcal{I}}_{(2)} \Big), \\
    \mathcal B_P(\tilde m, \theta) = &+ 4 C_2 (3+\xi) \tilde m^2 \sin{\theta} \Big(  2 \tilde{\mathcal{I}}_{(3)} + \tilde{\mathcal{I}}_{(2)} \Big) = \mathcal B_S(\tilde m, \theta), \\
    \mathcal C_P(\tilde m, \theta) = &- 8 N_c (3+\xi)^2 \left( \frac{C_2}{4 N_c} \right)^2 \Big( 1 - \tilde m^2 + 2 \tilde m^2 \cos{\theta} \Big) \tilde{\mathcal{I}}_{(3)} + \frac{1}{N_c} \left( \frac{C_2}{4 N_c} \right) \Big( \zeta^- (1 - \tilde m^2) - 2 \zeta^+ \tilde m^2 \cos{\theta} \Big) \tilde{\mathcal{I}}_{(3)}\nonumber\\
    &\quad - 16 N_c (3+\xi)^2 \left( \frac{C_2}{4 N_c} \right)^2 \Big( 1 + \tilde m^2 \cos{\theta} \Big) \tilde{\mathcal{I}}_{(2)} + \frac{2}{N_c} \left( \frac{C_2}{4 N_c} \right) \Big( \zeta^- - \zeta^+ \tilde m^2 \cos{\theta} \Big) \tilde{\mathcal{I}}_{(2)},
    \\
    \mathcal A_4(\tilde m, \theta) = &-4 C_2 (3+\xi) \Big( (1 - \tilde m^2) \tilde{\mathcal{I}}_{(3)} + \tilde{\mathcal{I}}_{(2)} \Big) ,\\
    \mathcal B_4(\tilde m, \theta) = &+ 4 C_2 (3+\xi) \tilde m^2 \sin{\theta} \Big( \tilde{\mathcal{I}}_{(3)} + \tilde{\mathcal{I}}_{(2)} \Big) ,\\
    \mathcal C_4(\tilde m, \theta) = &+ 16 N_c (3+\xi)^2 \left( \frac{C_2}{4 N_c} \right)^2 \tilde m^2 \sin{\theta} \Big( \tilde{\mathcal{I}}_{(3)} + \tilde{\mathcal{I}}_{(2)}  \Big) - \frac{2}{N_c} \left( \frac{C_2}{4 N_c} \right) \zeta^+ \tilde m^2 \sin{\theta} \Big( \tilde{\mathcal{I}}_{(3)} + \tilde{\mathcal{I}}_{(2)}  \Big) .
\end{align}
\end{subequations}

\section{Mean-field analysis}
\label{sec:meanfield}

In this appendix, we perform the mean-field analysis on a $U(1)$ NJL-type model with a $CP$-violating operator.
The starting action is the four-fermion sector in the Euclidean version of the action \labelcref{eq:UVQCDaction}.
By using the Hubbard-Stratinovich transformation, the action is written in the form of a Higgs-Yukawa type theory as
\begin{align}
S_\text{NJL}&= \int_x \bigg[\bar{\psi} \left( \gamma^\mu \partial_\mu + m e^{i\gamma_5 \theta/2} \right) \psi
+ \left( \sigma +  \frac{C_4}{G_P}\eta \right)\bar\psi \psi + \left( \eta + \frac{C_4}{G_S} \sigma \right) \bar\psi i\gamma^5 \psi\nonumber\\
&\quad
+ \frac{m\cos\theta}{G_S} \sigma  + \frac{m\sin\theta}{G_P} \eta
+ \frac{C_4}{G_SG_P} \sigma \eta
+ \frac{1}{2G_S}\sigma^2 + \frac{1}{2G_P}\eta^2\bigg] \,,
\end{align}
where $\sigma(x)$ and $\eta(x)$ are auxiliarry fields, see \cref{eq:HStrans}.
Replacing the auxiliary fields by homogeneous mean fields, $\sigma(x)=\bar\sigma$ and $\eta(x)=\bar\eta$, and integrating out the fermion fields, we obtain the effective potential for $\bar\sigma$ and $\bar\eta$:
\begin{align}
V_\text{eff}(\bar\sigma,\bar\eta) &= \frac{m\cos\theta}{G_S} \bar\sigma  + \frac{m\sin\theta}{G_P} \bar\eta
+ \frac{C_4}{G_SG_P} \bar\sigma \bar\eta
+ \frac{1}{2G_S}\bar\sigma^2 + \frac{1}{2G_P}\bar\eta^2\nonumber\\
&\quad
-\Tr\log\left[ i\Slash p 
+ \left( m\cos\theta +\bar\sigma +  \frac{C_4}{G_P}\bar\eta \right)+ \left( m\sin\theta  +\bar\eta + \frac{C_4}{G_S} \bar\sigma \right) i\gamma^5
\right]\nonumber\\
&=\frac{m\cos\theta}{G_S} \bar\sigma  + \frac{m\sin\theta}{G_P} \bar\eta
+ \frac{C_4}{G_SG_P} \bar\sigma\bar \eta
+ \frac{1}{2G_S}\bar\sigma^2 + \frac{1}{2G_P}\bar\eta^2\nonumber\\
&\quad
- 2\int \frac{\df^4p}{(2\pi)^4}\log\left[ p^2 
+ \left( m\cos\theta +\bar\sigma +  \frac{C_4}{G_P}\bar\eta \right)^2+ \left( m\sin\theta  +\bar\eta + \frac{C_4}{G_S} \bar\sigma \right)^2
\right]\nonumber\\
&=\frac{m\cos\theta}{G_S} \bar\sigma  + \frac{m\sin\theta}{G_P} \bar\eta
+ \frac{C_4}{G_SG_P} \bar\sigma \bar\eta
+ \frac{1}{2G_S}\bar\sigma^2 + \frac{1}{2G_P}\bar\eta^2\nonumber\\
&\quad
-\frac{1}{(4\pi)^2}
\Bigg[
\Lambda^2\left\{\left( m\cos\theta +\bar\sigma +  \frac{C_4}{G_P}\bar\eta \right)^2+ \left( m\sin\theta  +\bar\eta + \frac{C_4}{G_S} \bar\sigma \right)^2\right\}
\nonumber\\
&\qquad
+\Lambda^4\log\left( 1+ \tfrac{\left( m\cos\theta +\bar\sigma +  \frac{C_4}{G_P}\bar\eta \right)^2+ \left( m\sin\theta  +\bar\eta + \frac{C_4}{G_S} \bar\sigma \right)^2}{\Lambda^2} \right)
\nonumber\\
&\qquad
- \left\{\left( m\cos\theta +\bar\sigma +  \frac{C_4}{G_P}\bar\eta \right)^2+ \left( m\sin\theta  +\bar\eta + \frac{C_4}{G_S} \bar\sigma \right)^2\right\}^2\log\left(1+\tfrac{\Lambda^2}{\left( m\cos\theta +\bar\sigma +  \frac{C_4}{G_P}\bar\eta \right)^2+ \left( m\sin\theta  +\bar\eta + \frac{C_4}{G_S} \bar\sigma \right)^2} \right)
\Bigg]\,.
\label{eq:MFAfull}
\end{align}
Here, we have introduced a UV momentum cutoff, $|p|\leq \Lambda$.

\begin{figure}[t]
    \centering
    \includegraphics[width=0.45\linewidth]{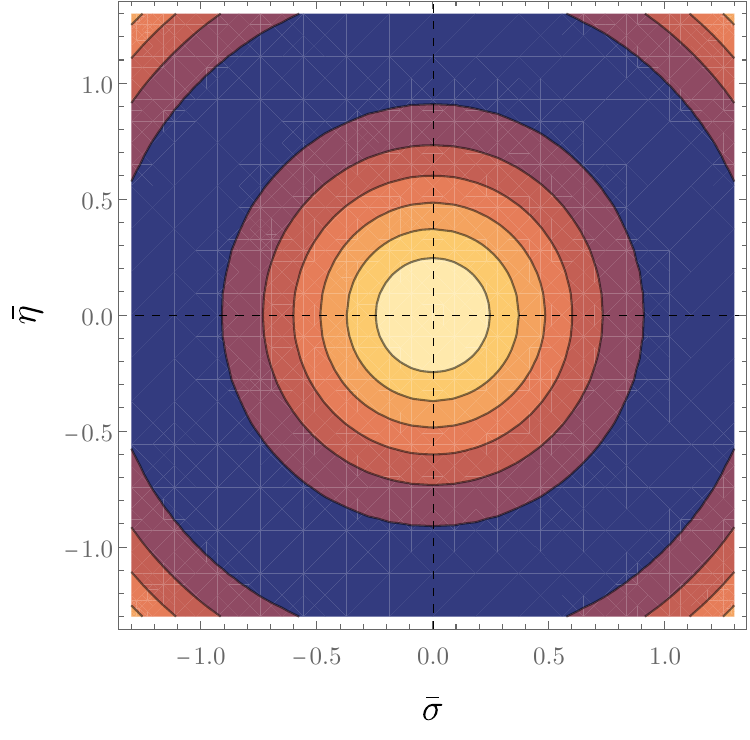}
    \hspace{5ex}
    \includegraphics[width=0.45\linewidth]{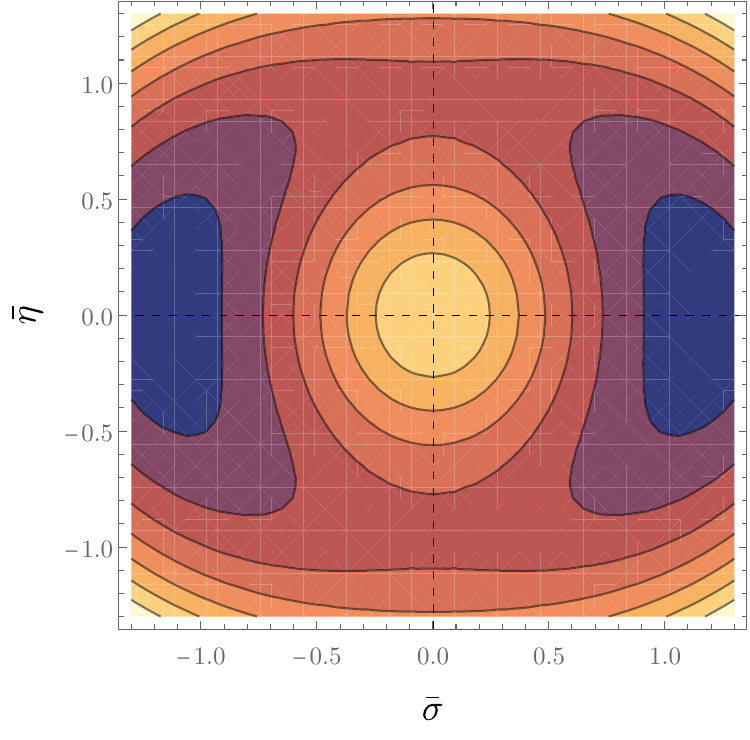}
    \includegraphics[width=0.45\linewidth]{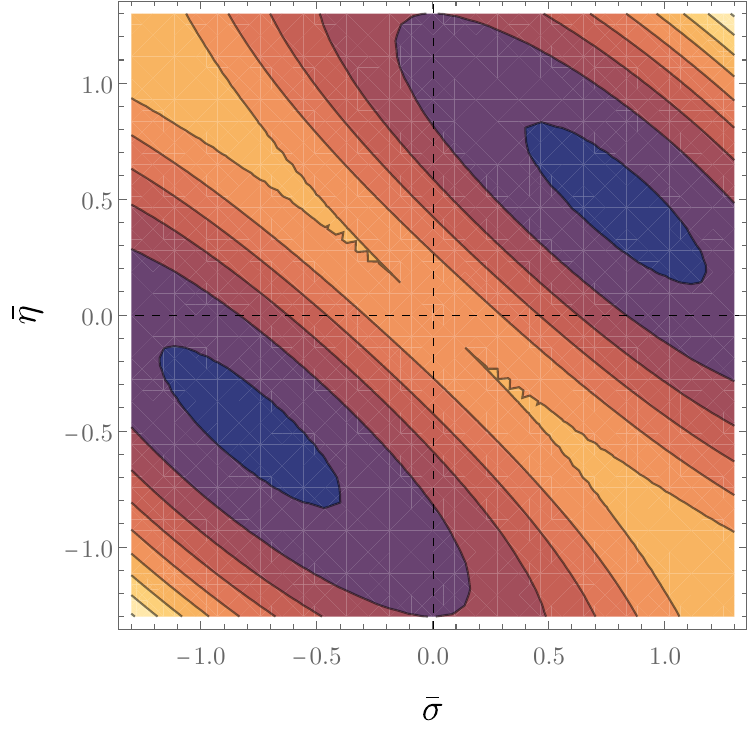}
    \hspace{5ex}
    \includegraphics[width=0.45\linewidth]{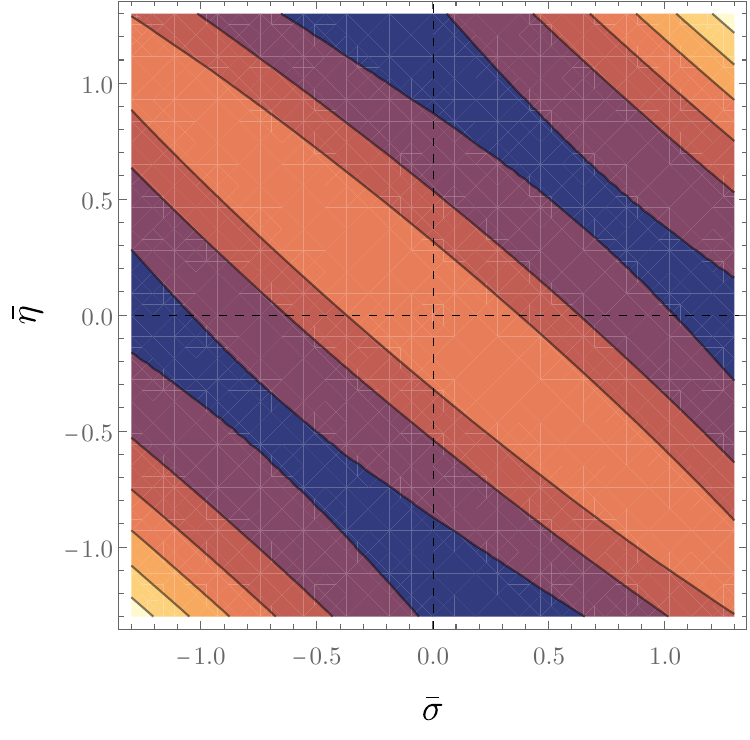}
    \caption{
    The effective potential \labelcref{eq:MFAfull} within the mean-field approximation for $m=\theta=0$. 
    Top-left: $G_S=G_P=(4\pi)^2$ and $C_4=0$. 
    Top-right: $G_S=(4\pi)^2$, $G_P=3(4\pi)^2/4$ and $C_4=0$. Bottom-left: $G_S=(4\pi)^2$, $G_P=3(4\pi)^2/4$ and $C_4=10\pi^2$. 
    Bottom-right: $G_S=(4\pi)^2$, $G_P=3(4\pi)^2/4$ and $C_4=15\pi^2$.
    }
    \label{fig:MFA:allzeo}
\end{figure}

First, we consider the simplified case with $m=\theta=C_4=0$.
In particular, when $G_S=G_P$, the system possesses a $U(1)$ chiral symmetry.
The critical coupling for spontaneous symmetry breaking is given by $G_S^*=G_P^*=(4\pi)^2\Lambda/2$.
This critical value corresponds to the fixed-point coupling in the fRG analysis. The difference between the critical coupling obtained here and the fixed-point value in \cref{eq:fixedPointsInGs} originates from the choice of regulator (cutoff scheme).
In \Cref{fig:MFA:allzeo}, we depict the effective potential by setting $m=0$ and $\theta=0$.
One sees that the local minimum (corresponding to the symmetric vacuum) is located at the origin $(\bar\sigma,\bar\eta)=(0,0)$.
In the case with $G_S=G_P$ and $C_4=0$, the contour plot exhibits the expected rotationally symmetric structure in the $(\bar\sigma,\bar\eta)$ plane, reflecting the underlying $U(1)$ chiral symmetry. 
Once $G_S\neq G_P$, this degeneracy is lifted and the potential becomes anisotropic, such that the scalar and pseudoscalar directions are no longer equivalent. 
Turning on a nonvanishing $C_4$ further introduces an explicit mixing between $\bar\sigma$ and $\bar\eta$, which is visible in the tilted valley structure of the effective potential. 
In this way, the contour plots provide a direct visualization of how the interaction proportional to $C_4$ correlates the scalar and pseudoscalar channels already at the mean-field level.

\begin{figure*}[t]
    \centering
    \includegraphics[width=0.45\linewidth]{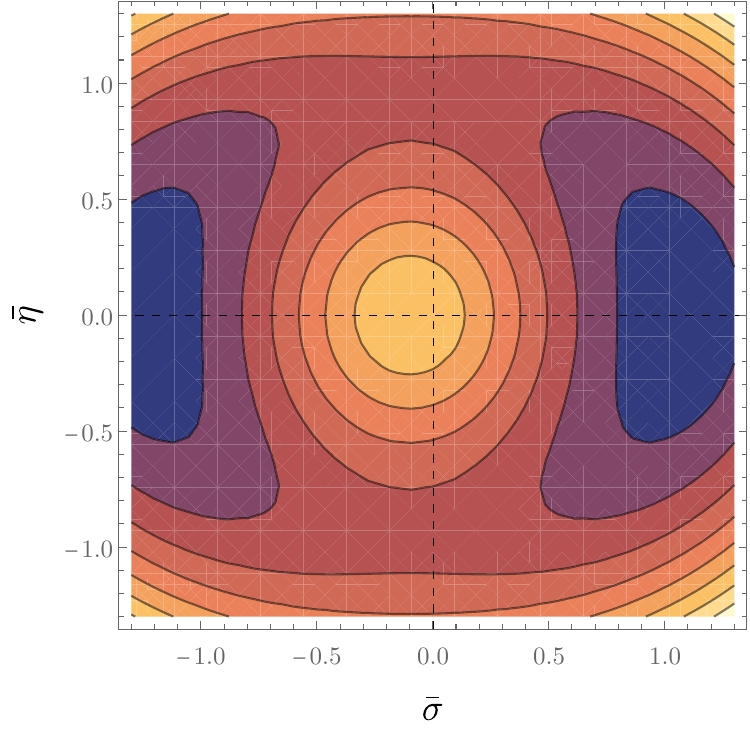}
    \hspace{5ex}
    \includegraphics[width=0.45\linewidth]{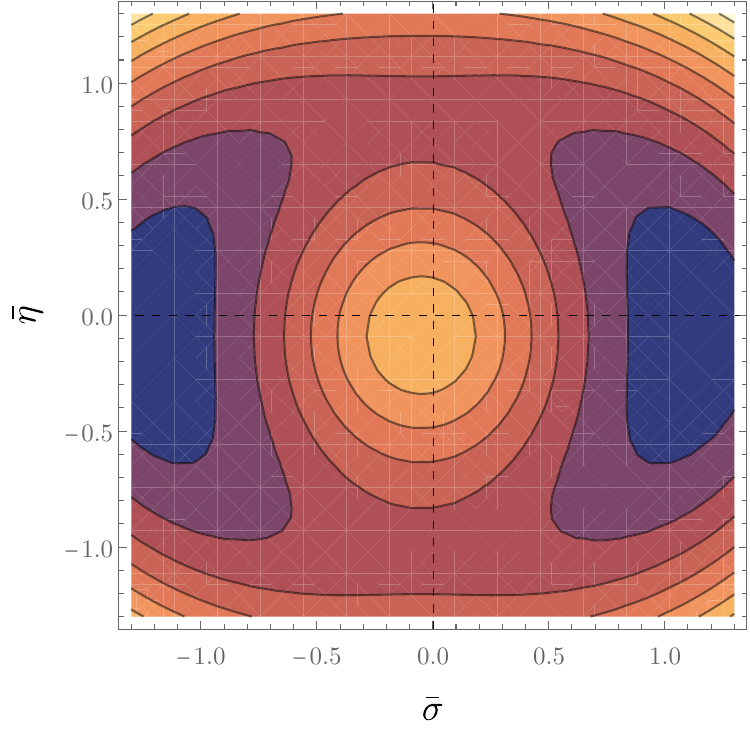}
    \includegraphics[width=0.45\linewidth]{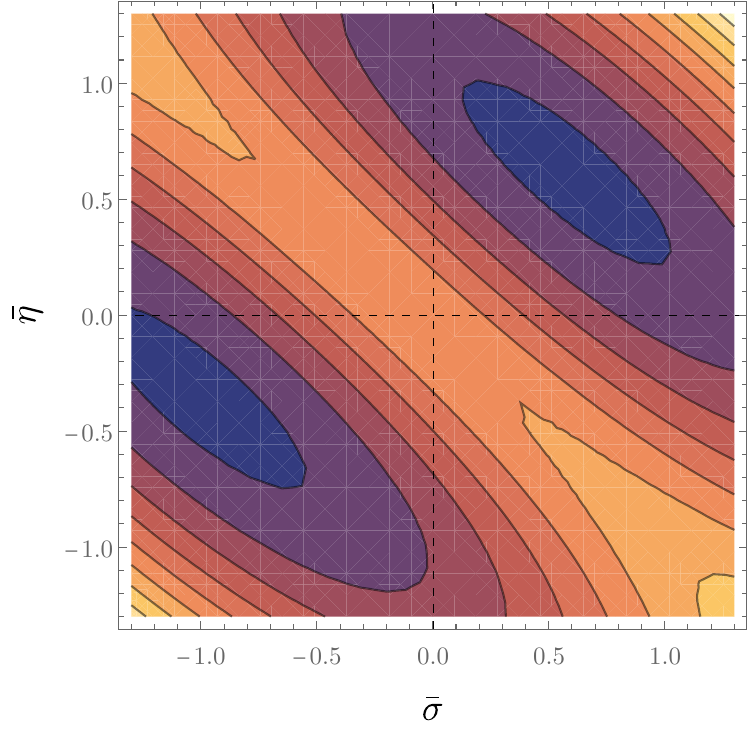}
    \hspace{5ex}
    \includegraphics[width=0.45\linewidth]{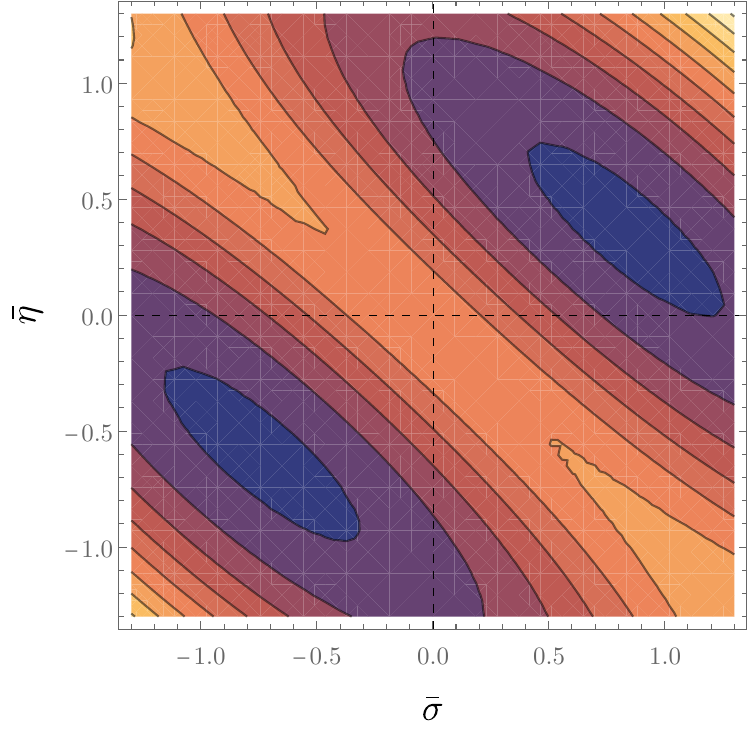}
    \caption{
    The effective potential \labelcref{eq:MFAfull} within the mean-field approximation. 
    Top-Left: $G_S=(4\pi)^2$, $G_P=3(4\pi)^2/4$, $C_4=0$, $m=0.1$ and $\theta=0$. 
    Top-right: $G_S=(4\pi)^2$, $G_P=3(4\pi)^2/4$, $C_4=0$, $m=0.1$ and $\theta=\pi/3$.
    Bottom-left: $G_S=(4\pi)^2$, $G_P=3(4\pi)^2/4$, $C_4=10\pi^2$, $m=0.1$ and $\theta=0$. 
    Bottom-right: $G_S=(4\pi)^2$, $G_P=3(4\pi)^2/4$, $C_4=15\pi^2$, $m=0$ and $\theta=\pi/3$.
    }
    \label{fig:MFA:zeroc4}
\end{figure*}

Next, we consider the case with nonvanishing current mass and phase. In the top panels of \Cref{fig:MFA:zeroc4}, we show the effective potential for $C_4=0$ and a fixed value $m=0.1$ for $\theta=0$ and $\theta=\pi/3$. 
For $\theta=0$, the local minimum, which would correspond to the symmetric vacuum in the chiral limit ($m=0$), is shifted to a nonzero value of $\bar{\sigma}$, while $\langle \bar{\eta} \rangle =0$. 
Turning on a nonvanishing $\theta$ shifts the local minimum toward a nonzero value of $\bar{\eta}$. 
In particular, for $\theta=\pi/2$, the local minimum is located at $\langle \bar{\eta} \rangle \neq 0$ and $\langle \bar{\sigma} \rangle =0$.
These qualitative features remain qualitatively unchanged when $C_4\neq 0$, in the sense that a nonvanishing $\theta$ still shifts the preferred vacuum direction from the scalar channel toward the pseudoscalar one. 
However, the role of $C_4$ is not merely quantitative: through the off-diagonal coupling between $\bar\sigma$ and $\bar\eta$, it modifies the orientation of the valleys of the effective potential and hence the detailed pattern of scalar--pseudoscalar mixing. 
The mean-field analysis therefore indicates that $\theta$ controls the direction of vacuum alignment, while $C_4$ controls how strongly the two channels are entangled.

\twocolumngrid
\bibliographystyle{JHEP} 
\bibliography{refs}
\end{document}